\documentclass{aa}

\makeatletter

\renewcommand*\aa@manuscriptname{%
  manuscript no.~aa62618-26%
  \hspace{\stretch{1}}%
  \copyright ESO \the\year
}

\renewcommand*\aa@textidlineempty{%
  {\slshape A\&A proofs:}\ manuscript no.~aa62618-26%
}
\makeatother

\usepackage{graphicx}
\usepackage{txfonts}
\usepackage{lipsum}
\usepackage{float}
\usepackage{subcaption}
\usepackage{lscape}
\usepackage{placeins}
\usepackage{natbib}
\usepackage{hyperref}
\usepackage{cleveref}
\usepackage{float}
\usepackage[compatibility=false]{caption}
\usepackage{xspace}
\usepackage{amsmath}
\usepackage{enumitem}
\usepackage[dvipsnames]{xcolor}
\usepackage{longtable}
\bibpunct{(}{)}{;}{a}{}{,}

\hypersetup{
colorlinks=true,
linkcolor=blue,
citecolor=blue,
filecolor=magenta,
urlcolor=blue
}

\newcommand{\fluxlinesunit}{\ensuremath{\text{erg s}^{-1}\text{ cm}^{-2}}\xspace}

\newcommand{\Hbeta}{\ensuremath{\text{H}\beta}\xspace}
\newcommand{\OIIIHbeta}{\ensuremath{\text{H}\beta}+[\text{O\,\textsc{iii}}]\xspace} 
\newcommand{\Halpha}{\ensuremath{\text{H}\alpha}\xspace}
\newcommand{\OIIIline}[1]{\ensuremath{[\mathrm{O\,III}]\,\lambda#1}}
\newcommand{\NIIline}{\ensuremath{[\mathrm{N\,II}]\,\xspace}}
\newcommand{\OIIline}[1]{\ensuremath{[\mathrm{O\,II}]\,\lambda#1}\xspace}

\newcommand{\z}[1]{\ensuremath{z_{\mathrm{#1}}}}
\newcommand{\massunit}{\ensuremath{\text{M}\odot}\xspace}       
\newcommand{\sfrunit}{\ensuremath{\text{M}\odot\,\text{yr}^{-1}}\xspace}
\newcommand{\ssfrunit}{\ensuremath{\text{yr}^{-1}}\xspace}
\newcommand{\EWrest}{\ensuremath{\text{EW}{\mathrm{rest}}}} 
\newcommand{\EWobs}{\ensuremath{\text{EW}{\mathrm{obs}}}}
\newcommand{\arcminsq}{\ensuremath{\text{arcmin}^2}\xspace}
\newcommand{\I}{\ensuremath{\text{I}_{\mathrm{F814W}}}\xspace}
\newcommand{\J}{\ensuremath{\text{J}_{\mathrm{F125W}}}\xspace}
\newcommand{\Hfilter}{\ensuremath{\text{H}_{\mathrm{F160W}}}\xspace}
\newcommand{\percent}{$\%$~}
\newcommand{\updown}[3]{\ensuremath{#1_{#3}^{#2}}\xspace}
\newcommand{\logmass}{\ensuremath{\log(M_{\star}/\mathrm{M}_{\odot})}\xspace}

\begin{document}

\title{Tracing the Epoch of Reionization through extreme emission line galaxies around Cosmic Noon}

\author{M. Delgado Mancheño\inst{1,2}  \corrauth{mariadelgadomancheno@gmail.com} \and
J. Gallego \inst{1,2}\and
S. Pascual\inst{1,2}\and
M. Tang \inst{3,4}\and
R. Pello\inst{5} \and
N. Laporte\inst{5} \and
A. Camps-Fariña\inst{1,6} \\
\and M. Chamorro-Cazorla\inst{1,2}
}

\institute{
Departamento de Física de la Tierra y Astrofísica, Universidad Complutense de Madrid, 28040 Madrid, Spain
\and
Instituto de Física de Partículas y del Cosmos (IPARCOS), Facultad de Ciencias Físicas, Universidad Complutense de Madrid, E-28040 Madrid, Spain
\and
Tsung-Dao Lee Institute, Shanghai Jiao Tong University, 1 Lisuo Road, Shanghai 201210, People’s Republic of China
\and
School of Physics and Astronomy, Shanghai Jiao Tong University, 800 Dongchuan Road, Shanghai 200240, People’s Republic of China
\and
Aix Marseille Univ., CNRS, CNES, LAM (Laboratoire d'Astrophysique de Marseille), Marseille, France
\and
ISDEFE, Beatriz de Bobadilla 3, E-28040 Madrid, Spain
}

\authorrunning{M. Delgado Mancheño et al.}
\date{Accepted for publication in A$\&$A (06 Nov 2026).}

\abstract
{Star-forming galaxies at z$>$6 are thought to be the main drivers of cosmic reionization, but the faintness of these systems makes their detailed study challenging even with JWST. Extreme Emission Line Galaxies (EELGs) are compact, low-mass systems with intense nebular emission that represent potential lower-redshift analogs of this population.}
{We aim to assemble and characterize a statistically significant sample of EELGs in the CANDELS/EGS field around the cosmic noon, and to assess their validity as low-mass analogs of the star-forming galaxies that drove the Epoch of Reionization (EoR).}
{We select EELG candidates from CANDELS multiband photometry using a $J$-band color-excess criterion, which identifies both \OIIIHbeta\ and \Halpha\ emitters. Physical properties are obtained from the CANDELS data-products, and the photometric selection is validated with spectroscopic follow-up from MMT/MMIRS, Keck/MOSFIRE, HST/WFC3 and JWST/NIRSpec data, and complemented by our own GTC/OSIRIS+ observational program.}
{We identify 102 \OIIIHbeta\ emitters at $z \sim 1.7$ and 127 \Halpha\ emitters at $z \sim 1.0$. These galaxies are compact ($\sim 1.6$\,kpc), of low stellar mass  ($\sim$ 10$^{8}$ \massunit) and strongly star-forming ($\sim$ 10$^{-8}$\,\ssfrunit). We estimate sample purities of  $84.3 \pm 1.5$\% \OIIIHbeta and $76.8 \pm 2.0$\percent \Halpha, consistent with the spectroscopic confirmation rates ($94\%$ for \OIIIHbeta\ and $63\%$ for \Halpha), and find no evidence for significant AGN contamination. A joint Bayesian fit, corrected for observational completeness via a truncated likelihood, yields a significantly positive slope, $\log(M_\star/\mathrm{M}_\odot) = (6.71 \pm 0.16) + (3.31 \pm 0.46)\,\log(1+z)$, suggesting a possible evolutionary trend in the characteristic mass of our sample across the probed cosmic time.}
{We spectroscopically confirm a highly pure, AGN-free sample of cosmic noon EELGs. While their mass scale evolves with redshift, their properties consistently match $z>5$ JWST extreme emitters, establishing them as intermediate-redshift analogs.}

\maketitle
\nolinenumbers

\section{Introduction}

The Epoch of Reionization (EoR) is a key milestone in the study of cosmic structure formation, as it marks the emergence of the first luminous objects, whether they are star-forming galaxies or active galactic nuclei (AGN) \citep{Fan_2006}. Star-forming galaxies are regarded as the leading candidates to supply the bulk of the ionizing photons that drove this transition \citep{llerena_physical_2024}. However, the physical properties of these sources (e.g. stellar mass, star-formation rate, morphology) and the relative contribution of the different source types are still poorly constrained \citep{Robertson_2015, Endsley_2023, Davis_2024,llerena_2026}.

The advent of the James Webb Space Telescope (JWST) has enabled the detection and spectroscopic study of galaxies at \z{}$>$6, revealing a young, low-mass population with intense nebular emission (e.g., \citealp{rinaldi2024}). Even for JWST, however, these sources are extremely faint and compact, which complicates the detailed characterization of their physical properties. A primary challenge is the degeneracy of photometric signatures, as similar color excesses can arise from distinct populations. While these excesses may trace the intense nebular emission of young galaxies driving reionization \citep{rinaldi2024, boyett2024a}, they can also be produced by lower-redshift interlopers. For instance, sources exhibiting a prominent Balmer $4000$ \AA\ break can mimic the colors of \z{}$>$7 galaxies \citep{Trussler_2023}. Additionally, Little Red Dots (LRDs)—which are compact, heavily dust-obscured sources—can successfully replicate the Lyman break \citep{pgpg_2024}. Because the extreme compactness of these interlopers is often comparable to the instrumental point spread function, the flux excesses commonly attributed to star formation introduce severe degeneracies in spectral energy distribution (SED) modeling \citep{Adams_2023, Trussler_2023, Naidu_2026}.  
This suggests searching for similar but closer galaxies in order to study their properties in greater detail and gain a deeper understanding of the population of galaxies that drove cosmic reionization \citep{tang2023,Daikuhara_2025A,davis2026}.

Several studies suggest the existence of lower-redshift galaxies that potentially share the same characteristics as high-redshift populations \citep{onodera_broadband_2020}. These systems are characterized by an optical spectrum dominated by unusually intense nebular emission lines, with rest-frame equivalent widths (EW) exceeding $100$\AA, low stellar masses (<$10^{9}$ \massunit), high specific star-formation rates (sSFR) of $10$--$100$ Gyr$^{-1}$, very young stellar populations ($<10$--$50$ Myr), subsolar metallicities, and extremely compact morphologies (\citealt{amorin_extreme_2015,Calabro_2017, boyett2024a, gimenez_alcazar_2026}). The study of extreme emission-line galaxies \citep[EELGs;][hereafter VDW]{vanderWel2011} is essential, as they serve as accessible, potential lower-redshift analogs. In fact, at very high redshift ($z>6$), typical low-mass galaxies are expected to be EELGs, playing a crucial role in the reionization of the Universe \citep{lumbreras_2022}. These properties make EELGs excellent candidates for studying the low-mass galaxies thought to be the main drivers of the EoR \citep{withers2023}.

The EELGs have been identified across a wide range of epochs, from the low redshift Universe (the "Green Peas", \citealt{amorin_extreme_2015}) through cosmic noon and up to the EoR itself. Although their number density increases with redshift \citep{boyett_o_2022,rosenwasser_2022}, they remain rare in the low redshift Universe. Moreover, EELGs at intermediate redshift may be related to the typical star-forming galaxies found at $z \sim 2$, the peak of the cosmic star-formation rate density \citep{Madau_2014,Maseda_2014, cohn_zfourge_2018}. Thus, characterizing this population is of particular importance.

The most widely used technique to identify EELGs at intermediate redshift is the broadband color-excess method (e.g., \citealt{MMT_Tang_2019,Huang_2015}). This approach is effective at identifying galaxies whose extreme line emission overlaps with a stellar continuum that is too faint to be detected through traditional spectroscopy \citep{withers2023}. VDW were among the first to select EELGs at $z \sim 1.7$, by searching for a color excess in the $J$ band relative to the adjacent $I$ and $H$ bands, produced by the \OIIIHbeta\ lines. Due to the fact that VDW includes the photometric uncertainties, the parameter space explored by their criterion is S/N dependent, which can sacrifice completeness of the sample. In exchange for biasing the sample toward more extreme EELGs, sample purity is preserved. Exploring further combinations of emission lines and z ranges would also allow for larger potential datasets and enable studies on the cosmic evolution of EELGs.

In this work we explored the multiwavelength imaging of the Extended Groth Strip (EGS) field from the Cosmic Assembly Near-IR Deep Extragalactic Legacy Survey (CANDELS) survey \citep{grogin_2011,koekemoer_candels_2011, Stefanon_2017} to assemble and characterize a sample of EELGs at cosmic noon. Our goal is to characterize the physical properties of a statistically significant EELG population at intermediate redshift and to assess its validity as a low-mass analog of the star-forming galaxies that drove the EoR.

The paper is organized as follows. Sect.~\ref{section:data_sample} describes the data, the sample selection and presents the spectroscopic data compiled from the literature alongside our own follow-up program with OSIRIS+ at the Gran Telescopio Canarias (GTC); Sect.~\ref{sec:validation} assesses the reliability and quantifies the residual contamination of the sample; Sect.~\ref{sec:physical_morf} characterizes the physical and morphological properties of the selected EELGs; Sect.~\ref{sec:connection_eor} investigates the cosmic downsizing scenario and explores the role of our EELGs as analogs to high-redshift galaxies; and Sect.~\ref{sec:conclusions} summarizes our main conclusions. Throughout this paper, we assume a flat $\Lambda$CDM cosmology with $H_0 = 70\,\mathrm{km\,s^{-1}\,Mpc^{-1}}$, $\Omega_m = 0.3$, and $\Omega_\Lambda = 0.7$, alongside a Chabrier initial mass function (IMF) \citep{Chabrier_2003}. All magnitudes are expressed in the AB system.

\section{Data and sample selection}
\label{section:data_sample}

\subsection{CANDELS survey}
\label{sec:candels_survey}

The CANDELS survey is designed to document the evolution of galaxies in the redshift range between 1.5 and 8 through deep imaging of more than 250000 galaxies with the Advanced Camera for Surveys (ACS), the Wide Field Camera 3 infrared channel (WFC3/IR), and the ultraviolet/optical channel (WFC3/UVIS) of the Hubble Space Telescope (HST). CANDELS covers the GOODS-N, GOODS-S, COSMOS, UDS, and EGS fields, reaching a total area of $\sim$800 \arcminsq \citep{koekemoer_candels_2011, grogin_2011}. In this work we focus on the EGS field, which provides the second largest individual area among the CANDELS fields, 206 \arcminsq, distributed along a relatively narrow strip of $6.5' \times 32.5'$. It is slightly deeper than the other CANDELS fields, reaching a $5\sigma$ point-source detection limit of $27.6$ AB mag in the F160W band. Furthermore, its high Galactic latitude minimizes both background contamination and dust extinction ($A_V = 0.02$\,mag), yielding an exceptionally clean photometry. EGS benefits from one of the most complete multiwavelength coverages, spanning from CFHT/MegaCam in the optical to Spitzer/IRAC in the mid-infrared. Furthermore, it contains 41,457 sources, the largest number of any CANDELS field, which maximizes the number of available EELG candidates \citep{Stefanon_2017}. We further restrict the sample to sources with $\mathrm{S/N} > 3$ in the \I, \J, and \Hfilter bands and remove those flagged as photometrically unreliable by the CANDELS quality flag, which typically identifies objects affected by blending, image artifacts, or proximity to bright neighbors. After applying the above criteria, our final parent sample comprises 23838 sources at redshift between 0.007 and 7.26.

\subsection{Photometric properties}
\label{sec:photometric_properties}

For the \z{phot}, we use the updated v2.0 catalog from \citet{Kodra_2023}. This catalog combines the outputs of multiple template-fitting codes into an optimized photometric redshift estimator, which lowers the fraction of catastrophic outliers and improves the characterization of faint sources. This approach achieves a high photometric accuracy with a scatter of $\sigma_{\mathrm{NMAD}} \sim 0.02\,(1+z)$, where the normalized median absolute deviation (NMAD) is defined as:
\begin{equation}
\sigma_{\mathrm{NMAD}} = 1.48 \times \mathrm{median} \left| \Delta z - \mathrm{median}(\Delta z) \right|,
\end{equation}
with $\Delta z = (\z{phot} - \z{spec}) / (1 + \z{spec})$, (following \citealp{lumbreras_2022}).

For the redshift values, we adopt the \texttt{zbest} value in the CANDELS catalog (see Sect.~\ref{sec:candels_survey}), which corresponds to the spectroscopic redshift when available and to the photometric redshift otherwise. These spectroscopic redshifts are compiled from various large spectroscopic surveys carried out across the five CANDELS fields, such as DEEP2 (Deep Extragalactic Evolutionary Probe 2; \citealt{newman_deep2_2013}) and MOSDEF (MOSFIRE Deep Evolution Field; \citealt{Kriek_2015}), as well as from observations using the DEIMOS spectrograph (DEep Imaging MultiObject Spectrograph; \citealt{Faber_2003}).

The physical properties of the galaxies that we use in this work include the redshift, stellar mass and star formation rate (SFR) and we use values from publicly available catalogs in the CANDELS collaboration. The parameters used throughout this work are taken from the v1.0 CANDELS-EGS catalog \citep{grogin_2011,koekemoer_candels_2011, Stefanon_2017}. Among all the SED-fitting methods available in the CANDELS catalog, only two (M11 and M14) include nebular emission, which is an essential ingredient when modeling EELGs given the large contribution of their rest-frame optical lines to the broadband fluxes \citep{Atek_2011}. We adopt the M14 set of physical parameters, obtained with \texttt{SpeedyMC} \citep{Acquaviva_2012} using BC03 templates, a Chabrier IMF, and solar metallicity. We favor this method because it provides the delayed-$\tau$ star formation history parametrization that has become standard in recent EELG studies \citep{onodera_broadband_2020, boyett2024a}, together with a more complete physical treatment of the nebular continuum and line emission. Although M11 covers a wider metallicity grid extending to subsolar values, which is, in principle, more representative of the metal-poor regime of EELGs, a direct comparison between the two methods on our sample yields a negligible median offset in stellar mass and a modest offset in SFR ($\Delta \log M_\star = -0.04$\,dex and $\Delta \log \mathrm{SFR} = +0.15$\,dex, in the sense M14$-$M11). This indicates that the choice of metallicity grid does not dominate the inferred physical properties.

To estimate the fluxes for the candidates, we used the photometry provided by the CANDELS catalog. The continuum level ($C_\lambda$) was computed by averaging the two adjacent filters, which are assumed to be line-free. The line flux was obtained from the difference between the observed flux density ($f_\lambda$) and the estimated continuum ($C_\lambda$), and the rest-frame equivalent width was then computed as
\begin{align}
    \EWrest &= \frac{F(\mathrm{line})}{C_\lambda \, (1+z_{\mathrm{best}})}, \label{eq:ew} \\
    F(\mathrm{line}) &= \left(f_{\mathrm{F125W}} - C_\lambda\right) W_J, \label{eq:fline} \\
    C_\lambda &= \frac{f_{\mathrm{F814W}} + f_{\mathrm{F160W}}}{2}, \label{eq:cont}
\end{align}
where $W_J = 2674.40$~\AA\ is the effective width of the F125W filter (calculated as in \citealt{Pascual_2001}). We estimated the EW uncertainties using a Monte Carlo approach. For each source, we drew $10,000$ realizations of the \I, \J, and \Hfilter fluxes from Gaussian distributions defined by their measured values and photometric uncertainties. The median of the resulting EW distribution was adopted as the best estimate, and the $1\sigma$ uncertainty was derived from the 16th and 84th percentiles.

We note that these values may be overestimated. Firstly, we assume that the flux excess is entirely produced by the bright emission line, without accounting for possible contributions from fainter adjacent lines \citep{llerena_physical_2024}. Second, our estimate of the continuum does not account for prominent spectral features that some of the sources may have, such as the Balmer break at $4000$\AA. Although in this case the continuum could be better represented by the \Hfilter filter, we have used the average described above to ensure a consistent comparison with VDW.

\begin{figure}[!t]
\centering
\includegraphics[width=9cm]{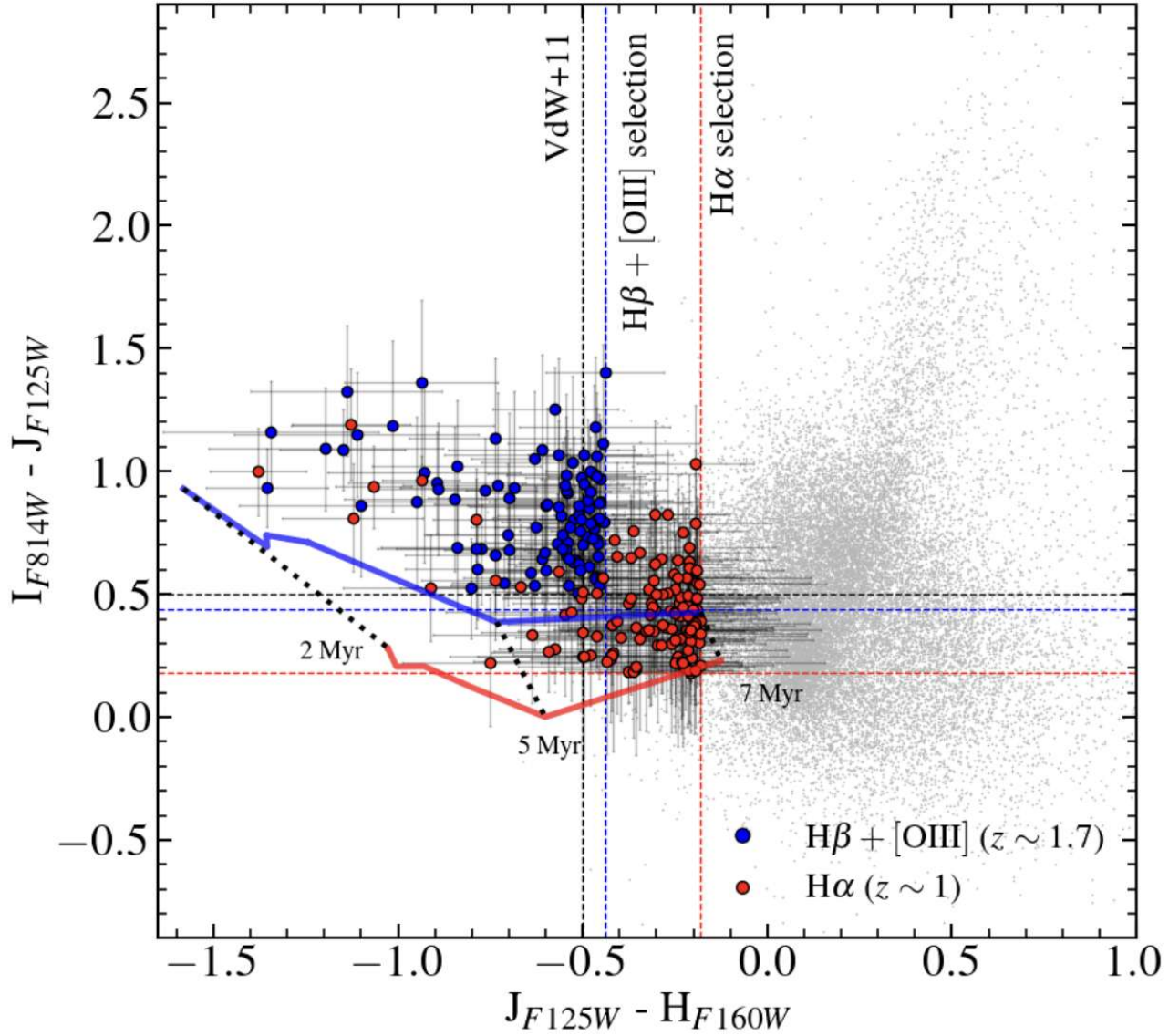}
\caption{Color-color diagram illustrating the selection criteria adopted for the EELG sample. Blue circles represent the selected \OIIIHbeta\ candidates, while red circles denote the \Halpha\ candidates. Vertical and horizontal dashed lines indicate the specific color thresholds used to isolate the different line-emitting populations, including the selection boundaries from VDW. The solid curves trace the color evolution of a single stellar population (SSP) as its age increases from 1 to 7\,Myr, computed with \textsc{Bagpipes} \citep{Carnall_2018} from \citet{Bruzual_2003} models including nebular emission, for $Z=0.4\,Z_\odot$, $\log U=-2.0$ and $A_V=0.2$ (Calzetti attenuation), at $z=1.7$ (blue) and $z=1.0$ (red).}
\label{selection}
\end{figure}

\begin{table*}[!t]
\caption{Spectroscopic data from the literature used in this work.}

\label{spectroscopic_table}
\centering
\begin{tabular}{ccccccc}
\hline
\hline
Program & Facility/Instrument & Filter & Spectral range & Resolution & Sample size & Reference \\
 & & & [$\mu$m] & & & \\
(1) & (2) & (3) & (4) & (5) & (6) & (7) \\
\hline
Tang+19 & MMT/MMIRS & J,H & 0.95--1.80 & 960, 1200 & 5 & \cite{MMT_Tang_2019} \\
Tang+22 & Keck/MOSFIRE & J & 1.15--1.30 & 3318 & 3 & \cite{Keck_Tang_2022} \\
MOSDEF & Keck/MOSFIRE & J,H & 1.14--1.82 & 3000, 3650 & 1 & \cite{Kriek_2015} \\
3D-HST & HST/WFC3 & G141 & 1.10--1.70 & 130 & 8 & \cite{Momcheva_2016} \\
DJA & JWST/NIRSpec & PRISM/CLEAR & 0.60--5.30 & 30--400 & 3 & \cite{Finkelstein_2023} \\
Kodra+23 & - & - & - & - & 5 & \cite{Kodra_2023} \\
\hline
\hline
\end{tabular}
\tablefoot{These data complement the spectroscopic redshifts of the CANDELS compilation \citep{Kodra_2023}; see text for details. Columns: (1) name of the spectroscopic program or survey; (2) telescope and instrument; (3) filter, grism, or disperser; (4) wavelength coverage; (5) spectral resolving power; (6) number of candidates from our sample observed in the program; (7) reference describing the program and data reduction.}
\end{table*}

\subsection{EELGs selection}
\label{sec:sample_selection}

We followed the color-excess selection method introduced by VDW, which identifies EELGs by their unusually bright fluxes in a single broadband filter relative to the adjacent ones. Sources that are at least 50\% brighter in the \J band than in both the \I and \Hfilter bands exhibit a color excess driven by the presence of one or more strong rest-frame optical emission lines, such as \OIIIHbeta\ or \Halpha depending on z. Fig.~\ref{selection} shows our parent catalog with S/N>3 (gray) in a color-color diagram, where EELGs are selected from the bulk of the galaxy population due to their broadband color excess. The blue and red circles correspond to the final \OIIIHbeta and \Halpha samples, obtained after applying the color-excess criteria (Eqs.~(\ref{eq:oiii}) and (\ref{eq:halpha})) together with the redshift restrictions described in this section.

Since most SED-derived properties are redshift-dependent, and in order to preserve consistency between the adopted photometric redshifts (v2.0; \citealt{Kodra_2023}) and the physical parameters, which are derived at the corresponding v1.0 redshifts \citep{Stefanon_2017}, we discarded all candidates with $|\Delta z|/(1+z) > 0.15$ between the two catalog versions. This corresponds to the standard threshold adopted to identify catastrophic outliers in CANDELS studies \citep{Dahlen_2013}, leading to the removal of 4 \OIIIHbeta\ and 19 \Halpha\ additional catastrophic candidates. In addition, we excluded sources whose emission line (\OIIIHbeta\ or \Halpha) does not fall within the \J-band filter, as determined by the adopted redshift \texttt{zbest} (spectroscopic where available, photometric otherwise; see Sect.~\ref{sec:photometric_properties}). This step removed another 25 and 491 sources, respectively.

In their original formulation, VDW included a term in the color-excess threshold that accounts for the photometric uncertainty of the color  to ensure a low contamination, in exchange for having a statistically less representative sample (see the dashed line in Fig.~\ref{selection}). In this work, we adopted the same color-excess criterion but removed the photometric-uncertainty requirement.
Although this approach might introduce a slight increase in the fraction of contaminants, we prioritize the completeness of the sample even if it lowers its purity to some extent. Field stars and blended sources can be a significant source of contamination. A visual inspection of the sample allows us to rule out contamination by stars. We then examine the photometric redshift probability distributions following the criterion of \citet{llerena_physical_2024}, which flags as a potential contaminant any source with less than 50\percent\ of its integrated photometric redshift probability within the target redshift window ($\int \mathcal{P} < 0.5$). This criterion identifies 13 potential contaminants: one in the \OIIIHbeta\ sample and 12 in the \Halpha\ sample. Eight of the \Halpha\ sources are compact and isolated, while the remaining four have nearby neighbors, none of which are flagged as blended in the CANDELS catalog. Moreover, for these sources, the probability outside the target window falls largely within the redshift window of the other emission line. They are therefore most likely EELGs affected by the \Halpha/\OIIIHbeta\ degeneracy, and we retain them in the final sample. The purity of both samples is quantified in Sect.~\ref{sec:purity}.

\subsubsection{\texorpdfstring{\OIIIHbeta}{[O III]+H-beta} emitters}

Applying our modified version, the selection criterion for \OIIIHbeta emitters reads:

\begin{equation}
\OIIIHbeta:
\left\{
\begin{aligned}
I - J &> 0.44 \\
J - H &< -0.44
\end{aligned}
\label{eq:oiii}
\right.
\end{equation}

The final sample consists of 102 \OIIIHbeta\ emitters spanning a photometric and spectroscopic redshift range of $z = 1.40$--$1.90$ ($\sim$0.5 objects per \arcminsq; see Fig.~\ref{selection}). Of these, 19 have spectroscopic confirmation (see Sect.~\ref{section:spectroscopic_data}).

\subsubsection{\texorpdfstring{\Halpha}{H-alpha} emitters}

We apply this same framework to select \Halpha\ emitters by translating the color-excess threshold into the equivalent flux excess produced by the \Halpha\ emission. From the criterion defined in Eq.~(\ref{eq:oiii}), we used the flux ratio of $f_2/f_1 \sim 1.5$, where $f_2$ corresponds to the \J-band flux, and $f_1$ represents the flux in either the \I or \Hfilter band. To determine the \Hbeta\ properties, we extract the observed equivalent width of the \OIIIHbeta\ blend calculated as explained in Sect.~\ref{sec:photometric_properties}, and derive the corresponding \EWobs(\Hbeta) by assuming the empirical line ratio of \Hbeta/(\OIIIHbeta) $\sim 1/8$ \citep{vanderWel2011}. Using the Case$\sim$B recombination ratio \Halpha /\Hbeta = 2.86 \citep{Osterbrock_2006}, we then estimated \EWobs(\Halpha) and translated it back to a color-excess threshold for the \J band. Given the way the selection was made, we assume that the \Halpha\ emitters selected here belong to the same population of EELGs as the \OIIIHbeta\ emitters.
The resulting criterion is:

\begin{equation}
\Halpha:
\left\{
\begin{aligned}
I - J &> 0.18 \\
J - H &< -0.18
\end{aligned}
\right.
\label{eq:halpha}
\end{equation}

We obtained a final sample of 127 \Halpha\ emitter candidates spanning a photometric and spectroscopic redshift range of $z = 0.67$--$1.14$ ($\sim$0.6 objects per \arcminsq; see Fig.~\ref{selection}), where 8 of them have spectroscopic confirmation.

\subsection{Spectroscopy data}
\label{section:spectroscopic_data}

To validate the photometric selection and to characterize the emission-line
properties of a subset of our candidates, we compile spectroscopic data
both from the literature (Sect.~\ref{sec:spectroscopic_literature}) and from our
own GTC/OSIRIS+ follow-up program (Sect.~\ref{sec:spectroscopic_osiris}).

\begin{table*}[t]
\renewcommand{\arraystretch}{1.3}

\centering
\caption{Summary of GTC/OSIRIS+ observations.}
\begin{tabular}{ccccccccc}
\hline\hline
Mask name & Number of targets & Date (UT) & R.A. & Decl. & Filter & Exposure time & Average seeing & Nexp \\
&  & &(hh:mm:ss)  & (dd:mm:ss) &  & (seconds)  & (")\\
(1) & (2) & (3) & (4)  & (5) & (6) & (7) & (8) & (9) \\
\hline
$egs1$ & 9 & 2025-04-25 & 14:19:24.00	 & 52:49:05.0 & I & 14400  & 1.2 & 12\\
\hline\hline
\end{tabular}
\tablefoot{Column (1): mask name; Column (2): number of science targets; Column (3): Observation date; Column (4): right ascension of the mask center; Column (5): declination of the mask center; Column (6): filter of the mask observed; Column (7): total exposure time; Column (8): average seeing during the observation. Column (9): number of exposures.}
\label{tab:observaciones}
\end{table*}

\subsection{Spectroscopy from the literature}
\label{sec:spectroscopic_literature}

In addition to the spectroscopic redshifts already available in the CANDELS catalog, we compiled spectroscopic data from various ground- and space-based facilities, summarized in Table~\ref{spectroscopic_table}, to validate our photometric selection and characterize the emission-line properties of the candidates, such as \OIIIline{4959,5007}, \Hbeta, and \Halpha. We identified our candidates in these programs by cross-matching with the target lists of \citet{MMT_Tang_2019, Keck_Tang_2022}, the MOSDEF survey \citep{Kriek_2015}, the 3D-HST survey \citep{Momcheva_2016}, and the Dawn JWST Archive (DJA)
\citep{Finkelstein_2023}.

Five of our candidates belong to the target sample of the MMT/MMIRS program of
\citet{MMT_Tang_2019}, for which public spectra are available. This instrument
has a field of view of $4\arcmin \times 6.9\arcmin$, and the observations were
carried out with slit masks using a slit length of $7.0\arcsec$ and a width of
$1.0\arcsec$. Integration times ranged from 1 to 8 hours, reaching S/N\,$>$\,3 in
the emission lines targeted by the program. A further  three of our candidates fall within the Keck/MOSFIRE sample of \citet{Keck_Tang_2022}, and one more
belongs to the MOSDEF survey \citep{Kriek_2015}, which used the same instrument.
The MOSFIRE spectrograph has an effective field of view of $3\arcmin \times 6\arcmin$, with a slit length of $7.1\arcsec$ and a width of $0.7\arcsec$. Integration times for these
observations ranged from 1 to 3 hours, reaching a 5$\sigma$ line-flux limit of
$\sim 1.5\times10^{-17}$\,\fluxlinesunit at $\sim 1.5\,\mu$m. For all nine candidates we
use the publicly released 1D reduced spectra. In what follows, we perform our own
emission-line fitting on these spectra (Sect.~\ref{sec:line_fitting_and_results}),
while for the remaining candidates we adopt the line fluxes reported in the
corresponding references.

An additional eight candidates have spectroscopic data from the 3D-HST survey \citep{Momcheva_2016}. The WFC3 camera has a field of view of
$136\arcsec \times 123\arcsec$, and observations were performed with slitless spectroscopy and integration times of $\sim 1$ hour. Line
detections in the WFC3/G141 grism reach a 1$\sigma$ flux limit of $\sim 8\times10^{-18}$\,\fluxlinesunit at $\sim 1.5\,\mu$m for typical
resolved sources. Finally, three candidates have been observed with JWST/NIRSpec, obtained from the DJA repository and processed with the \texttt{msaexp} pipeline \citep{msaexp_zenodo,Heintz_2024, deGraaff_2025}. These observations belong to the Cosmic Evolution Early Release Science Survey (CEERS) program \citep{Finkelstein_2023}, 
and the data were retrieved from the \texttt{nirspec\_graded\_v1} catalog. All three targets were observed exclusively with the PRISM/CLEAR configuration. The NIRSpec instrument has a field of view of $3.6\arcmin \times 3.4\arcmin$, and the observations were performed in multiobject spectroscopy (MOS) mode at low resolution to maximize the sensitivity to the faint continuum. The total integration time per candidate was $\sim 5$ hours, reaching S/N levels between 3 and 5 in the emission lines, our minimum
requirement for line detection.

\subsection{MOS spectroscopy with OSIRIS+ at GTC}
\label{sec:spectroscopic_osiris}

We carried out a spectroscopic follow-up program (GTC84-25A) aimed primarily at confirming the photometric redshifts of our EELG candidates. The OSIRIS+ spectrograph covers the wavelength range 0.365--1.05\,$\mu$m and provides a field of view of $7.8\times8.5$\,$\arcmin$ ($7.8\times7.8$\,$\arcmin$ unvignetted) \citep{Cepa_2002}\footnote{\url{https://www.gtc.iac.es/instruments/osiris+/osiris+.php}}. We used the MOS mode, which offers an effective field of $7.5\times6$\,$\arcmin$ and a set of grisms and volume-phase holographic gratings (VPHs). For this program, we employed the VPH R2500I grating, covering the range 7330-10000 \AA, which allows us to measure key emission lines such as \OIIline{3727}, \Hbeta, and/or \OIIIline{4959,5007} for galaxies in the targeted redshift range.

\subsubsection{Target selection}

We performed a spectroscopic follow-up on the parent EELG catalog defined in Sect.~\ref{sec:sample_selection} using OSIRIS+. We restricted the redshift range to galaxies for which  \OIIline{3727}, \Hbeta, and \OIIIline{4959,5007}  fall within the spectral coverage of the VPH R2500I grating. Objects whose expected emission lines would overlap with strong OH sky features were excluded \citep{atlas_OH}. From the remaining sample, we prioritized the brightest candidates in the I-band to ensure adequate signal-to-noise on the targeted lines. The observing sample comprised 9 galaxies in the interval 0.86<\z{phot}<1.73, placed on a single MOS mask.

\subsubsection{Observations}

One slit mask was designed using the official OSIRIS+ Mask Designer  tool (egs1). Each slitlet had a width of 1.2\arcsec\ and a  length of 20\arcsec.  A summary of the egs1 observations is given in Table~\ref{tab:observaciones}. This mask was observed on UT 2025 April 25--26, with 12~individual exposures of 1200\,s for a total of 14\,400\,s (4.00\,h) at an average airmass of $\sim$1.13.

Wavelength calibration frames (HgAr, Ne, and Xe arc lamps), dome spectral flats, and bias frames were obtained as part of the standard OSIRIS+ calibrations. Spectrophotometric standard stars were observed with the same R2500I grating used for the science exposures, through a $2.52\arcsec$ long slit. On the night, GD~153 was observed at the beginning (250\,s, airmass 1.30) and Ross~640 at the end (200\,s, airmass 1.18).

\subsubsection{Data reduction}

The data were reduced using the \texttt{PypeIt} pipeline v1.17.4  \footnote{\url{https://pypeit.readthedocs.io/en/stable/}} \citep{pypeit:zenodo}. The standard calibration steps: bias subtraction, flat-fielding, and wavelength calibration from HgAr, Ne, and Xe arc-lamp exposures were
performed with \texttt{run\_pypeit}, which also carries out slit tracing and cosmic-ray rejection. Sky subtraction was done on a per-slitlet basis using a B-spline model. The sky-subtracted two-dimensional spectra were then combined with \texttt{pypeit\_coadd\_2dspec}. One-dimensional spectra were extracted via boxcar summation over an aperture of $\pm 3$\,pixels, centered on positions identified by visual inspection of the coadded 2D frames.

\subsection{Emission-line measurements and OSIRIS+ results}
\label{sec:line_fitting_and_results}

Of the 22 candidates with available spectra (Sect.~\ref{section:spectroscopic_data}),
we performed our own line fitting on 10 of them (since they lacked published emission-line fits) --- the 8 from MMT/MMIRS and
Keck/MOSFIRE \citep{MMT_Tang_2019, Keck_Tang_2022} and the 2 from our OSIRIS+
follow-up --- and adopted the published line fluxes for the MOSDEF, 3D-HST and DJA
candidates \citep{Kriek_2015, Momcheva_2016, msaexp_zenodo, Heintz_2024, deGraaff_2025}.
Five other candidates have a spectroscopic redshift listed in the CANDELS catalog 
(\citealp{Kodra_2023}, see Sect.~\ref{sec:photometric_properties}). For these sources, \z{spec} value
 is the only spectroscopic information available to us. The resulting comparison sample used to validate
  our selection comprises 27 of the 229 candidates ($\sim$19\percent\ \OIIIHbeta\ and $\sim$6\percent\ \Halpha); 
  their spectroscopic and photometric properties are compared in Sect.~\ref{sec:comparison}.

\begin{figure*}[!t]
\centering
\includegraphics[width=\textwidth]{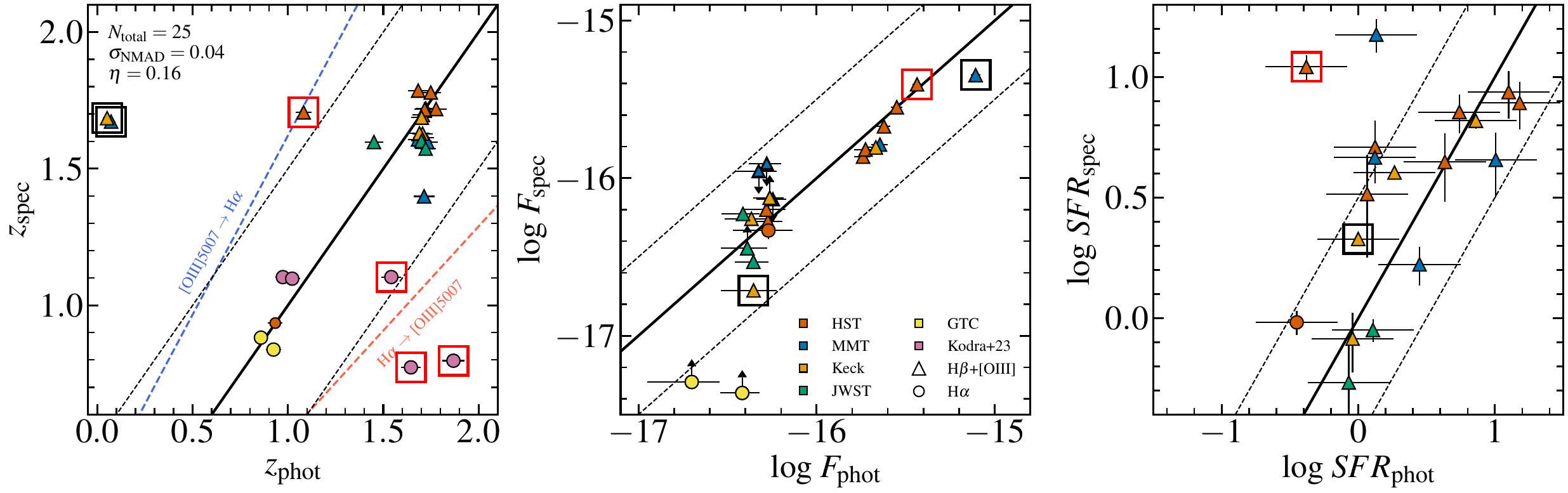}
\caption{Comparison between photometrically derived and spectroscopically measured properties for the EELG sample. \textit{Left panel}: Spectroscopic versus photometric redshift.  \textit{Middle panel}: Comparison of emission-line fluxes in \fluxlinesunit units. \textit{Right panel}: Comparison of star formation rates in \sfrunit units. In all panels, the solid black line denotes the ideal 1:1 relation. The black dashed lines indicate a difference of 0.5 in redshift and a factor of 3 in flux and SFR. Marker shapes indicate the primary emission lines used for the spectroscopic identification (triangles for \OIIIHbeta and circles for \Halpha), while marker colors denote the facility or instrument providing the spectroscopy (HST, MMT, Keck, JWST, GTC and the CANDELS compilation of \citealp{Kodra_2023}), as detailed in the legend. The outlier sources are highlighted with a red box symbol. It is important to note that no photometric SFR value has been computed for the MMT galaxy. The dashed blue line shows the expected relation between \z{phot} and \z{spec} if an \OIIIline{5007} line were misidentified as \Halpha\ by the photometric redshift fit. The dashed red line shows the corresponding relation if an \Halpha\ line were misidentified as \OIIIline{5007}.}
\label{comparison_z_flux_sfr}
\end{figure*}

We fit \Hbeta, \OIIIline{4959}, \OIIIline{5007} and \Halpha\ using the
\texttt{LevMarLSQFitter} routine from \texttt{astropy.modeling}
\citep{astropy:2013, astropy:2018, astropy:2022}, adopting different procedures
depending on the line peak S/N: a Gaussian plus flat continuum for $S/N > 5$,
direct integration for $2 < S/N < 5$, and a $2\sigma$ upper limit for $S/N < 2$
(similar to \citealp{MMT_Tang_2019}, see an example in Fig.~\ref{fitting_example}). When \OIIIline{4959}, \OIIIline{5007} and
\Hbeta\ are fully resolved, we fit them simultaneously, tying the velocity
dispersion across the lines and fixing the \OIIIline{5007}/\OIIIline{4959} ratio
to 3. Flux uncertainties were computed via a Monte Carlo procedure with 10,000
realizations, perturbing the spectra according to their noise vector. The
best-fit values and $1\sigma$ uncertainties are taken as the median and half the
16th--84th percentile range of the resulting distributions. Redshifts were fixed
to the published spectroscopic values during the fit, except for the two OSIRIS+
candidates, which lacked a prior spectroscopic redshift and for which we adopted
the photometric value.

We applied a dust attenuation correction exclusively to the only two galaxies in our sample where both \Halpha\ and \Hbeta\ were detected and their derived color excess was $E(B-V) > 0$. For these objects, we used a Balmer-decrement correction assuming a \citet{Calzetti_2000} attenuation law and an intrinsic line ratio of 2.86 \citep{Osterbrock_2006}. For the rest of the sample, no extinction correction was applied. This includes objects with $E(B-V) \leq 0$, whose dust content is expected to be negligible, and sources where only a single Balmer line was available. In cases where only \Hbeta\ was detected, the \Halpha\ flux was directly estimated from the theoretical ratio. The SFR was then computed with the \cite{Kennicutt_1998} calibration, $\mathrm{SFR}(\Halpha) = 7.9 \times 10^{-42}\,L(\Halpha)\ [\mathrm{erg\,s^{-1}}]$, scaling the values down by a factor of 1.7 to convert from the Salpeter IMF assumed by that calibration to a Chabrier IMF. For the uncorrected galaxies, the SFRs might be underestimated, an effect we discuss in detail in Sect.~\ref{sec:physical_morf}.

Applying this procedure to our GTC/OSIRIS+ campaign, emission lines were detected
in two of the nine observed candidates (see their spectra in Fig.~\ref{osiris_espectro}). For EGS-34094
($z_\mathrm{phot} {=} 0.86$) we detect a single line at
$\lambda_\mathrm{obs} {=} 9145.88 \pm 0.14$\,\AA, identified as \OIIIline{5007} at
$z_\mathrm{spec} = 0.82662 \pm 0.00003$. For EGS-32856 ($z_\mathrm{phot} {=} 0.92$) we detect the \OIIIline{4959,5007} doublet at $\lambda_\mathrm{obs} {=} 9721.9 \pm 0.5$\,\AA\
and $9815.65 \pm 0.07$\,\AA, respectively, yielding $z_\mathrm{spec} {=} 0.96045 \pm 0.00001$. In both objects the fainter lines (\OIIIline{4959} for EGS-34094, and \Hbeta) fall below the noise, as expected
from their intrinsically lower fluxes.

The remaining 7 candidates yielded no emission features above the noise, which
was largely dominated by OH sky-line residuals. Although the overall detection rate of
$\sim$ 22\percent reflects the extreme faintness of these sources, the two
confirmations demonstrate that the color-excess selection successfully
identifies emission-line galaxies. The full set of spectroscopic
measurements is reported in Table~\ref{tab:fluxes}.

\section{Validation and reliability of the sample}
\label{sec:validation}
Using our color-excess selection (Sect.~\ref{sec:sample_selection}), we obtain a sample of 102 \OIIIHbeta candidates at $z = 1.40$--$1.90$ and 127 \Halpha\ emitters at $z = 0.67$--$1.14$. We note that these redshift ranges are defined by the $z_{\rm best}$ value adopted for each source, which corresponds to the spectroscopic redshift when available and to the photometric redshift otherwise
(Sect.~\ref{sec:photometric_properties}). In this section, we assess its reliability through the available spectroscopy and quantify the residual contamination of the photometry.

\subsection{Comparison with spectroscopy}
\label{sec:comparison}

In Fig.~\ref{comparison_z_flux_sfr} (left panel), we compare our photometric redshifts from CANDELS against the spectroscopic confirmations, finding good overall agreement. Because our selection is based on the CANDELS \texttt{zbest} values, which correspond to the spectroscopic redshift when available (Sect.~\ref{sec:photometric_properties}), two sources with \z{phot}$\sim$0.05 were included in the sample solely because of their \z{spec}. Since our selection criterion is meant to be applied on the basis of the photometry alone, these sources would not have been selected otherwise. Consequently, these objects are omitted from the redshift statistics, but are retained in the sample as confirmed sources. They are highlighted with black square borders across all panels, distinguishing them from other redshift outliers, which are marked in red. The bulk of the sample closely follows the 1:1 relation, showing a small median bias of $\Delta z \simeq 0.01$, a value of $\sigma_{\mathrm{NMAD}} = 0.04$, and an outlier fraction of $\eta = 0.16$ . We identify four significant outliers, one in the \OIIIHbeta and three in the \Halpha subsample, compiled from  HST, and the CANDELS catalog. Given their high S/N>5  (\citealp{Momcheva_2016, MMT_Tang_2019, Keck_Tang_2022,Kodra_2023}), we rely on the robustness of these spectroscopic measurements. 

Our outlier fraction is a factor of $\sim$2 higher than that reported for the parent CANDELS catalog \citep[$\eta = 0.07$;][]{Stefanon_2017}, and our dispersion is a factor of $\sim$3 larger ($\sigma_{\rm NMAD} = 0.013$ in \citealt{Stefanon_2017}). This elevated scatter, combined with the systematic offset toward lower \z{phot}, is consistent with the extreme nature of our sources: strong emission lines characteristic of EELGs can contaminate broadband photometry and mimic the Balmer or Lyman breaks, leading to systematic \z{phot} errors. Indeed, $\sim$85\percent of the photometric-redshift outliers in the parent catalog are galaxies with intense emission lines \citep{Stefanon_2017}.

In Fig.~\ref{comparison_z_flux_sfr} (middle panel), we compare the integrated fluxes of \OIIIHbeta and \Halpha derived from our spectroscopic fitting against those estimated via the photometric excess method. The measurements show good agreement, with an average offset of 0.20 dex. Sources represented as upper or lower limits correspond to lines where no significant detection was achieved during the spectral fitting. Due to the faint nature and distance of the objects, non-detection of the emission lines in spectroscopy is to be expected. The photometric method, while more imprecise in the reported values, has the advantage of integrating all light in a wider spectral range which improves its detection rate for these objects. Nevertheless, the flux limits for these objects remain consistent with the photometric expectations. The observed discrepancies likely arise from a combination of inherent uncertainties in the photometric calibration, residual slit-loss corrections in the spectroscopic data, and the assumption of a flat continuum between the F814W and F160W filters, which may not hold if the continuum shape varies across this wavelength range. To statistically verify this comparison, we performed a Kolmogorov-Smirnov (K-S) test. The resulting $p$-value of $\sim0.7$ is well above our adopted significance threshold ($p < 0.05$) indicating that the two distributions are consistent with each other. Furthermore, the spectroscopic confirmation of \OIIIline{5007} and \Halpha in emission demonstrates that the flux excess observed in the broadband photometry is typically driven by strong emission lines, thereby validating the reported equivalent width values.

In Fig.~\ref{comparison_z_flux_sfr} (right panel), we compare the SFR averaged over 100~Myr obtained from the SED fitting to the SFR derived from \Halpha fluxes. The SED-derived values show a scatter of 0.52~dex around the spectroscopic measurements. Part of this dispersion is expected due to the comparison of two indicators sensitive to different timescales. Several studies of EELGs comparing the SFR derived from \Halpha with that from SED fitting averaged over 10~Myr report a typical scatter of $\sim 0.25$~dex \citep[e.g.,][]{amorin_extreme_2015,Calabro_2017,lumbreras_2022}. Subtracting this value in quadrature from our observed scatter ($\sigma_{\rm obs}^2 = \sigma_{\rm lit}^2 + \sigma_{\rm extra}^2$, where $\sigma_{\rm obs}^2$ is the variance of our observed scatter and $\sigma_{\rm lit}^2$ is the variance reported in the literature), we obtain an extra scatter of $\sigma_{\rm extra} \sim 0.46$~dex. We attribute this additional dispersion primarily to the difference in the traced timescales. Consequently, we expect that the observed scatter would be significantly reduced if a shorter timescale were considered for the SED-derived SFRs. We performed a K-S test using the SFR$_{phot}$ and SFR$_{spec}$ and we found a p-value of ~0.3. The fact that there is no statistically significant difference implies that both methods yield results that are consistent with each other despite the systematic bias.

The four redshift outliers are highlighted in red in Fig.~\ref{comparison_z_flux_sfr}. Since their SED-derived properties were computed at their photometric redshift, their stellar masses and SFRs are expected to be overestimated. To estimate the impact on these values, we examined the difference between the SED-derived SFR and the SFR(\Halpha) for these outliers and compared it with the scatter of the non-outliers. For the \Halpha outlier sub-sample, we calculated their SFR(\Halpha) using their $J$-band photometric flux excess and the calibration of \cite{Kennicutt_1998} (Sect.~\ref{sec:line_fitting_and_results}). We find that two of these outliers lie within the general noise level, whereas the remaining \Halpha outlier and the \OIIIHbeta source exhibit a scatter of $\sim$1\text{ dex}. Regarding the stellar mass, we rescaled the SED-fitting values to the spectroscopic redshift, assuming $\text{M}_*\propto D_L^2$. We find that the scatter for the outliers reaches $0.55\text{ dex}$, exceeding the general noise level by a factor of $\sim$7. Nevertheless, since these objects account for less than 2\percent\ of the total sample, their impact on the physical properties of the full sample is negligible and does not alter the conclusions of this work (Sect.~\ref{sec:physical_morf}). For this reason, we did not apply this correction in the rest of the analysis.

In summary, the multiwavelength and spectroscopic comparisons presented in this section validate the overall robustness of our sample selection and physical parameter estimation. While the extreme nature of our EELGs introduces a well-understood scatter in the photometric redshifts due to strong line contamination, the spectroscopic confirmations ensure the reliability of our final sample. Furthermore, the excellent statistical agreement found in both the emission line fluxes and the star-formation history distributions confirms that our photometric methods capture the essential physical properties of these systems.

\subsection{Quantifying sample purity}
\label{sec:purity}

Potential sources of contamination in EELG samples include AGN contribution, stars (e.g., white dwarfs or blue horizontal-branch stars), redder
galaxy populations (e.g., LRDs, quiescent galaxies), data-processing artifacts, and redshift confusion through lower-redshift line emitters whose features mimic the broadband excess of EELGs at the target redshift \citep[e.g.,][]{P_nin_2015, Davis_2024,Daikuhara_2025A}. To minimize contamination by interlopers, our selection first removed all candidates without a reliable photometric redshift in the relevant window, which corresponds to an initial rejection of $\sim$20\% of the \OIIIHbeta\ candidates and $\sim$77\% of the \Halpha\ candidates from the parent photometric sample (see Sect.~\ref{sec:sample_selection}). For each candidate, we compute the integrated probability that its photometric redshift falls within the target redshift window of the selected emission line. We define this individual purity as $p_i = P(\z{min}<z<\z{max}\,|\,Z_i)$, where $Z_i$ is the photometric redshift probability distribution obtained with the combination of different templates from
\texttt{EAZY}, \texttt{HyperZ}, \texttt{LePhare}, \texttt{WikZ} and \texttt{zphot}
\citep{Kodra_2023}. We define the overall sample purity ($\mathcal{P}$) as the average of these individual probabilities over the final sample of \OIIIHbeta\ and \Halpha\ emitters, respectively, as follows:
\begin{equation}
\mathcal{P} = \frac{1}{N} \sum_{i=1}^{N} p_i,
\label{eq:purity}
\end{equation}
where $N$ is the total number of sources in the sample. Using this approach, we
obtain a purity of $84.3 \pm 1.5$\percent\ for the \OIIIHbeta\ emitters and
$76.8 \pm 2.0$\percent\ for the \Halpha\ emitters.

To evaluate the robustness of our method and to rule out the possibility that these
high purities are a statistical consequence of the width of the redshift windows,
we conducted a control test using Monte Carlo simulations. We defined a purity-control sample that excluded our EELG candidates and randomly extracted a subsample of the same size as our actual detections. After computing the purity of 1000
iterations, we determined that the background probability due purely to
chance is $23 \pm 4$\percent\ for the \OIIIHbeta\ window and $20 \pm 3$\percent\
for the \Halpha\ window. This confirms that the selection isolates sources that
are concentrated within the target redshift windows. These values are consistent
with the purities reported for comparable broadband selected line-emitter
samples, such as $>$80\percent\ purity of the simulated $z<7$ samples in
\citet{Kauffmann_2020}.

The estimated purities are validated by our spectroscopic compilation and
follow-up. Out of the 229 sources, spectroscopic redshifts are
available for 27 sources (19 \OIIIHbeta\ and 8 \Halpha\ candidates). The redshift
comparison in Sect.~\ref{sec:comparison} identifies one outlier in the \OIIIHbeta subsample and three in the \Halpha subsample, corresponding to spectroscopic confirmation rates of 94.1\percent\ for \OIIIHbeta\ and 62.5\percent\ for \Halpha. For the \OIIIHbeta\ sample, this rate is in
excellent agreement with the $P(z)$-based purity, providing a direct statistical
validation of the purity calculation. For the \Halpha\ sample, the small number of
sources confirmed  spectroscopically prevents a rigorous test, but the confirmation rate remains
statistically consistent with the photometric estimate. Both purity levels are in
line with those reported for comparable broadband selected EELG samples
\citep[e.g.,][]{lumbreras_2022, jaiswar_mosel_2024}.

\subsection{AGN contamination}
\label{sec:agn}

\begin{figure}[!t]
\centering
\includegraphics[width=9cm]{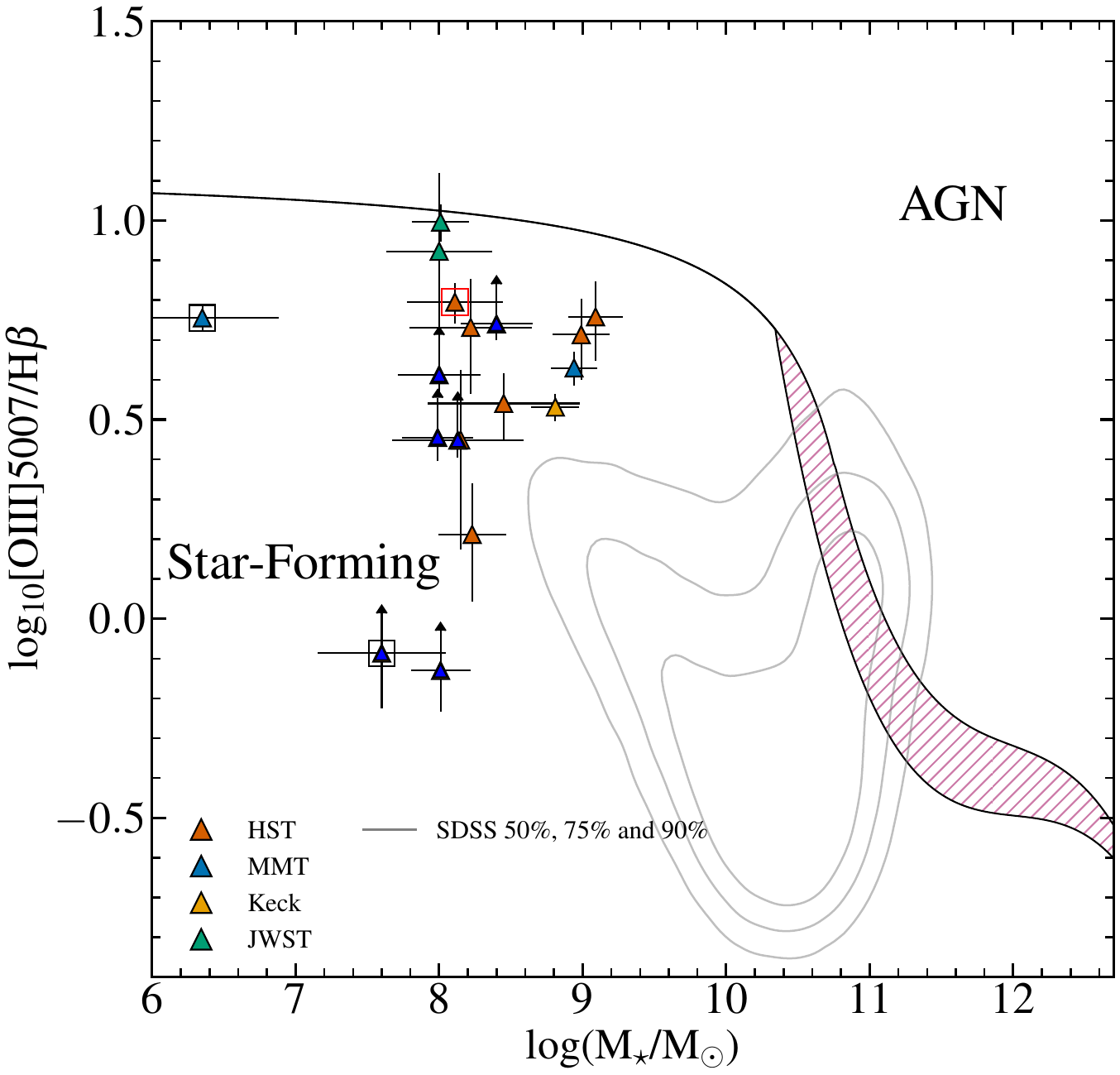}
\caption{Mass-Excitation (MEx) diagram. The EELG candidates with available spectroscopy are shown as triangles, color-coded by facility: HST (red), MMT (blue), Keck (orange) and JWST (green). Gray contours enclose 90\percent, 75\percent, and 50\percent of the Sloan Digital Sky Survey Data Release 19 (SDSS DR19; \citealp{SDSS_DR19_2025}). Lines indicate the demarcation lines  by \citet{Coil_2015} to separate star-forming and AGN galaxies at z$\sim$2. The region between the two lines corresponds to the composite regime.}
\label{diagnostic_diagram}
\end{figure}

Among the contaminants listed above, the potential contribution of AGNs deserves particular attention, 
as these objects can exhibit emission lines as strong as those of young starbursts \citep{lumbreras_2022, boyett2024a}. 
To assess this contamination, we evaluate the available Infrared Array Camera (IRAC) photometry using the mid-infrared color criteria proposed
by \citet{Donley_2012}. We find that only 43 sources have strictly positive fluxes in all four IRAC bands 
(3.6, 4.5, 5.8, and 8.0~$\mu$m). When applying the color-color diagnostic boundaries defined 
by \citet{Donley_2012}, six sources fall within the theoretical AGN limits; however, only one of 
them exhibits the strictly red IRAC SED required for a robust AGN classification. Nevertheless,
because this particular candidate has a signal-to-noise ratio of $S/N < 1$ in all IRAC bands,
we cannot conclusively confirm its AGN nature.

 \begin{figure*}[!t]
\centering
\includegraphics[width=\textwidth]{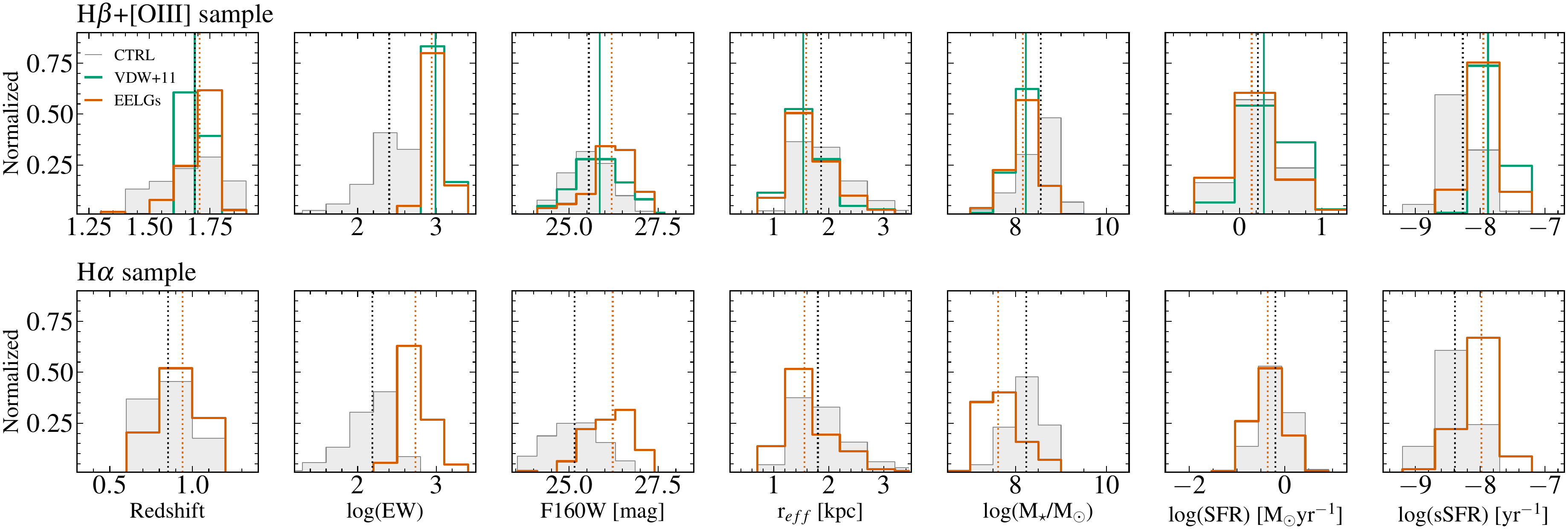}
\caption{Normalized distributions of redshift, rest-frame equivalent width (\AA), F160W magnitude (mag), effective radius (kpc), stellar mass, $\log\,\mathrm{SFR}$, and $\log\,\mathrm{sSFR}$. The \OIIIHbeta\ (first row) and \Halpha\ (second row) samples are shown in orange. The gray histograms in both rows represent the control sample, matched in redshift, F160W magnitude, and stellar mass. Additionally, the first row includes the VDW sample, plotted in green.}
\label{hist_completo}
\end{figure*}

As a complementary test, we use the spectroscopic subsample. Since \NIIline\ is too 
faint to be detected in our spectra, the standard Baldwin–Phillips–Terlevich (BPT) diagram (\citealp{Baldwin_1981, Kewley_2001,
Kauffmann_2003}) cannot be applied. Therefore, we adopt the Mass-Excitation (MEx) diagnostic, which
identifies the dominant ionization source based on the \OIIIline{5007}/\Hbeta\ ratio and the stellar mass. 
Specifically, we use the version originally derived by \citet{Juneau_2014} and later modified by \citet{Coil_2015} 
for galaxies at $z \sim 2$ with low masses and high SFRs. We note that the MEx diagnostic is generally considered 
less robust than other line-ratio diagnostics due to the additional uncertainties introduced by relying on stellar mass,
such as the redshift evolution of the mass--metallicity relation (\citealp{Maiolino_2008, Sanders_2021}).
We nonetheless adopt it because it can be applied to a larger fraction of our sample than alternatives requiring
additional emission lines \citep{llerena_physical_2024}. The AGN/SF demarcation on the MEx plane is shown as the black lines in Fig.~\ref{diagnostic_diagram}. As a
local reference, we additionally plot a sample of $\sim$300{,}000 galaxies from the Sloan Digital Sky Survey Data Release 19 (SDSS DR19; \citealp{SDSS_DR19_2025}) in the redshift range $0.02 < z < 0.2$, with $S/N > 3$ in the 
\Hbeta and \OIIIline{5007} emission lines. 

Since \OIIIline{5007} and \Hbeta\ cannot be resolved from broadband photometry, we restrict the MEx analysis to 
our spectroscopic subsample, for which line fluxes were measured. All our spectroscopically confirmed EELGs fall 
within the star-forming region of the diagram (see Fig.~\ref{diagnostic_diagram}), supporting the IRAC results and pointing to the absence of a dominant 
AGN population. Nevertheless, neither diagnostic can exclude the presence of heavily obscured AGNs. Indeed, literature 
values for the AGN fraction in EELG samples at different redshifts typically range between 5\percent and 10\percent (e.g., \citealp{Huang_2015, moral-castro_extreme_2024, bonatto2026}), a value that has been shown to depend on spectroscopic depth 
\citep{davis2026}.

To summarize, while we find evidence for one tentative AGN candidate, the overall sample remains likely star-forming. We therefore expect the residual contamination of the \OIIIHbeta\ and \Halpha\ samples (Sect.~\ref{sec:purity}) to be dominated by lower-redshift line emitters of similar star-forming nature, rather than by AGNs or evolved interlopers. Consequently, sample-averaged properties should not be systematically biased by this residual contamination.

\section{Physical and morphological properties}
\label{sec:physical_morf}
\subsection{Physical properties}

Our final sample consists of 102 \OIIIHbeta\ candidates at $z = 1.40$--$1.90$ and 127 \Halpha\ candidates at $z = 0.67$--$1.14$, identified in the EGS field. The \OIIIHbeta\ candidates span observed F160W magnitudes between 23.04 and 27.31~mag (median 26.20~mag), rest-frame equivalent widths between 572 and 2123~\AA\ (median 865~\AA), effective radii between 0.19 and 3.62~kpc (median 1.59~kpc), stellar masses $\log(M_\star/M_\odot) = 6.35$--$9.09$ (median 8.16), $\log\,\mathrm{SFR_{100Myr}} = -0.46$ to $1.18$~[\sfrunit] (median 0.15), and $\log\,\mathrm{sSFR_{100Myr}} = -8.56$ to $-7.52$~[\ssfrunit] (median -7.95). The corresponding distributions are shown in the first row of Fig.~\ref{hist_completo}.

The \Halpha\ candidates span F160W magnitudes between 23.74 and 27.26~mag (median 26.22~mag), rest-frame equivalent widths between 267 and 2506~\AA\ (median 542~\AA), effective radii between 0.98 and 4.64~kpc (median 1.56~kpc), stellar masses $\log(M_\star/M_\odot) = 6.87$--$8.90$ (median 7.61), $\log\,\mathrm{SFR_{100Myr}} = -1.32$ to $0.53$~[\sfrunit] (median -0.35), and $\log\,\mathrm{sSFR_{100Myr}} = -9.11$ to $-7.25$~[\ssfrunit] (median -7.98). The distributions are shown in the second row of Fig.~\ref{hist_completo}.

To compare the photometric and physical properties of the EELG candidates against those of the general galaxy population at the same epoch, we define a control sample drawn from the same fields for both the \OIIIHbeta\ and \Halpha\ samples. While these control galaxies fall within the same  redshift, F160W magnitude, and stellar mass range to the EELGs, they do not satisfy the color selection criteria defined in Eq.~(\ref{eq:oiii}) and Eq.~ (\ref{eq:halpha}). To avoid contamination of the control sample by marginal candidates, we additionally require their colors to lie at least 1$\sigma$ away from the selection thresholds. The resulting control sample comprises 3432 galaxies for \OIIIHbeta\ and 2490 for \Halpha\ (see Fig.~\ref{hist_completo}).

\begin{figure*}[!t]
    \centering
    \includegraphics[width=\textwidth]{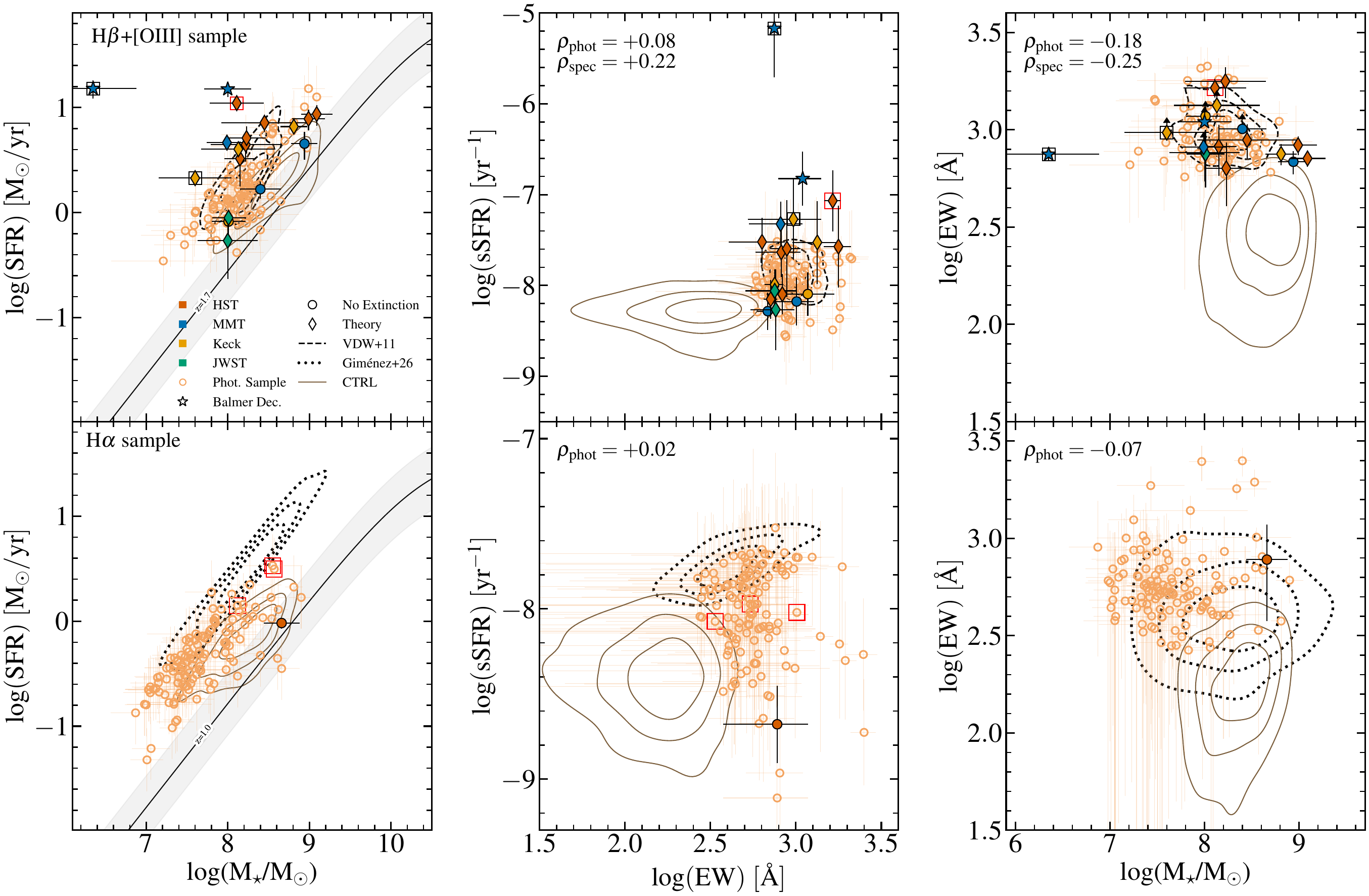}
    \caption{SFR$-$M$_*$, sSFR$-$\EWrest\ and \EWrest$-$Mass diagrams for the \OIIIHbeta\ sample (first row) and \Halpha\ (second row). Orange points are the photometric values. For the spectroscopic sample: Balmer decrement (star), theoretical ratios (diamond) or assuming no extinction (circle). The contours enclose 75\percent, 50\percent, and 25\percent of the sample from the control sample (solid), VDW (dashed), \citet{gimenez_alcazar_2026} (dotted). The solid line is the main sequence from \citealp{Popesso_2023} and the shaded region the 1$\sigma$-scatter.}
    \label{sfr_ew_ssfr_mass}
\end{figure*}

The K-S tests confirm that all photometric and physical properties of the EELGs and the control galaxies differ significantly for both samples ($p$-values $\sim 10^{-5}$--$10^{-71}$). To determine whether our EELGs are more compact than the control sample, we show the effective radius as a function of stellar mass for both populations in Appendix Fig.~\ref{radios}, alongside the mass--size relation reported by \citet{Nedkova_2021}. We measured the deviation between the EELG and control samples using the 16th and 84th percentiles, finding offsets of $-0.030$~dex for \OIIIHbeta\ and $-0.024$~dex for \Halpha. Furthermore, a Wilcoxon-Mann-Whitney test evaluating whether the median sizes of the EELGs are strictly smaller yields $p$-values of 0.003 and 0.031. Given these marginal physical deviations, we cannot robustly conclude that our sample is significantly more compact than the general galaxy population at the same epoch.

In Fig.~\ref{hist_completo}, we have also included the VDW sample by re-extracting
the physical and photometric properties of the sources from their study using the
CANDELS catalog (GOODS-S and UDS fields). This ensures that all properties are
derived using the same pipeline as our main sample. As shown in the figure, our
parameter space extends beyond the range explored by VDW. To assess whether this difference is significant, we first verify that our pipeline
reproduces their selection. Classifying the 69 VDW sources according to which
emission line falls into the \J band, we obtain $65\pm13$ \OIIIHbeta emitters,
$2\pm2$ \Halpha emitters, and 2 unclassified sources. Applying identical color
cuts to our sample, we recover $69\pm13$ \OIIIHbeta and $10\pm4$ \Halpha
emitters. Uncertainties combine Poisson noise with the field-to-field cosmic
variance, added in quadrature, the latter estimated with the Cosmic Variance
Calculator\footnote{\url{https://www.ph.unimelb.edu.au/~mtrenti/cvc/CosmicVariance.html}}
\citep{Trenti_2008} assuming 100\percent completeness for both samples. For the \OIIIHbeta{} samples, Poisson noise ($\sim$12\percent) and cosmic variance ($\sim$14\percent) contribute comparably, whereas for the smaller \Halpha samples the budget is dominated by Poisson noise ($\sim$32\percent). Regarding the \Halpha emitters, a direct comparison of the counts is not meaningful, since the VDW color criterion was specifically designed to target \OIIIHbeta in the $J$ band. This illustrates the intrinsic line degeneracy of broadband excess selections.  

In summary, our selected EELGs are systematically fainter, less massive, and more highly star-forming than the general galaxy population at similar redshifts, although they do not exhibit significantly more compact sizes. Furthermore, the excellent agreement between our recovered counts and the VDW sample when applying identical selection limits validates our methodology. This confirms that the broader physical parameter space explored by our full sample is a robust consequence of our survey depth and selection criteria.

\begin{figure*}[htbp]
\centering
\includegraphics[width=\textwidth]{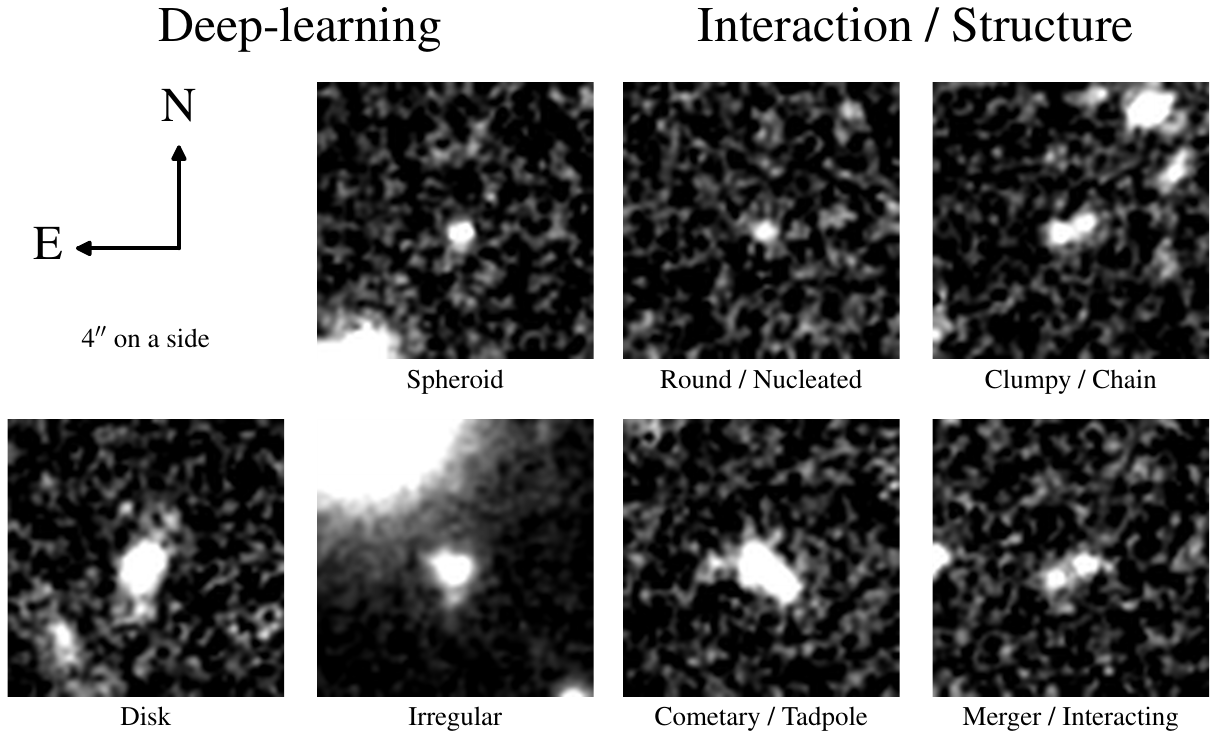}
\caption{Morphology of star-forming EELGs. Examples of classifications obtained in this work using deep learning and structural/interaction: Spheroid, Disk, Irregular (left), Round/Nucleated, Clumpy/Chain, Cometary/Tadpole, and Merger/Interacting (right).}
\label{morfology_example}
\end{figure*}

\begin{figure}[!t]
\centering
\includegraphics[width=9cm]{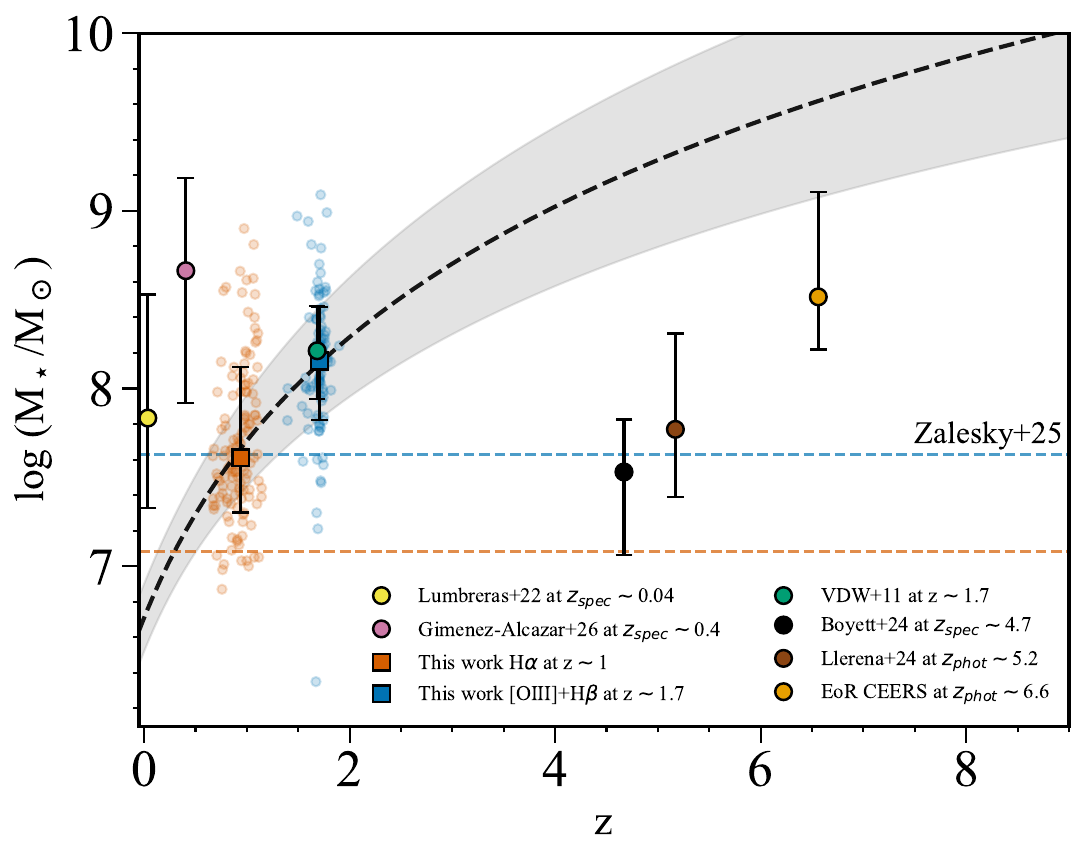}
\caption{Cosmic evolution of the characteristic stellar mass of EELGs as a function of redshift $z$. The blue and red filled squares represent the median values of our \OIIIHbeta\ and \Halpha\ subsamples, respectively. Colored circles show various literature comparison samples across different epochs (VDW; \citealt{lumbreras_2022, boyett2024a, llerena_physical_2024, gimenez_alcazar_2026}), including a compiled sample of EoR galaxies at $z >6 $ from CEERS \citep{Finkelstein_2023}. Vertical error bars denote the 16th-84th percentile range of the mass distribution for each sample. The black dashed line represents our joint Bayesian best fit, while the gray shaded region indicates the $1\sigma$ intrinsic dispersion of the population. The horizontal dotted lines mark the stellar mass completeness limit applied to our data \citep{Zalesky_2025}.}
\label{evolucion_cosmica}
\end{figure}

\begin{figure*}[!t]
\centering
\includegraphics[width=\textwidth]{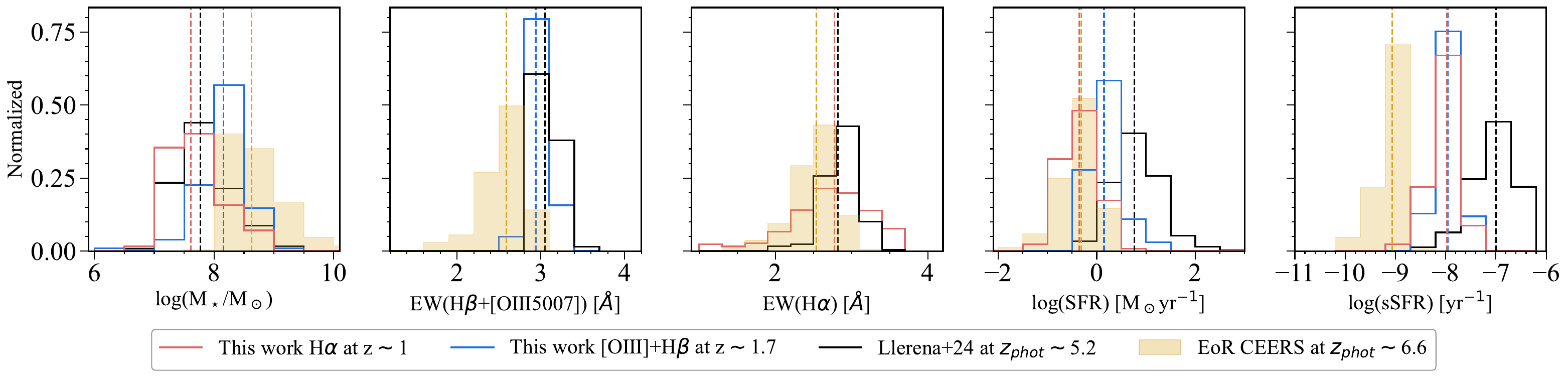}
\caption{Histograms of the physical parameters for our EELGs sample, the EELG sample from \citet{llerena_physical_2024}, and the CEERS sample in the EoR. The median values along with the 16th and 84th percentiles are reported in Table~\ref{tab:histogram}.}
\label{histogram_eor}
\end{figure*}

\subsection{Scaling relations between SFR, sSFR, EW, and stellar mass}

To characterize the star-forming properties of our EELG sample and the control sample, we examine three scaling relations and compare them with the reference contours from VDW and \citet{gimenez_alcazar_2026}. The three relations are presented in Fig.~\ref{sfr_ew_ssfr_mass} for both the \OIIIHbeta\ (top panels) and \Halpha\ (bottom panels) samples.

The left panels of Fig.~\ref{sfr_ew_ssfr_mass} show the SFR as a function of stellar mass, as well as the star-forming main sequence (MS) from \citet{Popesso_2023}. The stellar-mass error bars correspond to the uncertainty on the median logarithmic stellar mass reported in the CANDELS catalog, while the SED-derived SFRs, for which the catalog does not provide per-object uncertainties, are assigned a representative systematic uncertainty of $\sim$0.3~dex, consistent with the typical errors of SED-based SFR estimates at z$<$3 \citep{Barro_2019}. Defining the starburst regime as the region >$1\sigma$ above the MS, we find that $\sim$88\percent\ and $\sim$96\percent\ of the \OIIIHbeta\ and \Halpha\ photometric samples are experiencing a starburst phase. The spectroscopic subsample confirms this trend, with 94.7\percent\ of the spectroscopically confirmed sources also qualifying as starbursts. Our results are consistent with previous EELG studies: VDW report a starburst fraction of 96.9\percent\ for their sample at z$\sim$1.7, while essentially the full sample of \citet{gimenez_alcazar_2026} at z$\sim$0.4 falls into this regime. Although the fractions found here are marginally lower, they remain statistically comparable to those resulting from the stricter VDW criterion and the bright, massive selection of \citet{gimenez_alcazar_2026}. This demonstrates the effectiveness of our more complete selection. It successfully recovers a similar proportion of extreme starbursts while simultaneously extending our sensitivity to more moderate emitters of lower sSFR and higher stellar mass that lie closer to the main sequence.

The central panels of Fig.~\ref{sfr_ew_ssfr_mass} show the specific star-formation rate as a function of equivalent width. Neither the \OIIIHbeta\ nor the \Halpha\ photometric samples show a significant correlation, as confirmed by Spearman rank tests ($\rho$=0.08 and 0.02). Similarly, the spectroscopic sample exhibits no significant correlation ($\rho$=0.22). This lack of correlation can be attributed to the narrow range of EW probed by our selection: since our sample is restricted to high-EW emitters by construction, the dynamic range is insufficient to reveal the underlying trend \citep{lumbreras_2022}. Our results are consistent with VDW at $z \sim 1.7$, whose sample spans a similarly restricted EW range. In contrast, \citet{gimenez_alcazar_2026} at $z \sim 0.4$ do find a significant correlation. Although their study targets brighter and more massive starbursts, their selection actually spans a broader dynamic range that extends toward lower EW values. Conversely, our sample reaches more extreme EWs but remains confined to a narrower window, preventing the correlation from emerging.

The right panels of Fig.~\ref{sfr_ew_ssfr_mass} show the equivalent width as a function of stellar mass. As in the previous relation, neither the \OIIIHbeta\ nor the \Halpha\ photometric samples show a significant correlation (Spearman $\rho$= -0.18 and -0.07), again attributable to the restricted EW range imposed by our selection criteria; the spectroscopic \OIIIHbeta\ subsample likewise shows no significant anticorrelation ($\rho = -0.25$). Our results are consistent with VDW at $z \sim 1.7$, whose sample spans a comparably narrow EW range. The contours of \citet{gimenez_alcazar_2026} also show a lack of significant correlation. This consistency across different regimes from their bright starbursts to our more extreme EW emitters could be due to selection effects or the high sensitivity of EW(\Halpha) to the stellar population age; since EW(\Halpha) decays rapidly after the first 10 Myr of the star formation activity, it can drive the observed scatter, which dominates over the accumulated stellar mass \citep{boyett2024a}.

\subsection{Morphological properties of the EELG sample}

Understanding the morphology and environment of EELGs, as well as their connection to
other properties such as metallicity and specific star-formation rate, is
crucial for gaining insight into galaxy evolution and the physical mechanisms driving
intense star formation \citep{amorin_extreme_2015}. We adopt the deep-learning morphological classifications of \citet{Huertas_Company_2015}, trained to reproduce the visual scheme of \citet{Kartaltepe_2015}, which assigns each galaxy to one of five broad classes following
the Hubble sequence:

\begin{description}
\item[\textit{Disk}] galaxies showing a disk structure, with or without clear spiral arms.
\item[\textit{Spheroid}] centrally concentrated, smooth, and roughly round or ellipsoidal
objects, regardless of size, color, or surface brightness.
\item[\textit{Irregular}] galaxies that do not fit the other classes, including strongly
disturbed objects such as mergers, as well as disks or spheroids with mildly
perturbed morphologies.
\item[\textit{Point source}] clear point sources, unresolved compact galaxies, or objects
too small for their internal structure to be discerned.
\item[\textit{Unclassifiable}] objects that cannot be assigned to any class because of
image artifacts, spurious detections, or insufficient brightness to resolve any
structure.
\end{description}

Within our sample (see Fig.~\ref{morfology_example} and Table~\ref{tab:morfologia}), 223 galaxies ($\sim$97\percent) are classified as spheroids, with only 4 ($\sim$2\percent) disks and 2 ($\sim$1\percent) irregulars. A mosaic of the sample is shown in Fig.~\ref{mosaic}. To probe finer substructure not captured by the automated classification, we additionally performed our own visual inspection, following the interaction- and structure-based criteria commonly applied to blue compact dwarfs \citep{Cairos_2001}. Most systems (128 galaxies, $\sim$56\percent) appear compact and isolated (e.g., EGS-5883; Fig.~\ref{mosaic}, row~3, column~3), consistent with previous work (VDW, \citealt{lumbreras_2022, bonatto2026}). Some display clumpy or chain-like morphologies dominated by one or several high-surface-brightness clumps (20 galaxies, $\sim$9\percent; e.g., EGS-8850; Fig.~\ref{mosaic} row~5, column~4), similar to EELGs observed at intermediate and high redshift \citep{amorin_extreme_2015, llerena_physical_2024}. We also identify 26 systems ($\sim$11\percent) hosting one or more bright star-forming knots near the edge of an extended, low-surface-brightness component (e.g., EGS-5613;  Fig.~\ref{mosaic}, row~3, column~1), reminiscent of asymmetric or "cometary" starbursts. Finally, a substantial fraction of the sample (55 galaxies, $\sim$24\percent) shows signs of interaction or morphological disturbance, suggestive of mergers or interacting pairs (e.g., EGS-12787; Fig.~\ref{mosaic}, row~7, column~9). Overall, given the limited spatial resolution, the majority of the sample is best described as compact and centrally concentrated.

\section{Connection to the EoR}
\label{sec:connection_eor}

The current paradigm of galaxy evolution describes a cosmic downsizing scenario whereby the most massive galaxies completed their assembly and quenched their star formation at very early epochs, shifting the most efficient starburst activity toward progressively less massive galaxies at lower redshifts \citep{Cowie_1996}. Characterized by extremely high $\text{sSFR}$ and low metallicities, EELGs constitute ideal laboratories to test this scenario. To this end, we combined both our \OIIIHbeta and \Halpha EELG samples to perform a linear regression of the form $\log M = M_0 + \beta \log(1+z)$. Prior to the fitting procedure, we estimated the stellar mass completeness limit using the empirical method described by \citet{Zalesky_2025}, which is a modification of the \citet{Pozzetti_2010} approach. First, we computed a rescaled stellar mass ($M_{\mathrm{resc}}$) for each galaxy, defined as the mass it would have if its apparent magnitude were equal to the catalog's magnitude limit. This is calculated using the following equation: $\log_{10}(M_{\mathrm{resc}}/\mathrm{M}_{\odot})$ = \logmass + 0.4(m$_{\mathrm{F160W}} - 27.6)$. Subsequently, we selected the 30\percent\ of galaxies with the lowest stellar masses to represent the population closest to the completeness threshold. Finally, we defined the stellar mass limit as the 95th percentile of the $M_{\mathrm{resc}}$ values within this subsample. To perform the fit we employed a Bayesian linear regression with a Gaussian likelihood, sampling the posterior distributions of the parameters via Markov Chain Monte Carlo. Because both samples are subject to the stellar mass completeness limit derived above, fitting an ordinary Gaussian likelihood to the surviving galaxies would bias the inferred parameters. To account for this, we truncated the Gaussian likelihood at the completeness threshold, renormalizing the model density by the probability mass lying below it. The fit yields

\begin{equation}
\log M/M_\odot = (6.71 \pm 0.16) + (3.31 \pm 0.46) \log(1+z).
\end{equation}

which is shown as a dashed line in Fig.~\ref{evolucion_cosmica} with $1\sigma$ dispersion. The stellar mass of the EELG population therefore evolves significantly with redshift over the range probed by our data. To test the downsizing scenario more directly, we followed the original diagnostic of \citet{Cowie_1996} by comparing the upper end of the stellar mass distributions of both subsamples. Characterizing this upper limit by the 90th percentile, we find \logmass = 8.56 $\pm$ 0.07 for the \OIIIHbeta\ sample at $z \sim 1.7$, which decreases to $8.29 \pm 0.11$ for the \Halpha\ sample at $z \sim 1$. This drop of $0.27$~dex toward lower redshifts suggests that the most massive galaxies progressively cease to sustain the extreme emission-line equivalent widths that define the EELG population, as expected in a downsizing scenario. Overplotted on the best-fit relation, we show our \OIIIHbeta\ (blue) and \Halpha\ (red) subsamples, along with several literature EELG samples at various redshifts (VDW; \citealt{lumbreras_2022, boyett2024a, llerena_physical_2024,gimenez_alcazar_2026}). Additionally, we include a sample of non-EELGs from the CEERS survey at \z{phot}$> 6$ with rest-frame $\mathrm{EW}(\OIIIHbeta) < 680$~\AA\ (see \citealt{Finkelstein_2023} for further details on the data) to represent a general galaxy population in EoR. To select this sample, we followed a procedure similar to the control sample selection described in \citet{llerena_physical_2024}. We note a general lack of agreement with some of the literature samples, which we attribute to the different selection criteria and pipelines employed across studies. Indeed, our results are fully compatible with those of VDW, where both the selection method and data are similar to ours.

To enable a more direct comparison with the EoR, we present the parameter distributions in Fig.~\ref{histogram_eor}. These histograms include our two EELG samples, the EoR EELG sample from \citet{llerena_physical_2024}, and a reference sample of general galaxies (non-EELGs) drawn from the CEERS survey at similar redshifts (see Table~\ref{tab:histogram}). The stellar masses of our \Halpha\ and \OIIIHbeta\ samples are highly consistent with those reported by \citet{llerena_physical_2024}, showing offsets of only $-0.16$ and $+0.39$~dex, respectively. Thus, all three extreme emitter samples are clearly located within the dwarf galaxy regime. In contrast, the general galaxy population from CEERS is systematically more massive than all the EELG samples, exhibiting offsets exceeding $0.5$~dex. The equivalent width distributions provide the most robust comparison among the different catalogs, as they are measured directly from the photometric flux excess attributed to line emission. Our \Halpha\ and \OIIIHbeta\ samples exhibit excellent agreement with the EELGs from \citet{llerena_physical_2024}, with differences of merely $0.14$~dex and $0.11$~dex, respectively. The comparison with the CEERS population, however, behaves differently for each line. We observe an offset of $0.14$~dex for \Halpha, which depends predominantly on recent star formation \citep{rinaldi2024} and a larger offset of $0.40$~dex for \OIIIHbeta, since the latter is also highly sensitive to variations in metallicity and the ionization parameter \citep{Caputi_2024}. Regarding the sSFR, the values of our samples ($-7.98$ and $-7.95$~\ssfrunit\ for \Halpha\ and \OIIIHbeta, respectively) are comparable to those obtained by \citet{jaiswar_mosel_2024} for extreme emitters at $z>5.5$, reinforcing the analog nature of our objects. Nevertheless, compared to the $z\sim5$ EELGs from \citet{llerena_physical_2024}, we find values that are $\sim1$~dex lower, corresponding to stellar mass doubling times of $\sim100$~Myr versus $\sim10$~Myr. Although differences in selection criteria and SFR derivation methods prevent a strict quantitative comparison between both works, the same trend is observed within \citet{jaiswar_mosel_2024}, whose emitters at $z\sim3.4$ present lower sSFRs than those at $z>5.5$. The fact that this decline is reproduced across independent samples using different methodologies suggests an evolution of the specific star formation activity of extreme emitters with redshift. Comparing our EELGs with the CEERS sample yields typical doubling times of $\sim1$~Gyr for the latter, confirming that equivalent width selection effectively isolates galaxies undergoing strong starbursts. In summary, these results validate our EELGs as excellent analogs to the extreme emitters of the EoR, and therefore as ideal laboratories for studying the physical processes that dominated that epoch. Furthermore, both the mass-redshift relation derived from our EELGs and the secular decrease in sSFR toward lower redshifts point to a possible evolution of the properties of this population across cosmic time.

\begin{table}
   \caption{Parameter median with 16th and 84th percentiles for our sample of EELGs, \cite{llerena_physical_2024} EELGs and CEERS sample.}
   \centering
   \setlength{\tabcolsep}{2pt} 
   \renewcommand{\arraystretch}{1.4}
   \resizebox{\columnwidth}{!}{
   \begin{tabular}{l c c c c}
      \hline
      \hline
      \noalign{\smallskip}
                & EELGs & EELGs & EELGs  & non-EELGs  \\
                & \Halpha & \OIIIHbeta & Llerena+24& CEERS\\ 
      \hline
      \noalign{\smallskip}
      Number of galaxies & 127 & 102 & 736 & 774 \\
      \logmass                       & \updown{7.61}{0.51}{-0.31}    & \updown{8.16}{0.30}{-0.33} & \updown{7.77}{0.54}{-0.38} & \updown{8.63}{0.53}{-0.35} \\
      log(sSFR) [\ssfrunit]& \updown{-7.98}{0.22}{-0.34} & \updown{-7.95}{0.24}{-0.17}& \updown{-7.00}{0.40}{-0.49}& \updown{-9.06}{0.19}{-0.27} \\
      log(EW(\Halpha)) [\AA] & \updown{2.73}{0.16}{-0.13}    & -                         & \updown{2.87}{0.20}{-0.18} & \updown{2.59}{0.24}{-0.31} \\
      log(EW(\OIIIHbeta))  [\AA]& -                          & \updown{2.94}{0.14}{-0.09} & \updown{3.05}{0.17}{-0.14} & \updown{2.54}{0.21}{-0.29} \\
      \noalign{\smallskip}
      \hline
      \hline
   \end{tabular}
   }

   \label{tab:histogram}
\end{table}

\section{Summary and conclusions}
\label{sec:conclusions}
We have used CANDELS broadband photometry of the EGS field to assemble a sample of EELGs across the cosmic noon. Candidates were selected through a color-excess technique: for the \OIIIHbeta\ emitters we adopted a less strict version of the VDW criterion that prioritizes completeness over a strict contamination control, and we extended the same framework to define a new criterion for \Halpha\ emitters. Using the F814W, F125W and F160W filters we identified EELGs over $0.67 < z < 1.90$, obtaining their physical properties from the M14 SED-fitting set provided by the CANDELS catalog. We further constructed control samples restricted to the same ranges of stellar mass, magnitude, and redshift, and validated our selection with spectroscopic follow-up from MMT/MMIRS, Keck/MOSFIRE, HST/WFC3, JWST/NIRSpec and our own GTC/OSIRIS+ program. Our main results are:

\begin{itemize}

\item Our final sample comprises 102 \OIIIHbeta\ candidates at 1.40<z< 1.90 and 127 \Halpha\ candidates at 0.67<z< 1.14, identified in the EGS field. Both populations are characterized by a median stellar mass of \logmass $\sim$ 8 and a median of sSFR$\sim 10^{-8}$ \ssfrunit.

\item We estimate high sample purities of $84.3 \pm 1.5$\percent  for the \OIIIHbeta\ emitters and $76.8 \pm 2.0$\percent for the \Halpha\ emitters based on the integrated photometric-redshift probabilities. These estimates are validated by our spectroscopic subsample, which yields actual confirmation rates of $94.1\%$ and $62.5\%$ for the \OIIIHbeta\ and \Halpha\ samples, respectively.

\item The lack of robust IRAC-AGN detections \citep{Donley_2012}, with only one marginal, low-S/N candidate identified, and the location of all confirmed sources within the star-forming locus of the MEx diagram \citep{Coil_2015} indicate that AGN contamination is negligible. 

\item EELGs are significantly fainter, less massive, and more strongly star-forming than the general galaxy population at the same epoch, but they are not significantly more compact. 

\item The vast majority of the photometric sample (88\percent\ of the \OIIIHbeta\ and 96\percent\ of the \Halpha\ candidates) lies above the star-forming main sequence. This is further corroborated by the \OIIIHbeta\ spectroscopic subsample, where 94.7\percent\ of the confirmed sources also lie above the main sequence, ultimately confirming that EELGs occupy the starburst regime.

\item Our sample identifies the same physical class of mostly compact, strongly star-forming emission-line galaxies as the VDW reference. 

\item A joint Bayesian fit of the stellar mass as a function of redshift, corrected for completeness via a truncated likelihood, yields 
$\log(M_\star/\mathrm{M}_\odot) = (6.71 \pm 0.16) + (3.31 \pm 0.46)\,\log(1+z)$. 
This significant positive slope suggests a possible evolution of the mass in our sample over the cosmic timescale analyzed, further supported by 
the decrease in sSFR toward lower redshifts reported in EELG studies at $z>5$ \citep{llerena_physical_2024, jaiswar_mosel_2024}.

\item Our EELGs show stellar masses and equivalent widths consistent with those of the $z\sim5$ EELGs from \citet{llerena_physical_2024}, as well as sSFRs 
comparable to those of extreme emitters at $z>5.5$ \citep{jaiswar_mosel_2024}, and are clearly offset from the general galaxy population at the same epoch. 
This establishes them as intermediate-redshift analogs of the extreme emitters  of the EoR.
 
These results establish cosmic-noon EELGs as accessible laboratories for studying the physical conditions of the galaxies that drove reionization. However, deeper spectroscopic observations will be essential to confirm the nature of the full sample and to fully characterize these systems. Facilities such as JWST/NIRSpec and, in the coming decade, the MOSAIC multi-object spectrograph  \citep{mosaic_2026} on the Extremely Large Telescope (ELT)  will provide the sensitivity and multiplexing needed to extend this analysis to complete samples of faint EELGs.
\end{itemize}

\begin{acknowledgements}

The authors acknowledge the support from Agencia Estatal de Investigación del Ministerio de Ciencia e Innovación and the European Regional Development Fund (ERDF "A way of making Europe") through grants PID2021-123417OB-I00 and PCI2022-135023-2. MDM acknowledges the grant PRE2022-103940, funded by MCIN/AEI/10.13039/501100011033 and by ESF+.

This work is based on observations taken by the CANDELS Multi-Cycle Treasury Program with the NASA/ESA HST, which is operated by the Association of Universities for Research in Astronomy, Inc., under NASA contract NAS5-26555.

This work is based in part on observations made with the Gran Telescopio Canarias (GTC), installed at the Spanish Observatorio del Roque de los Muchachos of the Instituto de Astrofísica de Canarias, on the island of La Palma. The observations were carried out under program GTC84-25A.

This work is based in part on observations made with the NASA/ESA/CSA James
Webb Space Telescope. The data were obtained from the Mikulski Archive for
Space Telescopes at the Space Telescope Science Institute, which is operated by
the Association of Universities for Research in Astronomy, Inc., under NASA
contract NAS 5-03127 for JWST.

Some of the data products presented herein were retrieved from the Dawn JWST Archive (DJA). DJA is an initiative of the Cosmic Dawn Center (DAWN), which is funded by the Danish National Research Foundation under grant DNRF140.

This research made use of \texttt{PypeIt} version
1.17.4,\footnote{\url{https://pypeit.readthedocs.io/en/stable/}} a Python
package for semi-automated reduction of astronomical slit-based spectroscopy
\citep{pypeit:zenodo,pypeit:joss_pub}.

This work made use of Astropy: \footnote{https://www.astropy.org} a community-developed core Python package and an ecosystem of tools and resources for astronomy \citep{astropy:2013, astropy:2018, astropy:2022}.

\end{acknowledgements}

\bibliographystyle{aa}
\bibliography{biblio}

@article{vanderWel2011,
  author = {Van der Wel, A. and Straughn, A. N. and Rix, H.-W. and Finkelstein, S. L. and Koekemoer, A. M. and Weiner, B. J. and Wuyts, S. and Bell, E. F. and Faber, S. M. and Trump, J. R. and Koo, D. C. and Ferguson, H. C. and Scarlata, C. and Hathi, N. P. and Dunlop, J. S. and Newman, J. A. and Dickinson, M. and Jahnke, K. and Salmon, B. W. and de Mello, D. F.},
  title = {Extreme Emission-line Galaxies in CANDELS: Broadband-selected, Starbursting Dwarf Galaxies at z > 1},
  journal = {\apj},
  volume = {742},
  number = {2},
  pages = {111},
  year = {2011},
  doi = {10.1088/0004-637X/742/2/111}
}

@article{Pascual_2001,
   title={{H$\alpha$} emitting galaxies and the star formation rate density at $z \simeq 0.24$},
   volume={379},
   ISSN={1432-0746},
   url={http://dx.doi.org/10.1051/0004-6361:20011389},
   DOI={10.1051/0004-6361:20011389},
   number={3},
   journal={\aap},
   publisher={EDP Sciences},
   author={Pascual, S. and Gallego, J. and Arag{\'o}n-Salamanca, A. and Zamorano, J.},
   year={2001},
   month=dec, 
   pages={798--806}
}

@ARTICLE{Maseda_2014,
       author = {{Maseda}, Michael V. and {van der Wel}, Arjen and {Rix}, Hans-Walter and {da Cunha}, Elisabete and {Pacifici}, Camilla and {Momcheva}, Ivelina and {Brammer}, Gabriel B. and {Meidt}, Sharon E. and {Franx}, Marijn and {van Dokkum}, Pieter and {Fumagalli}, Mattia and {Bell}, Eric F. and {Ferguson}, Henry C. and {F{\"o}rster-Schreiber}, Natascha M. and {Koekemoer}, Anton M. and {Koo}, David C. and {Lundgren}, Britt F. and {Marchesini}, Danilo and {Nelson}, Erica J. and {Patel}, Shannon G. and {Skelton}, Rosalind E. and {Straughn}, Amber N. and {Trump}, Jonathan R. and {Whitaker}, Katherine E.},
        title = "{The Nature of Extreme Emission Line Galaxies at z = 1-2: Kinematics and Metallicities from Near-infrared Spectroscopy}",
      journal = {\apj},
         year = 2014,
        month = aug,
       volume = {791},
       number = {1},
          eid = {17},
        pages = {17},
          doi = {10.1088/0004-637X/791/1/17},
archivePrefix = {arXiv},
       eprint = {1406.3351},
 primaryClass = {astro-ph.GA},
       adsurl = {https://ui.adsabs.harvard.edu/abs/2014ApJ...791...17M}
}

@ARTICLE{Endsley_2023,
       author = {{Endsley}, Ryan and {Stark}, Daniel P. and {Whitler}, Lily and {Topping}, Michael W. and {Chen}, Zuyi and {Plat}, Ad{\`e}le and {Chisholm}, John and {Charlot}, St{\'e}phane},
        title = "{A JWST/NIRCam study of key contributors to reionization: the star-forming and ionizing properties of UV-faint z   7-8 galaxies}",
      journal = {\mnras},
         year = 2023,
        month = sep,
       volume = {524},
       number = {2},
        pages = {2312-2330},
          doi = {10.1093/mnras/stad1919},
archivePrefix = {arXiv},
       eprint = {2208.14999},
 primaryClass = {astro-ph.GA},
       adsurl = {https://ui.adsabs.harvard.edu/abs/2023MNRAS.524.2312E}
}

@ARTICLE{Pozzetti_2010,
       author = {{Pozzetti}, L. and {Bolzonella}, M. and {Zucca}, E. and {Zamorani}, G. and {Lilly}, S. and {Renzini}, A. and {Moresco}, M. and {Mignoli}, M. and {Cassata}, P. and {Tasca}, L. and {Lamareille}, F. and {Maier}, C. and {Meneux}, B. and {Halliday}, C. and {Oesch}, P. and {Vergani}, D. and {Caputi}, K. and {Kova{\v{c}}}, K. and {Cimatti}, A. and {Cucciati}, O. and {Iovino}, A. and {Peng}, Y. and {Carollo}, M. and {Contini}, T. and {Kneib}, J.-P. and {Le F{\'e}vre}, O. and {Mainieri}, V. and {Scodeggio}, M. and {Bardelli}, S. and {Bongiorno}, A. and {Coppa}, G. and {de la Torre}, S. and {de Ravel}, L. and {Franzetti}, P. and {Garilli}, B. and {Kampczyk}, P. and {Knobel}, C. and {Le Borgne}, J.-F. and {Le Brun}, V. and {Pell{\`o}}, R. and {Perez Montero}, E. and {Ricciardelli}, E. and {Silverman}, J.~D. and {Tanaka}, M. and {Tresse}, L. and {Abbas}, U. and {Bottini}, D. and {Cappi}, A. and {Guzzo}, L. and {Koekemoer}, A.~M. and {Leauthaud}, A. and {Maccagni}, D. and {Marinoni}, C. and {McCracken}, H.~J. and {Memeo}, P. and {Porciani}, C. and {Scaramella}, R. and {Scarlata}, C. and {Scoville}, N.},
        title = "{zCOSMOS - 10k-bright spectroscopic sample. The bimodality in the galaxy stellar mass function: exploring its evolution with redshift}",
      journal = {\aap},
         year = 2010,
        month = nov,
       volume = {523},
          eid = {A13},
        pages = {A13},
          doi = {10.1051/0004-6361/200913020},
archivePrefix = {arXiv},
       eprint = {0907.5416},
 primaryClass = {astro-ph.CO},
       adsurl = {https://ui.adsabs.harvard.edu/abs/2010A&A...523A..13P}
}

@article{Carnall_2018,
   title={Inferring the star formation histories of massive quiescent galaxies with bagpipes: evidence for multiple quenching mechanisms},
   volume={480},
   ISSN={1365-2966},
   url={http://dx.doi.org/10.1093/mnras/sty2169},
   DOI={10.1093/mnras/sty2169},
   number={4},
   journal={\mnras},
   publisher={Oxford University Press (OUP)},
   author={Carnall, A C and McLure, R J and Dunlop, J S and Davé, R},
   year={2018},
   month=Aug, pages={4379–4401} }

@ARTICLE{Cowie_1996,
       author = {{Cowie}, Lennox L. and {Songaila}, Antoinette and {Hu}, Esther M. and {Cohen}, J.~G.},
        title = "{New Insight on Galaxy Formation and Evolution From Keck Spectroscopy of the Hawaii Deep Fields}",
      journal = {\aj},
         year = 1996,
        month = sep,
       volume = {112},
        pages = {839},
          doi = {10.1086/118058},
archivePrefix = {arXiv},
       eprint = {astro-ph/9606079},
 primaryClass = {astro-ph},
       adsurl = {https://ui.adsabs.harvard.edu/abs/1996AJ....112..839C}
}

@ARTICLE{Momcheva_2016,
       author = {{Momcheva}, Ivelina G. and {Brammer}, Gabriel B. and {van Dokkum}, Pieter G. and {Skelton}, Rosalind E. and {Whitaker}, Katherine E. and {Nelson}, Erica J. and {Fumagalli}, Mattia and {Maseda}, Michael V. and {Leja}, Joel and {Franx}, Marijn and {Rix}, Hans-Walter and {Bezanson}, Rachel and {Da Cunha}, Elisabete and {Dickey}, Claire and {F{\"o}rster Schreiber}, Natascha M. and {Illingworth}, Garth and {Kriek}, Mariska and {Labb{\'e}}, Ivo and {Ulf Lange}, Johannes and {Lundgren}, Britt F. and {Magee}, Daniel and {Marchesini}, Danilo and {Oesch}, Pascal and {Pacifici}, Camilla and {Patel}, Shannon G. and {Price}, Sedona and {Tal}, Tomer and {Wake}, David A. and {van der Wel}, Arjen and {Wuyts}, Stijn},
        title = "{The 3D-HST Survey: Hubble Space Telescope WFC3/G141 Grism Spectra, Redshifts, and Emission Line Measurements for \raisebox{-0.5ex}\textasciitilde 100,000 Galaxies}",
      journal = {\apjs},
         year = 2016,
        month = aug,
       volume = {225},
       number = {2},
          eid = {27},
        pages = {27},
          doi = {10.3847/0067-0049/225/2/27},
archivePrefix = {arXiv},
       eprint = {1510.02106},
 primaryClass = {astro-ph.GA},
       adsurl = {https://ui.adsabs.harvard.edu/abs/2016ApJS..225...27M}
}

@ARTICLE{Finkelstein_2023,
       author = {{Finkelstein}, Steven L. and {Bagley}, Micaela B. and {Ferguson}, Henry C. and {Wilkins}, Stephen M. and {Kartaltepe}, Jeyhan S. and {Papovich}, Casey and {Yung}, L.~Y. Aaron and {Arrabal Haro}, Pablo and {Behroozi}, Peter and {Dickinson}, Mark and {Kocevski}, Dale D. and {Koekemoer}, Anton M. and {Larson}, Rebecca L. and {Le Bail}, Aur{\'e}lien and {Morales}, Alexa M. and {P{\'e}rez-Gonz{\'a}lez}, Pablo G. and {Burgarella}, Denis and {Dav{\'e}}, Romeel and {Hirschmann}, Michaela and {Somerville}, Rachel S. and {Wuyts}, Stijn and {Bromm}, Volker and {Casey}, Caitlin M. and {Fontana}, Adriano and {Fujimoto}, Seiji and {Gardner}, Jonathan P. and {Giavalisco}, Mauro and {Grazian}, Andrea and {Grogin}, Norman A. and {Hathi}, Nimish P. and {Hutchison}, Taylor A. and {Jha}, Saurabh W. and {Jogee}, Shardha and {Kewley}, Lisa J. and {Kirkpatrick}, Allison and {Long}, Arianna S. and {Lotz}, Jennifer M. and {Pentericci}, Laura and {Pierel}, Justin D.~R. and {Pirzkal}, Nor and {Ravindranath}, Swara and {Ryan}, Russell E. and {Trump}, Jonathan R. and {Yang}, Guang and {Bhatawdekar}, Rachana and {Bisigello}, Laura and {Buat}, V{\'e}ronique and {Calabr{\`o}}, Antonello and {Castellano}, Marco and {Cleri}, Nikko J. and {Cooper}, M.~C. and {Croton}, Darren and {Daddi}, Emanuele and {Dekel}, Avishai and {Elbaz}, David and {Franco}, Maximilien and {Gawiser}, Eric and {Holwerda}, Benne W. and {Huertas-Company}, Marc and {Jaskot}, Anne E. and {Leung}, Gene C.~K. and {Lucas}, Ray A. and {Mobasher}, Bahram and {Pandya}, Viraj and {Tacchella}, Sandro and {Weiner}, Benjamin J. and {Zavala}, Jorge A.},
        title = "{CEERS Key Paper. I. An Early Look into the First 500 Myr of Galaxy Formation with JWST}",
      journal = {\apjl},
         year = 2023,
        month = mar,
       volume = {946},
       number = {1},
          eid = {L13},
        pages = {L13},
          doi = {10.3847/2041-8213/acade4},
archivePrefix = {arXiv},
       eprint = {2211.05792},
 primaryClass = {astro-ph.GA},
       adsurl = {https://ui.adsabs.harvard.edu/abs/2023ApJ...946L..13F}
}

@article{Madau_2014,
   title={Cosmic Star-Formation History},
   volume={52},
   ISSN={1545-4282},
   url={http://dx.doi.org/10.1146/annurev-astro-081811-125615},
   DOI={10.1146/annurev-astro-081811-125615},
   number={1},
   journal={\araa},
   publisher={Annual Reviews},
   author={Madau, Piero and Dickinson, Mark},
   year={2014},
   month=Aug, pages={415–486} }

@ARTICLE{withers2023,
       author = {{Withers}, Sunna and {Muzzin}, Adam and {Ravindranath}, Swara and {Sarrouh}, Ghassan T.~E. and {Abraham}, Roberto and {Asada}, Yoshihisa and {Brada{\v{c}}}, Maru{\v{s}}a and {Brammer}, Gabriel and {Desprez}, Guillaume and {Iyer}, Kartheik and {Martis}, Nicholas and {Mowla}, Lamiya and {Noirot}, Ga{\"e}l and {Sawicki}, Marcin and {Strait}, Victoria and {Willott}, Chris J.},
        title = "{Spectroscopy from Photometry: A Population of Extreme Emission Line Galaxies at 1.7 {\ensuremath{\lesssim}} z {\ensuremath{\lesssim}} 6.7 Selected with JWST Medium Band Filters}",
      journal = {\apjl},
         year = 2023,
        month = nov,
       volume = {958},
       number = {1},
          eid = {L14},
        pages = {L14},
          doi = {10.3847/2041-8213/ad01c0},
archivePrefix = {arXiv},
       eprint = {2304.11181},
 primaryClass = {astro-ph.GA},
       adsurl = {https://ui.adsabs.harvard.edu/abs/2023ApJ...958L..14W}
}

@article{Huang_2015,
   title={CLASH: EXTREME EMISSION-LINE GALAXIES AND THEIR IMPLICATION ON SELECTION OF HIGH-REDSHIFT GALAXIES},
   volume={801},
   ISSN={1538-4357},
   url={http://dx.doi.org/10.1088/0004-637X/801/1/12},
   DOI={10.1088/0004-637x/801/1/12},
   number={1},
   journal={\apj},
   publisher={American Astronomical Society},
   author={Huang, Xingxing and Zheng, Wei and Wang, Junxian and Ford, Holland and Lemze, Doron and Moustakas, John and Shu, Xinwen and Van der Wel, Arjen and Zitrin, Adi and Frye, Brenda L. and Postman, Marc and Bartelmann, Matthias and Benítez, Narciso and Bradley, Larry and Broadhurst, Tom and Coe, Dan and Donahue, Megan and Infante, Leopoldo and Kelson, Daniel and Koekemoer, Anton and Lahav, Ofer and Medezinski, Elinor and Moustakas, Leonidas and Rosati, Piero and Seitz, Stella and Umetsu, Keiichi},
   year={2015},
   month=Feb, pages={12} }

@ARTICLE{Calabro_2017,
       author = {{Calabr{\`o}}, A. and {Amor{\'\i}n}, R. and {Fontana}, A. and {P{\'e}rez-Montero}, E. and {Lemaux}, B.~C. and {Ribeiro}, B. and {Bardelli}, S. and {Castellano}, M. and {Contini}, T. and {De Barros}, S. and {Garilli}, B. and {Grazian}, A. and {Guaita}, L. and {Hathi}, N.~P. and {Koekemoer}, A.~M. and {Le F{\`e}vre}, O. and {Maccagni}, D. and {Pentericci}, L. and {Schaerer}, D. and {Talia}, M. and {Tasca}, L.~A.~M. and {Zucca}, E.},
        title = "{Characterization of star-forming dwarf galaxies at 0.1 {\ensuremath{\lesssim}}z {\ensuremath{\lesssim}} 0.9 in VUDS: probing the low-mass end of the mass-metallicity relation}",
      journal = {\aap},
         year = 2017,
        month = may,
       volume = {601},
          eid = {A95},
        pages = {A95},
          doi = {10.1051/0004-6361/201629762},
archivePrefix = {arXiv},
       eprint = {1701.04418},
 primaryClass = {astro-ph.GA},
       adsurl = {https://ui.adsabs.harvard.edu/abs/2017A&A...601A..95C}
}

@INPROCEEDINGS{Faber_2003,
       author = {{Faber}, Sandra M. and {Phillips}, Andrew C. and {Kibrick}, Robert I. and {Alcott}, Barry and {Allen}, Steven L. and {Burrous}, Jim and {Cantrall}, T. and {Clarke}, De and {Coil}, Alison L. and {Cowley}, David J. and {Davis}, Marc and {Deich}, William T.~S. and {Dietsch}, Ken and {Gilmore}, David K. and {Harper}, Carol A. and {Hilyard}, David F. and {Lewis}, Jeffrey P. and {McVeigh}, Molly and {Newman}, Jeffrey and {Osborne}, Jack and {Schiavon}, Ricardo and {Stover}, Richard J. and {Tucker}, Dean and {Wallace}, Vernon and {Wei}, Mingzhi and {Wirth}, Gregory and {Wright}, Christopher A.},
        title = "{The DEIMOS spectrograph for the Keck II Telescope: integration and testing}",
    booktitle = {Instrument Design and Performance for Optical/Infrared Ground-based Telescopes},
         year = 2003,
       editor = {{Iye}, Masanori and {Moorwood}, Alan F.~M.},
       series = {SPIE Conference Series},
       volume = {4841},
        month = mar,
        pages = {1657-1669},
          doi = {10.1117/12.460346},
       adsurl = {https://ui.adsabs.harvard.edu/abs/2003SPIE.4841.1657F}
}

@ARTICLE{Sanders_2021,
       author = {{Sanders}, Ryan L. and {Shapley}, Alice E. and {Jones}, Tucker and {Reddy}, Naveen A. and {Kriek}, Mariska and {Siana}, Brian and {Coil}, Alison L. and {Mobasher}, Bahram and {Shivaei}, Irene and {Dav{\'e}}, Romeel and {Azadi}, Mojegan and {Price}, Sedona H. and {Leung}, Gene and {Freeman}, William R. and {Fetherolf}, Tara and {de Groot}, Laura and {Zick}, Tom and {Barro}, Guillermo},
        title = "{The MOSDEF Survey: The Evolution of the Mass-Metallicity Relation from z = 0 to z 3.3}",
      journal = {\apj},
         year = 2021,
        month = jun,
       volume = {914},
       number = {1},
          eid = {19},
        pages = {19},
          doi = {10.3847/1538-4357/abf4c1},
archivePrefix = {arXiv},
       eprint = {2009.07292},
 primaryClass = {astro-ph.GA},
       adsurl = {https://ui.adsabs.harvard.edu/abs/2021ApJ...914...19S}
}

@ARTICLE{Kriek_2015,
       author = {{Kriek}, Mariska and {Shapley}, Alice E. and {Reddy}, Naveen A. and {Siana}, Brian and {Coil}, Alison L. and {Mobasher}, Bahram and {Freeman}, William R. and {de Groot}, Laura and {Price}, Sedona H. and {Sanders}, Ryan and {Shivaei}, Irene and {Brammer}, Gabriel B. and {Momcheva}, Ivelina G. and {Skelton}, Rosalind E. and {van Dokkum}, Pieter G. and {Whitaker}, Katherine E. and {Aird}, James and {Azadi}, Mojegan and {Kassis}, Marc and {Bullock}, James S. and {Conroy}, Charlie and {Dav{\'e}}, Romeel and {Kere{\v{s}}}, Du{\v{s}}an and {Krumholz}, Mark},
        title = "{The MOSFIRE Deep Evolution Field (MOSDEF) Survey: Rest-frame Optical Spectroscopy for \raisebox{-0.5ex}\textasciitilde1500 H-selected Galaxies at 1.37 < z < 3.8}",
      journal = {\apjs},
         year = 2015,
        month = jun,
       volume = {218},
       number = {2},
          eid = {15},
        pages = {15},
          doi = {10.1088/0067-0049/218/2/15},
archivePrefix = {arXiv},
       eprint = {1412.1835},
 primaryClass = {astro-ph.GA},
       adsurl = {https://ui.adsabs.harvard.edu/abs/2015ApJS..218...15K}
}

@INPROCEEDINGS{Acquaviva_2012,
       author = {{Acquaviva}, Viviana and {Gawiser}, Eric and {Guaita}, Lucia},
        title = "{SED fitting with MCMC: methodology and application to large galaxy surveys}",
    booktitle = {The Spectral Energy Distribution of Galaxies - SED 2011},
         year = 2012,
       editor = {{Tuffs}, Richard J. and {Popescu}, Cristina C.},
       series = {IAU Symposium},
       volume = {284},
        month = aug,
        pages = {42-45},
          doi = {10.1017/S1743921312008691},
archivePrefix = {arXiv},
       eprint = {1111.4243},
 primaryClass = {astro-ph.CO},
       adsurl = {https://ui.adsabs.harvard.edu/abs/2012IAUS..284...42A}
}

@article{Atek_2011,
   title={VERY STRONG EMISSION-LINE GALAXIES IN THE WFC3 INFRARED SPECTROSCOPIC PARALLEL SURVEY AND IMPLICATIONS FOR HIGH-REDSHIFT GALAXIES<sup>,</sup>},
   volume={743},
   ISSN={1538-4357},
   url={http://dx.doi.org/10.1088/0004-637X/743/2/121},
   DOI={10.1088/0004-637x/743/2/121},
   number={2},
   journal={\apj},
   publisher={American Astronomical Society},
   author={Atek, H. and Siana, B. and Scarlata, C. and Malkan, M. and McCarthy, P. and Teplitz, H. and Henry, A. and Colbert, J. and Bridge, C. and Bunker, A. J. and Dressler, A. and Fosbury, R. A. E. and Hathi, N. P. and Martin, C. and Ross, N. R. and Shim, H.},
   year={2011},
   month=Nov, pages={121} }

@ARTICLE{Coil_2015,
       author = {{Coil}, Alison L. and {Aird}, James and {Reddy}, Naveen and {Shapley}, Alice E. and {Kriek}, Mariska and {Siana}, Brian and {Mobasher}, Bahram and {Freeman}, William R. and {Price}, Sedona H. and {Shivaei}, Irene},
        title = "{The MOSDEF Survey: Optical Active Galactic Nucleus Diagnostics at z \raisebox{-0.5ex}\textasciitilde 2.3}",
      journal = {\apj},
         year = 2015,
        month = mar,
       volume = {801},
       number = {1},
          eid = {35},
        pages = {35},
          doi = {10.1088/0004-637X/801/1/35},
archivePrefix = {arXiv},
       eprint = {1409.6522},
 primaryClass = {astro-ph.GA},
       adsurl = {https://ui.adsabs.harvard.edu/abs/2015ApJ...801...35C}
}

@ARTICLE{Dahlen_2013,
       author = {{Dahlen}, Tomas and {Mobasher}, Bahram and {Faber}, Sandra M. and {Ferguson}, Henry C. and {Barro}, Guillermo and {Finkelstein}, Steven L. and {Finlator}, Kristian and {Fontana}, Adriano and {Gruetzbauch}, Ruth and {Johnson}, Seth and {Pforr}, Janine and {Salvato}, Mara and {Wiklind}, Tommy and {Wuyts}, Stijn and {Acquaviva}, Viviana and {Dickinson}, Mark E. and {Guo}, Yicheng and {Huang}, Jiasheng and {Huang}, Kuang-Han and {Newman}, Jeffrey A. and {Bell}, Eric F. and {Conselice}, Christopher J. and {Galametz}, Audrey and {Gawiser}, Eric and {Giavalisco}, Mauro and {Grogin}, Norman A. and {Hathi}, Nimish and {Kocevski}, Dale and {Koekemoer}, Anton M. and {Koo}, David C. and {Lee}, Kyoung-Soo and {McGrath}, Elizabeth J. and {Papovich}, Casey and {Peth}, Michael and {Ryan}, Russell and {Somerville}, Rachel and {Weiner}, Benjamin and {Wilson}, Grant},
        title = "{A Critical Assessment of Photometric Redshift Methods: A CANDELS Investigation}",
      journal = {\apj},
         year = 2013,
        month = oct,
       volume = {775},
       number = {2},
          eid = {93},
        pages = {93},
          doi = {10.1088/0004-637X/775/2/93},
archivePrefix = {arXiv},
       eprint = {1308.5353},
 primaryClass = {astro-ph.CO},
       adsurl = {https://ui.adsabs.harvard.edu/abs/2013ApJ...775...93D}
}

@ARTICLE{deGraaff_2025,
       author = {{de Graaff}, Anna and {Brammer}, Gabriel and {Weibel}, Andrea and {Lewis}, Zach and {Maseda}, Michael V. and {Oesch}, Pascal A. and {Bezanson}, Rachel and {Boogaard}, Leindert A. and {Cleri}, Nikko J. and {Cooper}, Olivia R. and {Gottumukkala}, Rashmi and {Greene}, Jenny E. and {Hirschmann}, Michaela and {Hviding}, Raphael E. and {Katz}, Harley and {Labb{\'e}}, Ivo and {Leja}, Joel and {Matthee}, Jorryt and {McConachie}, Ian and {Miller}, Tim B. and {Naidu}, Rohan P. and {Price}, Sedona H. and {Rix}, Hans-Walter and {Setton}, David J. and {Suess}, Katherine A. and {Wang}, Bingjie and {Whitaker}, Katherine E. and {Williams}, Christina C.},
        title = "{RUBIES: A complete census of the bright and red distant Universe with JWST/NIRSpec}",
      journal = {\aap},
         year = 2025,
        month = may,
       volume = {697},
          eid = {A189},
        pages = {A189},
          doi = {10.1051/0004-6361/202452186},
archivePrefix = {arXiv},
       eprint = {2409.05948},
 primaryClass = {astro-ph.GA},
       adsurl = {https://ui.adsabs.harvard.edu/abs/2025A&A...697A.189D}
}

@ARTICLE{Heintz_2024,
       author = {{Heintz}, Kasper E. and {Watson}, Darach and {Brammer}, Gabriel and {Vejlgaard}, Simone and {Hutter}, Anne and {Strait}, Victoria B. and {Matthee}, Jorryt and {Oesch}, Pascal A. and {Jakobsson}, P{\'a}ll and {Tanvir}, Nial R. and {Laursen}, Peter and {Naidu}, Rohan P. and {Mason}, Charlotte A. and {Killi}, Meghana and {Jung}, Intae and {Hsiao}, Tiger Yu-Yang and {Abdurro'uf} and {Coe}, Dan and {Arrabal Haro}, Pablo and {Finkelstein}, Steven L. and {Toft}, Sune},
        title = "{Strong damped Lyman-{\ensuremath{\alpha}} absorption in young star-forming galaxies at redshifts 9 to 11}",
      journal = {Science},
         year = 2024,
        month = may,
       volume = {384},
       number = {6698},
        pages = {890-894},
          doi = {10.1126/science.adj0343},
archivePrefix = {arXiv},
       eprint = {2306.00647},
 primaryClass = {astro-ph.GA},
       adsurl = {https://ui.adsabs.harvard.edu/abs/2024Sci...384..890H}
}

@article{Huertas_Company_2015,
   title={A CATALOG OF VISUAL-LIKE MORPHOLOGIES IN THE 5 CANDELS FIELDS USING DEEP LEARNING},
   volume={221},
   ISSN={1538-4365},
   url={http://dx.doi.org/10.1088/0067-0049/221/1/8},
   DOI={10.1088/0067-0049/221/1/8},
   number={1},
   journal={\apjs},
   publisher={American Astronomical Society},
   author={Huertas-Company, M. and Gravet, R. and Cabrera-Vives, G. and Pérez-González, P. G. and Kartaltepe, J. S. and Barro, G. and Bernardi, M. and Mei, S. and Shankar, F. and Dimauro, P. and Bell, E. F. and Kocevski, D. and Koo, D. C. and Faber, S. M. and Mcintosh, D. H.},
   year={2015},
   month=oct, pages={8} }

@ARTICLE{jaiswar_mosel_2024,
       author = {{Jaiswar}, Ravi and {Gupta}, Anshu and {da Cunha}, Elisabete and {Trott}, Cathryn M. and {Harshan}, Anishya and {Battisti}, Andrew and {Forrest}, Ben},
        title = "{MOSEL survey: Unwrapping the Epoch of Reionisation through mimic galaxies at Cosmic Noon}",
      journal = {\pasa},
         year = 2024,
        month = jan,
       volume = {41},
          eid = {e040},
        pages = {e040},
          doi = {10.1017/pasa.2024.37},
archivePrefix = {arXiv},
       eprint = {2405.04870},
 primaryClass = {astro-ph.GA},
       adsurl = {https://ui.adsabs.harvard.edu/abs/2024PASA...41...40J}
}

@article{cohn_zfourge_2018,
	title = {zfourge: {Extreme} 5007 Å {Emission} {May} {Be} a {Common} {Early}-lifetime {Phase} for {Star}-forming {Galaxies} at z {\textgreater} 2.5},
	volume = {869},
	issn = {0004-637X, 1538-4357},
	shorttitle = {zfourge},
	url = {https://iopscience.iop.org/article/10.3847/1538-4357/aaed3d},
	doi = {10.3847/1538-4357/aaed3d},
	language = {en},
	number = {2},
	urldate = {2024-11-01},
	journal = {\apj},
	author = {Cohn, Jonathan H. and Leja, Joel and Tran, Kim-Vy H. and Forrest, Ben and Johnson, Benjamin D. and Tillman, Megan and Alcorn, Leo and Conroy, Charlie and Glazebrook, Karl and Kacprzak, Glenn G. and Kelson, Daniel D. and Nanayakkara, Themiya and Papovich, Casey and Van Dokkum, Pieter G. and Yuan, Tiantian},
	month = dec,
	year = {2018},
	pages = {141},
}

@article{koekemoer_candels_2011,
	title = {{CANDELS}: {THE} {COSMIC} {ASSEMBLY} {NEAR}-{INFRARED} {DEEP} {EXTRAGALACTIC} {LEGACY} {SURVEY}—{THE} \textit{{HUBBLE} {SPACE} {TELESCOPE}} {OBSERVATIONS}, {IMAGING} {DATA} {PRODUCTS}, {AND} {MOSAICS}},
	volume = {197},
	issn = {0067-0049, 1538-4365},
	shorttitle = {{CANDELS}},
	url = {https://iopscience.iop.org/article/10.1088/0067-0049/197/2/36},
	doi = {10.1088/0067-0049/197/2/36},
	language = {en},
	number = {2},
	urldate = {2024-11-01},
	journal = {\apjs},
	author = {Koekemoer, Anton M. and Faber, S. M. and Ferguson, Henry C. and Grogin, Norman A. and Kocevski, Dale D. and Koo, David C. and Lai, Kamson and Lotz, Jennifer M. and Lucas, Ray A. and McGrath, Elizabeth J. and Ogaz, Sara and Rajan, Abhijith and Riess, Adam G. and Rodney, Steve A. and Strolger, Louis and Casertano, Stefano and Castellano, Marco and Dahlen, Tomas and Dickinson, Mark and Dolch, Timothy and Fontana, Adriano and Giavalisco, Mauro and Grazian, Andrea and Guo, Yicheng and Hathi, Nimish P. and Huang, Kuang-Han and Van Der Wel, Arjen and Yan, Hao-Jing and Acquaviva, Viviana and Alexander, David M. and Almaini, Omar and Ashby, Matthew L. N. and Barden, Marco and Bell, Eric F. and Bournaud, Frédéric and Brown, Thomas M. and Caputi, Karina I. and Cassata, Paolo and Challis, Peter J. and Chary, Ranga-Ram and Cheung, Edmond and Cirasuolo, Michele and Conselice, Christopher J. and Cooray, Asantha Roshan and Croton, Darren J. and Daddi, Emanuele and Davé, Romeel and De Mello, Duilia F. and De Ravel, Loic and Dekel, Avishai and Donley, Jennifer L. and Dunlop, James S. and Dutton, Aaron A. and Elbaz, David and Fazio, Giovanni G. and Filippenko, Alexei V. and Finkelstein, Steven L. and Frazer, Chris and Gardner, Jonathan P. and Garnavich, Peter M. and Gawiser, Eric and Gruetzbauch, Ruth and Hartley, Will G. and Häussler, Boris and Herrington, Jessica and Hopkins, Philip F. and Huang, Jia-Sheng and Jha, Saurabh W. and Johnson, Andrew and Kartaltepe, Jeyhan S. and Khostovan, Ali A. and Kirshner, Robert P. and Lani, Caterina and Lee, Kyoung-Soo and Li, Weidong and Madau, Piero and McCarthy, Patrick J. and McIntosh, Daniel H. and McLure, Ross J. and McPartland, Conor and Mobasher, Bahram and Moreira, Heidi and Mortlock, Alice and Moustakas, Leonidas A. and Mozena, Mark and Nandra, Kirpal and Newman, Jeffrey A. and Nielsen, Jennifer L. and Niemi, Sami and Noeske, Kai G. and Papovich, Casey J. and Pentericci, Laura and Pope, Alexandra and Primack, Joel R. and Ravindranath, Swara and Reddy, Naveen A. and Renzini, Alvio and Rix, Hans-Walter and Robaina, Aday R. and Rosario, David J. and Rosati, Piero and Salimbeni, Sara and Scarlata, Claudia and Siana, Brian and Simard, Luc and Smidt, Joseph and Snyder, Diana and Somerville, Rachel S. and Spinrad, Hyron and Straughn, Amber N. and Telford, Olivia and Teplitz, Harry I. and Trump, Jonathan R. and Vargas, Carlos and Villforth, Carolin and Wagner, Cory R. and Wandro, Pat and Wechsler, Risa H. and Weiner, Benjamin J. and Wiklind, Tommy and Wild, Vivienne and Wilson, Grant and Wuyts, Stijn and Yun, Min S.},
	month = dec,
	year = {2011},
	pages = {36},
}

@article{boyett_o_2022,
	title = {The [{O} {III}]λ5007 equivalent width distribution at z ∼ 2: the redshift evolution of the extreme emission line galaxies},
	volume = {513},
	copyright = {https://creativecommons.org/licenses/by/4.0/},
	issn = {0035-8711, 1365-2966},
	shorttitle = {The [{O} {III}]λ5007 equivalent width distribution at z ∼ 2},
	url = {https://academic.oup.com/mnras/article/513/3/4451/6573888},
	doi = {10.1093/mnras/stac1109},
	language = {en},
	number = {3},
	urldate = {2024-11-01},
	journal = {\mnras},
	author = {Boyett, Kristan N K and Stark, Daniel P and Bunker, Andrew J and Tang, Mengtao and Maseda, Michael V},
	month = may,
	year = {2022},
	pages = {4451--4463},
}

@article{onodera_broadband_2020,
	title = {Broadband {Selection}, {Spectroscopic} {Identification}, and {Physical} {Properties} of a {Population} of {Extreme} {Emission}-line {Galaxies} at 3 {\textless} z {\textless} 3.7*},
	volume = {904},
	issn = {0004-637X, 1538-4357},
	url = {https://iopscience.iop.org/article/10.3847/1538-4357/abc174},
	doi = {10.3847/1538-4357/abc174},
	language = {en},
	number = {2},
	urldate = {2024-11-01},
	journal = {\apj},
	author = {Onodera, Masato and Shimakawa, Rhythm and Suzuki, Tomoko L. and Tanaka, Ichi and Harikane, Yuichi and Hayashi, Masao and Kodama, Tadayuki and Koyama, Yusei and Nakajima, Kimihiko and Shibuya, Takatoshi},
	month = dec,
	year = {2020},
	pages = {180},
}

@ARTICLE{rosenwasser_2022,
       author = {{Rosenwasser}, Benjamin E. and {Taylor}, Anthony J. and {Barger}, Amy J. and {Cowie}, Lennox L. and {Hu}, Esther M. and {Jones}, Logan H. and {Songaila}, Antoinette},
        title = "{Subaru/Hyper Suprime-Cam z-Broadband Excess Selection of Extreme Emission Line Galaxies at z < 1}",
      journal = {\apj},
         year = 2022,
        month = mar,
       volume = {928},
       number = {1},
          eid = {78},
        pages = {78},
          doi = {10.3847/1538-4357/ac5117},
archivePrefix = {arXiv},
       eprint = {2201.10562},
 primaryClass = {astro-ph.GA},
       adsurl = {https://ui.adsabs.harvard.edu/abs/2022ApJ...928...78R}
}

@ARTICLE{moral-castro_extreme_2024,
       author = {{del Moral-Castro}, I. and {V{\'\i}lchez}, J.~M. and {Iglesias-P{\'a}ramo}, J. and {Arroyo-Polonio}, A.},
        title = "{Extreme-emission-line galaxies in the MUSE Hubble Ultra Deep Field Survey}",
      journal = {\aap},
         year = 2024,
        month = aug,
       volume = {688},
          eid = {A28},
        pages = {A28},
          doi = {10.1051/0004-6361/202348185},
archivePrefix = {arXiv},
       eprint = {2404.17415},
 primaryClass = {astro-ph.GA},
       adsurl = {https://ui.adsabs.harvard.edu/abs/2024A&A...688A..28D}
}

@ARTICLE{Kennicutt_1998,
       author = {{Kennicutt}, Jr., Robert C.},
        title = "{The Global Schmidt Law in Star-forming Galaxies}",
      journal = {\apj},
         year = 1998,
        month = may,
       volume = {498},
       number = {2},
        pages = {541-552},
          doi = {10.1086/305588},
archivePrefix = {arXiv},
       eprint = {astro-ph/9712213},
 primaryClass = {astro-ph},
       adsurl = {https://ui.adsabs.harvard.edu/abs/1998ApJ...498..541K}
}

@article{lumbreras_2022,
  title = {J-PLUS: Uncovering a large population of extreme [OIII] emitters in the local Universe},
  shorttitle = {J-PLUS},
  author = {{Lumbreras-Calle}, A. and {L{\'o}pez-Sanjuan}, C. and Sobral, D. and {Fern{\'a}ndez-Ontiveros}, J. A. and V{\'i}lchez, J. M. and {Hern{\'a}n-Caballero}, A. and Akhlaghi, M. and {D{\'i}az-Garc{\'i}a}, L. A. and Alcaniz, J. and Angulo, R. E. and Cenarro, A. J. and {Crist{\'o}bal-Hornillos}, D. and Dupke, R. A. and Ederoclite, A. and {Hern{\'a}ndez-Monteagudo}, C. and {Mar{\'i}n-Franch}, A. and Moles, M. and Sodr{\'e}, L. and Rami{\'o}, H. V{\'a}zquez and Varela, J.},
  year = 2022,
  month = dec,
  volume = {668},
  eprint = {2112.06938},
  primaryclass = {astro-ph},
  pages = {A60},
  issn = {0004-6361, 1432-0746},
  doi = {10.1051/0004-6361/202142898},
  urldate = {2026-04-17},
  archiveprefix = {arXiv},
  langid = {english},
  journal = {A\&A}
}

@ARTICLE{Robertson_2015,
       author = {{Robertson}, Brant E. and {Ellis}, Richard S. and {Furlanetto}, Steven R. and {Dunlop}, James S.},
        title = "{Cosmic Reionization and Early Star-forming Galaxies: A Joint Analysis of New Constraints from Planck and the Hubble Space Telescope}",
      journal = {\apjl},
         year = 2015,
        month = apr,
       volume = {802},
       number = {2},
          eid = {L19},
        pages = {L19},
          doi = {10.1088/2041-8205/802/2/L19},
archivePrefix = {arXiv},
       eprint = {1502.02024},
 primaryClass = {astro-ph.CO},
       adsurl = {https://ui.adsabs.harvard.edu/abs/2015ApJ...802L..19R}
}

@ARTICLE{Trussler_2023,
       author = {{Trussler}, James A.~A. and {Adams}, Nathan J. and {Conselice}, Christopher J. and {Ferreira}, Leonardo and {Austin}, Duncan and {Bhatawdekar}, Rachana and {Caruana}, Joseph and {Frye}, Brenda L. and {Harvey}, Tom and {Lovell}, Christopher C. and {Pascale}, Massimo and {Roper}, William J. and {Verma}, Aprajita and {Vijayan}, Aswin P. and {Wilkins}, Stephen M.},
        title = "{Seeing sharper and deeper: JWST's first glimpse of the photometric and spectroscopic properties of galaxies in the epoch of reionization}",
      journal = {\mnras},
         year = 2023,
        month = aug,
       volume = {523},
       number = {3},
        pages = {3423-3440},
          doi = {10.1093/mnras/stad1629},
archivePrefix = {arXiv},
       eprint = {2207.14265},
 primaryClass = {astro-ph.GA},
       adsurl = {https://ui.adsabs.harvard.edu/abs/2023MNRAS.523.3423T}
}

@ARTICLE{Caputi_2024,
       author = {{Caputi}, K.~I. and {Rinaldi}, P. and {Iani}, E. and {P{\'e}rez-Gonz{\'a}lez}, P.~G. and {{\"O}stlin}, G. and {Colina}, L. and {Greve}, T.~R. and {N{\o}rgaard-Nielsen}, H.~U. and {Wright}, G.~S. and {{\'A}lvarez-M{\'a}rquez}, J. and {Eckart}, A. and {Hjorth}, J. and {Labiano}, A. and {Le F{\`e}vre}, O. and {Walter}, F. and {van der Werf}, P. and {Boogaard}, L. and {Costantin}, L. and {Crespo G{\'o}mez}, A. and {Gillman}, S. and {Jermann}, I. and {Langeroodi}, D. and {Melinder}, J. and {Peissker}, F. and {G{\"u}del}, M. and {Henning}, Th. and {Lagage}, P.~O. and {Ray}, T.~P.},
        title = "{MIDIS: The Relation between Strong (H{\ensuremath{\beta}} + [O III]) Emission, Star Formation, and Burstiness around the Epoch of Reionization}",
      journal = {\apj},
         year = 2024,
        month = jul,
       volume = {969},
       number = {2},
          eid = {159},
        pages = {159},
          doi = {10.3847/1538-4357/ad4eb2},
archivePrefix = {arXiv},
       eprint = {2311.12691},
 primaryClass = {astro-ph.GA},
       adsurl = {https://ui.adsabs.harvard.edu/abs/2024ApJ...969..159C}
}

@ARTICLE{bonatto2026,
       author = {{Bonatto}, L. and {Amor{\'\i}n}, R. and {Gim{\'e}nez-Alc{\'a}zar}, A. and {Fern{\'a}ndez-Ontiveros}, J.~A. and {Hern{\'a}n-Caballero}, A. and {Su{\'a}rez}, S. and {V{\'\i}lchez}, J.~M. and {P{\'e}rez-Montero}, E. and {Llerena}, M. and {S{\'a}nchez Almeida}, J.},
        title = "{The k-MENDEL sample of local analogs to reionization galaxies. Spectral identification of EELGs and properties of green peas in DESI}",
      journal = {arXiv e-prints},
         year = 2026,
        month = apr,
          eid = {arXiv:2604.09516},
        pages = {arXiv:2604.09516},
          doi = {10.48550/arXiv.2604.09516},
archivePrefix = {arXiv},
       eprint = {2604.09516},
 primaryClass = {astro-ph.GA},
       adsurl = {https://ui.adsabs.harvard.edu/abs/2026arXiv260409516B}
}

@ARTICLE{davis2026,
       author = {{Davis}, Kelcey and {Brooks}, Madisyn and {Trump}, Jonathan R. and {Fern{\'a}ndez}, Vital and {Hutchison}, Taylor A. and {Larson}, Rebecca L. and {Taylor}, Anthony J. and {McGrath}, Elizabeth J. and {Barro}, Guillermo and {Koekemoer}, Anton M. and {Arrabal Haro}, Pablo and {Dickinson}, Mark and {Backhaus}, Bren E. and {Cleri}, Nikko J. and {Finkelstein}, Steven L. and {Ganapathy}, Ananya and {Simons}, Raymond C. and {Amor{\'\i}n}, Ricardo O. and {de la Vega}, Alexander and {Grogin}, Norman A. and {Hirschmann}, Michaela and {Hu}, Weida and {Johnson}, Jarrett L. and {Kartaltepe}, Jeyhan S. and {Kocevski}, Dale and {Llerena}, Mario and {Lucas}, Ray A. and {Marshall}, Madeline A. and {Pacucci}, Fabio and {Pentericci}, Laura and {Upton Sanderbeck}, Phoebe R.},
        title = "{Extreme Emission Line Galaxies in CEERS Are Powered by Star Formation, not AGN}",
      journal = {arXiv e-prints},
         year = 2026,
        month = feb,
          eid = {arXiv:2602.23310},
        pages = {arXiv:2602.23310},
          doi = {10.48550/arXiv.2602.23310},
archivePrefix = {arXiv},
       eprint = {2602.23310},
 primaryClass = {astro-ph.GA},
       adsurl = {https://ui.adsabs.harvard.edu/abs/2026arXiv260223310D}
}

@ARTICLE{boyett2024a,
       author = {{Boyett}, Kit and {Bunker}, Andrew J. and {Curtis-Lake}, Emma and {Chevallard}, Jacopo and {Cameron}, Alex J. and {Jones}, Gareth C. and {Saxena}, Aayush and {Charlot}, St{\'e}phane and {Curti}, Mirko and {Wallace}, Imaan E.~B. and {Arribas}, Santiago and {Carniani}, Stefano and {Willott}, Chris and {Alberts}, Stacey and {Eisenstein}, Daniel J. and {Hainline}, Kevin and {Hausen}, Ryan and {Johnson}, Benjamin D. and {Rieke}, Marcia and {Robertson}, Brant and {Stark}, Daniel P. and {Tacchella}, Sandro and {Williams}, Christina C. and {Chen}, Zuyi and {Egami}, Eiichi and {Endsley}, Ryan and {Kumari}, Nimisha and {Laseter}, Isaac and {Looser}, Tobias J. and {Maseda}, Michael V. and {Scholtz}, Jan and {Shivaei}, Irene and {Simmonds}, Charlotte and {Smit}, Renske and {{\"U}bler}, Hannah and {Witstok}, Joris},
        title = "{Extreme emission line galaxies detected in JADES JWST/NIRSpec - I. Inferred galaxy properties}",
      journal = {\mnras},
         year = 2024,
        month = dec,
       volume = {535},
       number = {2},
        pages = {1796-1828},
          doi = {10.1093/mnras/stae2430},
archivePrefix = {arXiv},
       eprint = {2401.16934},
 primaryClass = {astro-ph.GA},
       adsurl = {https://ui.adsabs.harvard.edu/abs/2024MNRAS.535.1796B}
}

@ARTICLE{Popesso_2023,
       author = {{Popesso}, P. and {Concas}, A. and {Cresci}, G. and {Belli}, S. and {Rodighiero}, G. and {Inami}, H. and {Dickinson}, M. and {Ilbert}, O. and {Pannella}, M. and {Elbaz}, D.},
        title = "{The main sequence of star-forming galaxies across cosmic times}",
      journal = {\mnras},
         year = 2023,
        month = feb,
       volume = {519},
       number = {1},
        pages = {1526-1544},
          doi = {10.1093/mnras/stac3214},
archivePrefix = {arXiv},
       eprint = {2203.10487},
 primaryClass = {astro-ph.GA},
       adsurl = {https://ui.adsabs.harvard.edu/abs/2023MNRAS.519.1526P}
}

@ARTICLE{pypeit:joss_pub,
       author = {{Prochaska}, J. Xavier and {Hennawi}, Joseph F. and {Westfall}, Kyle B. and
         {Cooke}, Ryan J. and {Wang}, Feige and {Hsyu}, Tiffany and
         {Davies}, Frederick B. and {Farina}, Emanuele Paolo and {Pelliccia}, Debora},
        title = "{PypeIt: The Python Spectroscopic Data Reduction Pipeline}",
      journal = {JOSS},
         year = {2020b},
        month = dec,
       volume = {5},
       number = {56},
        pages = {2308},
          doi = {10.21105/joss.02308},
       adsurl = {https://ui.adsabs.harvard.edu/abs/2020JOSS....5.2308P}
}

@MISC{pypeit:zenodo,
       author = {{Prochaska}, J. Xavier and {Hennawi}, Joseph and {Cooke}, Ryan and
         {Westfall}, Kyle and {Wang}, Feige and {EmAstro} and {Tiffanyhsyu} and
         {Wasserman}, Asher and {Villaume}, Alexa and {Marijana777} and
         {Schindler}, JT and {Young}, David and {Simha}, Sunil and
         {Wilde}, Matt and {Tejos}, Nicolas and {Isbell}, Jacob and
         {Fl{\"o}rs}, Andreas and {Sandford}, Nathan and {Vasovi{\'c}}, Zlatan and
         {Betts}, Edward and {Holden}, Brad},
        title = "{pypeit/PypeIt: Release 1.0.0}",
  howpublished = {Zenodo},
         year = {2020a},
        month = apr,
          doi = {10.5281/zenodo.3743493},
      version = {v1.0.0},
    publisher = {Zenodo},
       adsurl = {https://ui.adsabs.harvard.edu/abs/2020zndo...3743493P}
}

@ARTICLE{Calzetti_2000,
       author = {{Calzetti}, Daniela and {Armus}, Lee and {Bohlin}, Ralph C. and {Kinney}, Anne L. and {Koornneef}, Jan and {Storchi-Bergmann}, Thaisa},
        title = "{The Dust Content and Opacity of Actively Star-forming Galaxies}",
      journal = {\apj},
         year = 2000,
        month = apr,
       volume = {533},
       number = {2},
        pages = {682-695},
          doi = {10.1086/308692},
archivePrefix = {arXiv},
       eprint = {astro-ph/9911459},
 primaryClass = {astro-ph},
       adsurl = {https://ui.adsabs.harvard.edu/abs/2000ApJ...533..682C}
}

@BOOK{Osterbrock_2006,
       author = {{Osterbrock}, Donald E. and {Ferland}, Gary J.},
        title = "{Astrophysics of gaseous nebulae and active galactic nuclei}",
         year = 2006,
         publisher = {University Science Books},
       adsurl = {https://ui.adsabs.harvard.edu/abs/2006agna.book.....O}
}

@misc{msaexp_zenodo,
       author = {{Brammer}, Gabriel},
        title = "{msaexp: NIRSpec analyis tools}",
         year = 2023,
        month = sep,
          eid = {10.5281/zenodo.7299500},
          doi = {10.5281/zenodo.7299500},
      version = {0.6.17},
    publisher = {Zenodo},
       adsurl = {https://ui.adsabs.harvard.edu/abs/2022zndo...7299500B}
}

@ARTICLE{amorin_extreme_2015,
       author = {{Amor{\'\i}n}, R. and {P{\'e}rez-Montero}, E. and {Contini}, T. and {V{\'\i}lchez}, J.~M. and {Bolzonella}, M. and {Tasca}, L.~A.~M. and {Lamareille}, F. and {Zamorani}, G. and {Maier}, C. and {Carollo}, C.~M. and {Kneib}, J.-P. and {Le F{\`e}vre}, O. and {Lilly}, S. and {Mainieri}, V. and {Renzini}, A. and {Scodeggio}, M. and {Bardelli}, S. and {Bongiorno}, A. and {Caputi}, K. and {Cucciati}, O. and {de la Torre}, S. and {de Ravel}, L. and {Franzetti}, P. and {Garilli}, B. and {Iovino}, A. and {Kampczyk}, P. and {Knobel}, C. and {Kova{\v{c}}}, K. and {Le Borgne}, J.-F. and {Le Brun}, V. and {Mignoli}, M. and {Pell{\`o}}, R. and {Peng}, Y. and {Presotto}, V. and {Ricciardelli}, E. and {Silverman}, J.~D. and {Tanaka}, M. and {Tresse}, L. and {Vergani}, D. and {Zucca}, E.},
        title = "{Extreme emission-line galaxies out to z \raisebox{-0.5ex}\textasciitilde 1 in zCOSMOS. I. Sample and characterization of global properties}",
      journal = {\aap},
         year = 2015,
        month = jun,
       volume = {578},
          eid = {A105},
        pages = {A105},
          doi = {10.1051/0004-6361/201322786},
archivePrefix = {arXiv},
       eprint = {1403.3441},
 primaryClass = {astro-ph.GA},
       adsurl = {https://ui.adsabs.harvard.edu/abs/2015A&A...578A.105A}
}

@ARTICLE{llerena_physical_2024,
       author = {{Llerena}, M. and {Amor{\'\i}n}, R. and {Pentericci}, L. and {Arrabal Haro}, P. and {Backhaus}, B.~E. and {Bagley}, M.~B. and {Calabr{\`o}}, A. and {Cleri}, N.~J. and {Davis}, K. and {Dickinson}, M. and {Finkelstein}, S.~L. and {Gawiser}, E. and {Grogin}, N.~A. and {Hathi}, N.~P. and {Hirschmann}, M. and {Kartaltepe}, J.~S. and {Koekemoer}, A.~M. and {McGrath}, E.~J. and {Mobasher}, B. and {Napolitano}, L. and {Papovich}, C. and {Pirzkal}, N. and {Trump}, J.~R. and {Wilkins}, S.~M. and {Yung}, L.~Y.~A.},
        title = "{Physical properties of extreme emission-line galaxies at z {\ensuremath{\sim}} 4─9 from the JWST CEERS survey}",
      journal = {\aap},
         year = 2024,
        month = nov,
       volume = {691},
          eid = {A59},
        pages = {A59},
          doi = {10.1051/0004-6361/202449904},
archivePrefix = {arXiv},
       eprint = {2403.05362},
 primaryClass = {astro-ph.GA},
       adsurl = {https://ui.adsabs.harvard.edu/abs/2024A&A...691A..59L}
}

@article{newman_deep2_2013,
	title = {{THE} {DEEP2} {GALAXY} {REDSHIFT} {SURVEY}: {DESIGN}, {OBSERVATIONS}, {DATA} {REDUCTION}, {AND} {REDSHIFTS}},
	volume = {208},
	copyright = {http://iopscience.iop.org/info/page/text-and-data-mining},
	issn = {0067-0049, 1538-4365},
	shorttitle = {{THE} {DEEP2} {GALAXY} {REDSHIFT} {SURVEY}},
	url = {https://iopscience.iop.org/article/10.1088/0067-0049/208/1/5},
	doi = {10.1088/0067-0049/208/1/5},
	language = {en},
	number = {1},
	urldate = {2024-11-04},
	journal = {\apjs},
	author = {Newman, Jeffrey A. and Cooper, Michael C. and Davis, Marc and Faber, S. M. and Coil, Alison L. and Guhathakurta, Puragra and Koo, David C. and Phillips, Andrew C. and Conroy, Charlie and Dutton, Aaron A. and Finkbeiner, Douglas P. and Gerke, Brian F. and Rosario, David J. and Weiner, Benjamin J. and Willmer, C. N. A. and Yan, Renbin and Harker, Justin J. and Kassin, Susan A. and Konidaris, N. P. and Lai, Kamson and Madgwick, Darren S. and Noeske, K. G. and Wirth, Gregory D. and Connolly, A. J. and Kaiser, N. and Kirby, Evan N. and Lemaux, Brian C. and Lin, Lihwai and Lotz, Jennifer M. and Luppino, G. A. and Marinoni, C. and Matthews, Daniel J. and Metevier, Anne and Schiavon, Ricardo P.},
	month = aug,
	year = {2013},
	pages = {5},
}

@article{Kodra_2023,
doi = {10.3847/1538-4357/ac9f12},
url = {https://dx.doi.org/10.3847/1538-4357/ac9f12},
year = {2023},
month = {jan},
publisher = {The American Astronomical Society},
volume = {942},
number = {1},
pages = {36},
author = {Dritan Kodra and Brett H. Andrews and Jeffrey A. Newman and Steven L. Finkelstein and Adriano Fontana and Nimish Hathi and Mara Salvato and Tommy Wiklind and Stijn Wuyts and Adam Broussard and Nima Chartab and Christopher Conselice and M. C. Cooper and Avishai Dekel and Mark Dickinson and Henry C. Ferguson and Eric Gawiser and Norman A. Grogin and Kartheik Iyer and Jeyhan Kartaltepe and Susan Kassin and Anton M. Koekemoer and David C. Koo and Ray A. Lucas and Kameswara Bharadwaj Mantha and Daniel H. McIntosh and Bahram Mobasher and Camilla Pacifici and Pablo G. Pérez-González and Paola Santini},
title = {Optimized Photometric Redshifts for the Cosmic Assembly Near-infrared Deep Extragalactic Legacy Survey (CANDELS)},
journal = {\apj}
}

@ARTICLE{Fan_2006,
       author = {{Fan}, Xiaohui and {Carilli}, C.~L. and {Keating}, B.},
        title = "{Observational Constraints on Cosmic Reionization}",
      journal = {\araa},
         year = 2006,
        month = sep,
       volume = {44},
       number = {1},
        pages = {415-462},
          doi = {10.1146/annurev.astro.44.051905.092514},
archivePrefix = {arXiv},
       eprint = {astro-ph/0602375},
 primaryClass = {astro-ph},
       adsurl = {https://ui.adsabs.harvard.edu/abs/2006ARA&A..44..415F}
}

@ARTICLE{Stefanon_2017,
       author = {{Stefanon}, Mauro and {Yan}, Haojing and {Mobasher}, Bahram and {Barro}, Guillermo and {Donley}, Jennifer L. and {Fontana}, Adriano and {Hemmati}, Shoubaneh and {Koekemoer}, Anton M. and {Lee}, BoMee and {Lee}, Seong-Kook and {Nayyeri}, Hooshang and {Peth}, Michael and {Pforr}, Janine and {Salvato}, Mara and {Wiklind}, Tommy and {Wuyts}, Stijn and {Ashby}, Matthew L.~N. and {Castellano}, Marco and {Conselice}, Christopher J. and {Cooper}, Michael C. and {Cooray}, Asantha R. and {Dolch}, Timothy and {Ferguson}, Henry and {Galametz}, Audrey and {Giavalisco}, Mauro and {Guo}, Yicheng and {Willner}, Steven P. and {Dickinson}, Mark E. and {Faber}, Sandra M. and {Fazio}, Giovanni G. and {Gardner}, Jonathan P. and {Gawiser}, Eric and {Grazian}, Andrea and {Grogin}, Norman A. and {Kocevski}, Dale and {Koo}, David C. and {Lee}, Kyoung-Soo and {Lucas}, Ray A. and {McGrath}, Elizabeth J. and {Nandra}, Kirpal and {Newman}, Jeffrey A. and {van der Wel}, Arjen},
        title = "{CANDELS Multi-wavelength Catalogs: Source Identification and Photometry in the CANDELS Extended Groth Strip}",
      journal = {\apjs},
         year = 2017,
        month = apr,
       volume = {229},
       number = {2},
          eid = {32},
        pages = {32},
          doi = {10.3847/1538-4365/aa66cb},
archivePrefix = {arXiv},
       eprint = {1703.05768},
 primaryClass = {astro-ph.GA},
       adsurl = {https://ui.adsabs.harvard.edu/abs/2017ApJS..229...32S}
}

@article{Kartaltepe_2015,
   title={CANDELS VISUAL CLASSIFICATIONS: SCHEME, DATA RELEASE, AND FIRST RESULTS},
   volume={221},
   ISSN={1538-4365},
   url={http://dx.doi.org/10.1088/0067-0049/221/1/11},
   DOI={10.1088/0067-0049/221/1/11},
   number={1},
   journal={\apjs},
   publisher={American Astronomical Society},
   author={Kartaltepe, Jeyhan S. and Mozena, Mark and Kocevski, Dale and McIntosh, Daniel H. and Lotz, Jennifer and Bell, Eric F. and Faber, Sandy and Ferguson, Harry and Koo, David and Bassett, Robert and Bernyk, Maksym and Blancato, Kirsten and Bournaud, Frederic and Cassata, Paolo and Castellano, Marco and Cheung, Edmond and Conselice, Christopher J. and Croton, Darren and Dahlen, Tomas and Mello, Duilia F. de and DeGroot, Laura and Donley, Jennifer and Guedes, Javiera and Grogin, Norman and Hathi, Nimish and Hilton, Matt and Hollon, Brett and Koekemoer, Anton and Liu, Nick and Lucas, Ray A. and Martig, Marie and McGrath, Elizabeth and McPartland, Conor and Mobasher, Bahram and Morlock, Alice and O’Leary, Erin and Peth, Mike and Pforr, Janine and Pillepich, Annalisa and Rosario, David and Soto, Emmaris and Straughn, Amber and Telford, Olivia and Sunnquist, Ben and Trump, Jonathan and Weiner, Benjamin and Wuyts, Stijn and Inami, Hanae and Kassin, Susan and Lani, Caterina and B. Poole, Gregory and Rizer, Zachary},
   year={2015},
   month=oct, pages={11} }

@ARTICLE{gimenez_alcazar_2026,
       author = {{Gim{\'e}nez-Alc{\'a}zar}, A. and {Amor{\'\i}n}, R. and {V{\'\i}lchez}, J.~M. and {Hern{\'a}n-Caballero}, A. and {Gonz{\'a}lez-Otero}, M. and {Arroyo-Polonio}, A. and {Iglesias-P{\'a}ramo}, J. and {Lumbreras-Calle}, A. and {Fern{\'a}ndez-Ontiveros}, J.~A. and {L{\'o}pez-Sanjuan}, C. and {Bonatto}, L. and {Gonz{\'a}lez Delgado}, R.~M. and {Kehrig}, C. and {Torralba}, A. and {Rahna}, P.~T. and {Jim{\'e}nez-Teja}, Y. and {M{\'a}rquez}, I. and {Breda}, I. and {{\'A}lvarez-Candal}, A. and {Abramo}, R. and {Alcaniz}, J. and {Benitez}, N. and {Bonoli}, S. and {Carneiro}, S. and {Cenarro}, J. and {Crist{\'o}bal-Hornillos}, D. and {Dupke}, R. and {Ederoclite}, A. and {Hern{\'a}ndez-Monteagudo}, C. and {Mar{\'\i}n-Franch}, A. and {Mendes de Oliveira}, C. and {Moles}, M. and {Sodr{\'e}}, Jr., L. and {Taylor}, K. and {Varela}, J. and {V{\'a}zquez Rami{\'o}}, H.},
        title = "{J-PAS: First identification, physical properties, and ionization efficiency of extreme emission line galaxies}",
      journal = {\aap},
         year = 2026,
        month = feb,
       volume = {706},
          eid = {A261},
        pages = {A261},
          doi = {10.1051/0004-6361/202557358},
archivePrefix = {arXiv},
       eprint = {2512.08484},
 primaryClass = {astro-ph.GA},
       adsurl = {https://ui.adsabs.harvard.edu/abs/2026A&A...706A.261G}
}

@article{P_nin_2015,
   author={Pénin, Aurélie and Cuby, Jean-Gabriel and Clément, Benjamin and Hibon, Pascale and Kneib, Jean-Paul and Cassata, Paolo and Ilbert, Olivier},
   title={Narrowband selected high-redshift galaxy candidates contaminated by lower redshift [OIII] ultra-strong emitter line galaxies},
   journal={\aap},
   volume={577},
   pages={A74},
   year={2015},
   month=may,
   DOI={10.1051/0004-6361/201425183},
   url={http://dx.doi.org/10.1051/0004-6361/201425183},
   ISSN={1432-0746},
   publisher={EDP Sciences}
}

@ARTICLE{rinaldi2024,
       author = {{Rinaldi}, P. and {Caputi}, K.~I. and {Iani}, E. and {Costantin}, L. and {Gillman}, S. and {Perez Gonzalez}, P.~G. and {{\"O}stlin}, G. and {Colina}, L. and {Greve}, T.~R. and {N{\o}rgard-Nielsen}, H.~U. and {Wright}, G.~S. and {{\'A}lvarez-M{\'a}rquez}, J. and {Eckart}, A. and {Garc{\'\i}a-Mar{\'\i}n}, M. and {Hjorth}, J. and {Ilbert}, O. and {Kendrew}, S. and {Labiano}, A. and {Le F{\`e}vre}, O. and {Pye}, J. and {Tikkanen}, T. and {Walter}, F. and {van der Werf}, P. and {Ward}, M. and {Annunziatella}, M. and {Azzollini}, R. and {Bik}, A. and {Boogaard}, L. and {Bosman}, S.~E.~I. and {Crespo G{\'o}mez}, A. and {Jermann}, I. and {Langeroodi}, D. and {Melinder}, J. and {Meyer}, R.~A. and {Moutard}, T. and {Peissker}, F. and {van Dishoeck}, E. and {G{\"u}del}, M. and {Henning}, Th. and {Lagage}, P.-O. and {Ray}, T. and {Vandenbussche}, B. and {Waelkens}, C. and {Dayal}, Pratika},
        title = "{MIDIS: Unveiling the Role of Strong H{\ensuremath{\alpha}} Emitters During the Epoch of Reionization with JWST}",
      journal = {\apj},
         year = 2024,
        month = jul,
       volume = {969},
       number = {1},
          eid = {12},
        pages = {12},
          doi = {10.3847/1538-4357/ad4147},
archivePrefix = {arXiv},
       eprint = {2309.15671},
 primaryClass = {astro-ph.GA},
       adsurl = {https://ui.adsabs.harvard.edu/abs/2024ApJ...969...12R}
}

@INPROCEEDINGS{Maiolino_2008,
       author = {{Maiolino}, R. and {Nagao}, T. and {Grazian}, A. and {Cocchia}, F. and {Marconi}, A. and {Mannucci}, F. and {Cimatti}, A. and {Pipino}, A. and {Fontana}, A. and {Granato}, G.~L. and {Matteucci}, F. and {Pentericci}, L. and {Risaliti}, G. and {Salvati}, M. and {Silva}, L.},
        title = "{The Mass-Metallicity Relation at z>3}",
    booktitle = {Formation and Evolution of Galaxy Disks},
         year = 2008,
       editor = {{Funes}, J.~G. and {Corsini}, E.~M.},
       series = {ASP Conference Series},
       volume = {396},
        month = oct,
        pages = {409},
       adsurl = {https://ui.adsabs.harvard.edu/abs/2008ASPC..396..409M}
}

@ARTICLE{tang2023,
       author = {{Tang}, Mengtao and {Stark}, Daniel P. and {Chen}, Zuyi and {Mason}, Charlotte and {Topping}, Michael and {Endsley}, Ryan and {Senchyna}, Peter and {Plat}, Ad{\`e}le and {Lu}, Ting-Yi and {Whitler}, Lily and {Robertson}, Brant and {Charlot}, St{\'e}phane},
        title = "{JWST/NIRSpec spectroscopy of z = 7-9 star-forming galaxies with CEERS: new insight into bright Ly{\ensuremath{\alpha}} emitters in ionized bubbles}",
      journal = {\mnras},
         year = 2023,
        month = dec,
       volume = {526},
       number = {2},
        pages = {1657-1686},
          doi = {10.1093/mnras/stad2763},
archivePrefix = {arXiv},
       eprint = {2301.07072},
 primaryClass = {astro-ph.GA},
       adsurl = {https://ui.adsabs.harvard.edu/abs/2023MNRAS.526.1657T}
}

@ARTICLE{Kauffmann_2020,
       author = {{Kauffmann}, O.~B. and {Le F{\`e}vre}, O. and {Ilbert}, O. and {Chevallard}, J. and {Williams}, C.~C. and {Curtis-Lake}, E. and {Colina}, L. and {P{\'e}rez-Gonz{\'a}lez}, P.~G. and {Pye}, J.~P. and {Caputi}, K.~I.},
        title = "{Simulating JWST deep extragalactic imaging surveys and physical parameter recovery}",
      journal = {\aap},
         year = 2020,
        month = aug,
       volume = {640},
          eid = {A67},
        pages = {A67},
          doi = {10.1051/0004-6361/202037450},
archivePrefix = {arXiv},
       eprint = {2004.06128},
 primaryClass = {astro-ph.GA},
       adsurl = {https://ui.adsabs.harvard.edu/abs/2020A&A...640A..67K}
}

@ARTICLE{Naidu_2026,
       author = {{Naidu}, Rohan P. and {Oesch}, Pascal A. and {Brammer}, Gabriel and {Weibel}, Andrea and {Li}, Yijia and {Matthee}, Jorryt and {Chisolm}, John and {Pollock}, Clara L. and {Heintz}, Kasper E. and {Johnson}, Benjamin D. and {Shen}, Xuejian and {Hviding}, Raphael E. and {Leja}, Joel and {Tacchella}, Sandro and {Ganguly}, Arpita and {Witten}, Callum and {Atek}, Hakim and {Belli}, Siro and {Bose}, Sownak and {Bouwens}, Rychard and {Dayal}, Pratika and {Decarli}, Roberto and {de Graaff}, Anna and {Fudamoto}, Yoshinobu and {Giovinazzo}, Emma and {Greene}, Jenny E. and {Illingworth}, Garth and {Inoue}, Akio K. and {Kane}, Sarah G. and {Labbe}, Ivo and {Leonova}, Ecaterina and {Marques-Chaves}, Rui and {Meyer}, Roman A. and {Nelson}, Erica J. and {Roberts-Borsani}, Guido and {Schaerer}, Daniel and {Simcoe}, Robert A. and {Stefanon}, Mauro and {Sugahara}, Yuma and {Toft}, Sune and {van der Wel}, Arjen and {van Dokkum}, Pieter and {Walter}, Fabian and {Watson}, Darrach and {Weaver}, John R. and {Whitaker}, Katherine E.},
        title = "{A Cosmic Miracle: A Remarkably Luminous Galaxy at zspec = 14.44 Confirmed with JWST}",
      journal = {OJAp},
         year = 2026,
        month = jan,
       volume = {9},
        pages = {56033},
          doi = {10.33232/001c.156033},
archivePrefix = {arXiv},
       eprint = {2505.11263},
 primaryClass = {astro-ph.GA},
       adsurl = {https://ui.adsabs.harvard.edu/abs/2026OJAp....956033N}
}

@ARTICLE{Adams_2023,
       author = {{Adams}, N.~J. and {Conselice}, C.~J. and {Ferreira}, L. and {Austin}, D. and {Trussler}, J.~A.~A. and {Juod{\v{z}}balis}, I. and {Wilkins}, S.~M. and {Caruana}, J. and {Dayal}, P. and {Verma}, A. and {Vijayan}, A.~P.},
        title = "{Discovery and properties of ultra-high redshift galaxies (9 < z < 12) in the JWST ERO SMACS 0723 Field}",
      journal = {\mnras},
         year = 2023,
        month = jan,
       volume = {518},
       number = {3},
        pages = {4755-4766},
          doi = {10.1093/mnras/stac3347},
archivePrefix = {arXiv},
       eprint = {2207.11217},
 primaryClass = {astro-ph.GA},
       adsurl = {https://ui.adsabs.harvard.edu/abs/2023MNRAS.518.4755A}
}

@ARTICLE{Davis_2024,
       author = {{Davis}, Kelcey and {Trump}, Jonathan R. and {Simons}, Raymond C. and {McGrath}, Elizabeth J. and {Wilkins}, Stephen M. and {Arrabal Haro}, Pablo and {Bagley}, Micaela B. and {Dickinson}, Mark and {Fern{\'a}ndez}, Vital and {Amor{\'\i}n}, Ricardo O. and {Backhaus}, Bren E. and {Cleri}, Nikko J. and {Llerena}, Mario and {Brunker}, Samantha W. and {Barro}, Guillermo and {Bisigello}, Laura and {Brooks}, Madisyn and {Costantin}, Luca and {de La Vega}, Alexander and {Dekel}, Avishai and {Finkelstein}, Steven L. and {Hathi}, Nimish P. and {Hirschmann}, Michaela and {Kartaltepe}, Jeyhan S. and {Koekemoer}, Anton M. and {Lucas}, Ray A. and {Papovich}, Casey and {P{\'e}rez-Gonz{\'a}lez}, Pablo G. and {Pirzkal}, Nor and {Rodighiero}, Giulia and {Rose}, Caitlin and {Yung}, L.~Y. Aaron and {Ceers Collaborators}},
        title = "{A Census from JWST of Extreme Emission-line Galaxies Spanning the Epoch of Reionization in CEERS}",
      journal = {\apj},
         year = 2024,
        month = oct,
       volume = {974},
       number = {1},
          eid = {42},
        pages = {42},
          doi = {10.3847/1538-4357/ad6865},
archivePrefix = {arXiv},
       eprint = {2312.07799},
 primaryClass = {astro-ph.GA},
       adsurl = {https://ui.adsabs.harvard.edu/abs/2024ApJ...974...42D}
}

@ARTICLE{Daikuhara_2025A,
       author = {{Daikuhara}, Kazuki and {Morishita}, Takahiro and {Kodama}, Tadayuki and {Chary}, Ranga-Ram and {Akiyama}, Masayuki and {P{\'e}rez-Mart{\'\i}nez}, Jose. M.},
        title = "{Nature of High Equivalent Width Emitters in the Epoch of Reionization Revealed by JWST Medium-band Imaging}",
      journal = {\apj},
         year = 2025,
        month = aug,
       volume = {989},
       number = {1},
          eid = {71},
        pages = {71},
          doi = {10.3847/1538-4357/ade148},
archivePrefix = {arXiv},
       eprint = {2507.13456},
 primaryClass = {astro-ph.GA},
       adsurl = {https://ui.adsabs.harvard.edu/abs/2025ApJ...989...71D}
}

@ARTICLE{Donley_2012,
       author = {{Donley}, J.~L. and {Koekemoer}, A.~M. and {Brusa}, M. and {Capak}, P. and {Cardamone}, C.~N. and {Civano}, F. and {Ilbert}, O. and {Impey}, C.~D. and {Kartaltepe}, J.~S. and {Miyaji}, T. and {Salvato}, M. and {Sanders}, D.~B. and {Trump}, J.~R. and {Zamorani}, G.},
        title = "{Identifying Luminous Active Galactic Nuclei in Deep Surveys: Revised IRAC Selection Criteria}",
      journal = {\apj},
         year = 2012,
        month = apr,
       volume = {748},
       number = {2},
          eid = {142},
        pages = {142},
          doi = {10.1088/0004-637X/748/2/142},
archivePrefix = {arXiv},
       eprint = {1201.3899},
 primaryClass = {astro-ph.CO},
       adsurl = {https://ui.adsabs.harvard.edu/abs/2012ApJ...748..142D}
}

@ARTICLE{Kauffmann_2003,
       author = {{Kauffmann}, Guinevere and {Heckman}, Timothy M. and {Tremonti}, Christy and {Brinchmann}, Jarle and {Charlot}, St{\'e}phane and {White}, Simon D.~M. and {Ridgway}, Susan E. and {Brinkmann}, Jon and {Fukugita}, Masataka and {Hall}, Patrick B. and {Ivezi{\'c}}, {\v{Z}}eljko and {Richards}, Gordon T. and {Schneider}, Donald P.},
        title = "{The host galaxies of active galactic nuclei}",
      journal = {\mnras},
         year = 2003,
        month = dec,
       volume = {346},
       number = {4},
        pages = {1055-1077},
          doi = {10.1111/j.1365-2966.2003.07154.x},
archivePrefix = {arXiv},
       eprint = {astro-ph/0304239},
 primaryClass = {astro-ph},
       adsurl = {https://ui.adsabs.harvard.edu/abs/2003MNRAS.346.1055K}
}

@ARTICLE{Kewley_2001,
       author = {{Kewley}, L.~J. and {Dopita}, M.~A. and {Sutherland}, R.~S. and {Heisler}, C.~A. and {Trevena}, J.},
        title = "{Theoretical Modeling of Starburst Galaxies}",
      journal = {\apj},
         year = 2001,
        month = jul,
       volume = {556},
       number = {1},
        pages = {121-140},
          doi = {10.1086/321545},
archivePrefix = {arXiv},
       eprint = {astro-ph/0106324},
 primaryClass = {astro-ph},
       adsurl = {https://ui.adsabs.harvard.edu/abs/2001ApJ...556..121K}
}

@ARTICLE{Baldwin_1981,
       author = {{Baldwin}, J.~A. and {Phillips}, M.~M. and {Terlevich}, R.},
        title = "{Classification parameters for the emission-line spectra of extragalactic objects.}",
      journal = {\pasp},
         year = 1981,
        month = feb,
       volume = {93},
        pages = {5-19},
          doi = {10.1086/130766},
       adsurl = {https://ui.adsabs.harvard.edu/abs/1981PASP...93....5B}
}

@ARTICLE{Barro_2019,
       author = {{Barro}, Guillermo and {P{\'e}rez-Gonz{\'a}lez}, Pablo G. and {Cava}, Antonio and {Brammer}, Gabriel and {Pandya}, Viraj and {Eliche Moral}, Carmen and {Esquej}, Pilar and {Dom{\'\i}nguez-S{\'a}nchez}, Helena and {Alcalde Pampliega}, Belen and {Guo}, Yicheng and {Koekemoer}, Anton M. and {Trump}, Jonathan R. and {Ashby}, Matthew L.~N. and {Cardiel}, Nicolas and {Castellano}, Marco and {Conselice}, Christopher J. and {Dickinson}, Mark E. and {Dolch}, Timothy and {Donley}, Jennifer L. and {Espino Briones}, N{\'e}stor and {Faber}, Sandra M. and {Fazio}, Giovanni G. and {Ferguson}, Henry and {Finkelstein}, Steve and {Fontana}, Adriano and {Galametz}, Audrey and {Gardner}, Jonathan P. and {Gawiser}, Eric and {Giavalisco}, Mauro and {Grazian}, Andrea and {Grogin}, Norman A. and {Hathi}, Nimish P. and {Hemmati}, Shoubaneh and {Hern{\'a}n-Caballero}, Antonio and {Kocevski}, Dale and {Koo}, David C. and {Kodra}, Dritan and {Lee}, Kyoung-Soo and {Lin}, Lihwai and {Lucas}, Ray A. and {Mobasher}, Bahram and {McGrath}, Elizabeth J. and {Nandra}, Kirpal and {Nayyeri}, Hooshang and {Newman}, Jeffrey A. and {Pforr}, Janine and {Peth}, Michael and {Rafelski}, Marc and {Rodr{\'\i}guez-Munoz}, Lucia and {Salvato}, Mara and {Stefanon}, Mauro and {van der Wel}, Arjen and {Willner}, Steven P. and {Wiklind}, Tommy and {Wuyts}, Stijn},
        title = "{The CANDELS/SHARDS Multiwavelength Catalog in GOODS-N: Photometry, Photometric Redshifts, Stellar Masses, Emission-line Fluxes, and Star Formation Rates}",
      journal = {\apjs},
         year = 2019,
        month = aug,
       volume = {243},
       number = {2},
          eid = {22},
        pages = {22},
          doi = {10.3847/1538-4365/ab23f2},
archivePrefix = {arXiv},
       eprint = {1908.00569},
 primaryClass = {astro-ph.GA},
       adsurl = {https://ui.adsabs.harvard.edu/abs/2019ApJS..243...22B}
}

@ARTICLE{Chabrier_2003,
       author = {{Chabrier}, Gilles},
        title = "{Galactic Stellar and Substellar Initial Mass Function}",
      journal = {\pasp},
         year = 2003,
        month = jul,
       volume = {115},
       number = {809},
        pages = {763-795},
          doi = {10.1086/376392},
archivePrefix = {arXiv},
       eprint = {astro-ph/0304382},
 primaryClass = {astro-ph},
       adsurl = {https://ui.adsabs.harvard.edu/abs/2003PASP..115..763C}
}

@ARTICLE{Trenti_2008,
       author = {{Trenti}, M. and {Stiavelli}, M.},
        title = "{Cosmic Variance and Its Effect on the Luminosity Function Determination in Deep High-z Surveys}",
      journal = {\apj},
         year = 2008,
        month = apr,
       volume = {676},
       number = {2},
        pages = {767-780},
          doi = {10.1086/528674},
archivePrefix = {arXiv},
       eprint = {0712.0398},
 primaryClass = {astro-ph},
       adsurl = {https://ui.adsabs.harvard.edu/abs/2008ApJ...676..767T}
}

@ARTICLE{SDSS_DR19_2025,
       author = {{SDSS Collaboration} and {Aghakhanloo}, Mojgan and {Aird}, James and {Almeida}, Andr{\'e}s and {Amrita}, Singh and {Anders}, Friedrich and {Anderson}, Scott F. and {Arseneau}, Stefan and {Gonz{\'a}lez {\'A}vila}, Consuelo and {Aviram}, Shir and {Aydar}, Catarina and {Badenes}, Carles and {Barrera-Ballesteros}, Jorge K. and {Bauer}, Franz E. and {Behmard}, Aida and {Berg}, Michelle and {Besser}, F. and {Moni Bidin}, Christian and {Bizyaev}, Dmitry and {Blanc}, Guillermo and {Blanton}, Michael R. and {Bovy}, Jo and {Brandt}, William Nielsen and {Brownstein}, Joel R. and {Buchner}, Johannes and {Bulbul}, Esra and {Burchett}, Joseph N. and {Carigi}, Leticia and {Carlberg}, Joleen K. and {Casey}, Andrew R. and {Chakraborty}, Priyanka and {Chanam{\'e}}, Julio and {Chandra}, Vedant and {Chiappini}, Cristina and {Chilingarian}, Igor and {Comparat}, Johan and {Covey}, Kevin and {Crumpler}, Nicole and {Cunha}, Katia and {D'Onghia}, Elena and {Dai}, Xinyu and {Darling}, Jeremy and {Davis}, Megan and {De Lee}, Nathan and {Deacon}, Niall and {M{\'e}ndez Delgado}, Jos{\'e} Eduardo and {Demasi}, Sebastian and {Demianenko}, Mariia and {Demke}, Delvin and {Donor}, John and {Drory}, Niv and {Villa Durango}, Monica Alejandra and {Dwelly}, Tom and {Egorov}, Oleg and {Egorova}, Evgeniya and {El-Badry}, Kareem and {Eracleous}, Mike and {Fan}, Xiaohui and {Farr}, Emily and {Finkbeiner}, Douglas P. and {Fries}, Logan and {Frinchaboy}, Peter and {Gentile Fusillo}, Nicola Pietro and {Serrano F{\'e}lix}, Luis Daniel and {G{\"a}nsicke}, Boris T. and {Galligan}, Emma and {Garc{\'\i}a}, Pablo and {Gelfand}, Joseph and {Grabowski}, Katie and {Grebel}, Eva and {Green}, Paul J. and {Greve}, Hannah and {Grier}, Catherine and {Griffith}, Emily J. and {Guetzoyan}, Paloma and {Gupta}, Pramod and {Hackshaw}, Zoe and {Hall}, Patrick B. and {Hawkins}, Keith and {Heged{\H{u}}s}, Viola and {Hekker}, Saskia and {Herbst}, T.~M. and {Hermes}, J.~J. and {Hern{\'a}ndez-Garc{\'\i}a}, Lorena and {Hiremath}, Pranavi and {Hogg}, David W. and {Holtzman}, Jon and {Horne}, Keith and {Horta}, Danny and {Huang}, Yang and {Hutchinson}, Brian and {H{\"a}berle}, Maximilian and {Ibarra-Medel}, Hector Javier and {Ji}, Alexander P. and {Jofre}, Paula and {Johnson}, James W. and {Johnson}, Jennifer and {Johnston}, Evelyn J. and {Kaldor}, Mary and {Katkov}, Ivan and {Khalatyan}, Arman and {Khoperskov}, Sergey and {Klessen}, Ralf and {Kluge}, Matthias and {Koekemoer}, Anton M. and {Kollmeier}, Juna A. and {Kounkel}, Marina and {Kreckel}, Kathryn and {Krishnarao}, Dhanesh and {Krumpe}, Mirko and {Lacerna}, Ivan and {Laporte}, Chervin and {L{\'e}pine}, S{\'e}bastien and {Li}, Jing and {Liang}, Fu-Heng and {Limberg}, Guilherme and {Liu}, Xin and {Loebman}, Sarah and {Long}, Knox and {Lu}, Yuxi(Lucy) and {Lucey}, Madeline and {Lugo-Aranda}, Alejandra Z. and {Mart{\'\i}nez-Aldama}, Mary Loli and {McKinnon}, Kevin A. and {Medan}, Ilija and {Merloni}, Andrea and {Morrison}, Sean and {Myers}, Natalie and {M{\'e}sz{\'a}ros}, Szabolcs and {M{\"u}ller-Horn}, Johanna and {Nepal}, Samir and {Ness}, Melissa and {Nidever}, David and {Nitschelm}, Christian and {Oravetz}, Audrey and {Otto}, Jonah and {Adamane Pallathadka}, Gautham and {Pan}, Kaike and {P{\'e}rez Paolino}, Facundo and {Negrete Pe{\~n}aloza}, Castalia Alenka and {Pinsonneault}, Marc and {Taghizadeh-Popp}, Manuchehr and {Price-Whelan}, Adrian and {Pulatova}, Nadiia and {Queiroz}, A.~B.~A. and {Raddick}, Jordan and {Rankine}, Amy L. and {Rix}, Hans-Walter and {Rom{\'a}n-Z{\'u}{\~n}iga}, Carlos G. and {Fern{\'a}ndez Rosso}, Daniela and {Runnoe}, Jessie and {Saad}, Serat Mahmud and {Salvato}, Mara and {Sanchez}, Sebastian F. and {Sattler}, Natascha and {Saydjari}, Andrew K. and {Sayres}, Conor and {Schlaufman}, Kevin C. and {Schneider}, Donald P. and {Schwope}, Axel and {Seaton}, Lucas M. and {Seeburger}, Rhys and {Serna}, Javier and {Sharma}, Sanjib and {Shen}, Yue and {Sinha}, Amaya and {Sizemore}, Logan and {Sniegowska}, Marzena and {Song}, Ying-Yi and {Souto}, Diogo and {Stassun}, Keivan G. and {Steinmetz}, Matthias and {Stone}, Zachary and {Stone-Martinez}, Alexander and {Stringfellow}, Guy S. and {Mata S{\'a}nchez}, Aurora and {S{\'a}nchez-Gallego}, Jos{\'e} and {Tan}, Jonathan C. and {Tayar}, Jamie and {Thai}, Riley and {Thakar}, Ani and {Thibodeaux}, Pierre and {Ting}, Yuan-Sen and {Tkachenko}, Andrew and {Trakhtenbrot}, Benny and {Fern{\'a}ndez-Trincado}, Jos{\'e} G. and {Troup}, Nicholas and {Trump}, Jonathan R. and {Ulloa}, Natalie and {Van Der Marel}, Roeland P. and {Vera}, Pablo and {Villanova}, Sandro and {Villase{\~n}or}, Jaime I. and {Wang}, Ji and {Way}, Zachary and {Weijmans}, Anne-Marie and {Wheeler}, Adam and {Wilson}, John C. and {Wofford}, Aida and {Wong}, Tony},
        title = "{The Nineteenth Data Release of the Sloan Digital Sky Survey}",
      journal = {\apjs},
         year = 2026,
        month = jul,
       volume = {285},
       number = {1},
          eid = {9},
        pages = {9},
          doi = {10.3847/1538-4365/ae4697},
archivePrefix = {arXiv},
       eprint = {2507.07093},
 primaryClass = {astro-ph.GA},
       adsurl = {https://ui.adsabs.harvard.edu/abs/2026ApJS..285....9S}
}

@ARTICLE{Juneau_2014,
       author = {{Juneau}, St{\'e}phanie and {Bournaud}, Fr{\'e}d{\'e}ric and {Charlot}, St{\'e}phane and {Daddi}, Emanuele and {Elbaz}, David and {Trump}, Jonathan R. and {Brinchmann}, Jarle and {Dickinson}, Mark and {Duc}, Pierre-Alain and {Gobat}, Raphael and {Jean-Baptiste}, Ingrid and {Le Floc'h}, {\'E}meric and {Lehnert}, M.~D. and {Pacifici}, Camilla and {Pannella}, Maurilio and {Schreiber}, Corentin},
        title = "{Active Galactic Nuclei Emission Line Diagnostics and the Mass-Metallicity Relation up to Redshift z \raisebox{-0.5ex}\textasciitilde 2: The Impact of Selection Effects and Evolution}",
      journal = {\apj},
         year = 2014,
        month = jun,
       volume = {788},
       number = {1},
          eid = {88},
        pages = {88},
          doi = {10.1088/0004-637X/788/1/88},
archivePrefix = {arXiv},
       eprint = {1403.6832},
 primaryClass = {astro-ph.GA},
       adsurl = {https://ui.adsabs.harvard.edu/abs/2014ApJ...788...88J}
}

@ARTICLE{grogin_2011,
       author = {{Grogin}, Norman A. and {Kocevski}, Dale D. and {Faber}, S.~M. and {Ferguson}, Henry C. and {Koekemoer}, Anton M. and {Riess}, Adam G. and {Acquaviva}, Viviana and {Alexander}, David M. and {Almaini}, Omar and {Ashby}, Matthew L.~N. and {Barden}, Marco and {Bell}, Eric F. and {Bournaud}, Fr{\'e}d{\'e}ric and {Brown}, Thomas M. and {Caputi}, Karina I. and {Casertano}, Stefano and {Cassata}, Paolo and {Castellano}, Marco and {Challis}, Peter and {Chary}, Ranga-Ram and {Cheung}, Edmond and {Cirasuolo}, Michele and {Conselice}, Christopher J. and {Roshan Cooray}, Asantha and {Croton}, Darren J. and {Daddi}, Emanuele and {Dahlen}, Tomas and {Dav{\'e}}, Romeel and {de Mello}, Du{\'\i}lia F. and {Dekel}, Avishai and {Dickinson}, Mark and {Dolch}, Timothy and {Donley}, Jennifer L. and {Dunlop}, James S. and {Dutton}, Aaron A. and {Elbaz}, David and {Fazio}, Giovanni G. and {Filippenko}, Alexei V. and {Finkelstein}, Steven L. and {Fontana}, Adriano and {Gardner}, Jonathan P. and {Garnavich}, Peter M. and {Gawiser}, Eric and {Giavalisco}, Mauro and {Grazian}, Andrea and {Guo}, Yicheng and {Hathi}, Nimish P. and {H{\"a}ussler}, Boris and {Hopkins}, Philip F. and {Huang}, Jia-Sheng and {Huang}, Kuang-Han and {Jha}, Saurabh W. and {Kartaltepe}, Jeyhan S. and {Kirshner}, Robert P. and {Koo}, David C. and {Lai}, Kamson and {Lee}, Kyoung-Soo and {Li}, Weidong and {Lotz}, Jennifer M. and {Lucas}, Ray A. and {Madau}, Piero and {McCarthy}, Patrick J. and {McGrath}, Elizabeth J. and {McIntosh}, Daniel H. and {McLure}, Ross J. and {Mobasher}, Bahram and {Moustakas}, Leonidas A. and {Mozena}, Mark and {Nandra}, Kirpal and {Newman}, Jeffrey A. and {Niemi}, Sami-Matias and {Noeske}, Kai G. and {Papovich}, Casey J. and {Pentericci}, Laura and {Pope}, Alexandra and {Primack}, Joel R. and {Rajan}, Abhijith and {Ravindranath}, Swara and {Reddy}, Naveen A. and {Renzini}, Alvio and {Rix}, Hans-Walter and {Robaina}, Aday R. and {Rodney}, Steven A. and {Rosario}, David J. and {Rosati}, Piero and {Salimbeni}, Sara and {Scarlata}, Claudia and {Siana}, Brian and {Simard}, Luc and {Smidt}, Joseph and {Somerville}, Rachel S. and {Spinrad}, Hyron and {Straughn}, Amber N. and {Strolger}, Louis-Gregory and {Telford}, Olivia and {Teplitz}, Harry I. and {Trump}, Jonathan R. and {van der Wel}, Arjen and {Villforth}, Carolin and {Wechsler}, Risa H. and {Weiner}, Benjamin J. and {Wiklind}, Tommy and {Wild}, Vivienne and {Wilson}, Grant and {Wuyts}, Stijn and {Yan}, Hao-Jing and {Yun}, Min S.},
        title = "{CANDELS: The Cosmic Assembly Near-infrared Deep Extragalactic Legacy Survey}",
      journal = {\apjs},
         year = 2011,
        month = dec,
       volume = {197},
       number = {2},
          eid = {35},
        pages = {35},
          doi = {10.1088/0067-0049/197/2/35},
archivePrefix = {arXiv},
       eprint = {1105.3753},
 primaryClass = {astro-ph.CO},
       adsurl = {https://ui.adsabs.harvard.edu/abs/2011ApJS..197...35G}
}

@ARTICLE{astropy:2013,
       author = {{Astropy Collaboration} and {Robitaille}, Thomas P. and {Tollerud}, Erik J. and {Greenfield}, Perry and {Droettboom}, Michael and {Bray}, Erik and {Aldcroft}, Tom and {Davis}, Matt and {Ginsburg}, Adam and {Price-Whelan}, Adrian M. and {Kerzendorf}, Wolfgang E. and {Conley}, Alexander and {Crighton}, Neil and {Barbary}, Kyle and {Muna}, Demitri and {Ferguson}, Henry and {Grollier}, Fr{\'e}d{\'e}ric and {Parikh}, Madhura M. and {Nair}, Prasanth H. and {Unther}, Hans M. and {Deil}, Christoph and {Woillez}, Julien and {Conseil}, Simon and {Kramer}, Roban and {Turner}, James E.~H. and {Singer}, Leo and {Fox}, Ryan and {Weaver}, Benjamin A. and {Zabalza}, Victor and {Edwards}, Zachary I. and {Azalee Bostroem}, K. and {Burke}, D.~J. and {Casey}, Andrew R. and {Crawford}, Steven M. and {Dencheva}, Nadia and {Ely}, Justin and {Jenness}, Tim and {Labrie}, Kathleen and {Lim}, Pey Lian and {Pierfederici}, Francesco and {Pontzen}, Andrew and {Ptak}, Andy and {Refsdal}, Brian and {Servillat}, Mathieu and {Streicher}, Ole},
        title = "{Astropy: A community Python package for astronomy}",
      journal = {\aap},
         year = 2013,
        month = oct,
       volume = {558},
          eid = {A33},
        pages = {A33},
          doi = {10.1051/0004-6361/201322068},
archivePrefix = {arXiv},
       eprint = {1307.6212},
 primaryClass = {astro-ph.IM},
       adsurl = {https://ui.adsabs.harvard.edu/abs/2013A&A...558A..33A}
}

@ARTICLE{astropy:2018,
       author = {{Astropy Collaboration} and {Price-Whelan}, A.~M. and {Sip{\H{o}}cz}, B.~M. and {G{\"u}nther}, H.~M. and {Lim}, P.~L. and {Crawford}, S.~M. and {Conseil}, S. and {Shupe}, D.~L. and {Craig}, M.~W. and {Dencheva}, N. and {Ginsburg}, A. and {VanderPlas}, J.~T. and {Bradley}, L.~D. and {P{\'e}rez-Su{\'a}rez}, D. and {de Val-Borro}, M. and {Aldcroft}, T.~L. and {Cruz}, K.~L. and {Robitaille}, T.~P. and {Tollerud}, E.~J. and {Ardelean}, C. and {Babej}, T. and {Bach}, Y.~P. and {Bachetti}, M. and {Bakanov}, A.~V. and {Bamford}, S.~P. and {Barentsen}, G. and {Barmby}, P. and {Baumbach}, A. and {Berry}, K.~L. and {Biscani}, F. and {Boquien}, M. and {Bostroem}, K.~A. and {Bouma}, L.~G. and {Brammer}, G.~B. and {Bray}, E.~M. and {Breytenbach}, H. and {Buddelmeijer}, H. and {Burke}, D.~J. and {Calderone}, G. and {Cano Rodr{\'\i}guez}, J.~L. and {Cara}, M. and {Cardoso}, J.~V.~M. and {Cheedella}, S. and {Copin}, Y. and {Corrales}, L. and {Crichton}, D. and {D'Avella}, D. and {Deil}, C. and {Depagne}, {\'E}. and {Dietrich}, J.~P. and {Donath}, A. and {Droettboom}, M. and {Earl}, N. and {Erben}, T. and {Fabbro}, S. and {Ferreira}, L.~A. and {Finethy}, T. and {Fox}, R.~T. and {Garrison}, L.~H. and {Gibbons}, S.~L.~J. and {Goldstein}, D.~A. and {Gommers}, R. and {Greco}, J.~P. and {Greenfield}, P. and {Groener}, A.~M. and {Grollier}, F. and {Hagen}, A. and {Hirst}, P. and {Homeier}, D. and {Horton}, A.~J. and {Hosseinzadeh}, G. and {Hu}, L. and {Hunkeler}, J.~S. and {Ivezi{\'c}}, {\v{Z}}. and {Jain}, A. and {Jenness}, T. and {Kanarek}, G. and {Kendrew}, S. and {Kern}, N.~S. and {Kerzendorf}, W.~E. and {Khvalko}, A. and {King}, J. and {Kirkby}, D. and {Kulkarni}, A.~M. and {Kumar}, A. and {Lee}, A. and {Lenz}, D. and {Littlefair}, S.~P. and {Ma}, Z. and {Macleod}, D.~M. and {Mastropietro}, M. and {McCully}, C. and {Montagnac}, S. and {Morris}, B.~M. and {Mueller}, M. and {Mumford}, S.~J. and {Muna}, D. and {Murphy}, N.~A. and {Nelson}, S. and {Nguyen}, G.~H. and {Ninan}, J.~P. and {N{\"o}the}, M. and {Ogaz}, S. and {Oh}, S. and {Parejko}, J.~K. and {Parley}, N. and {Pascual}, S. and {Patil}, R. and {Patil}, A.~A. and {Plunkett}, A.~L. and {Prochaska}, J.~X. and {Rastogi}, T. and {Reddy Janga}, V. and {Sabater}, J. and {Sakurikar}, P. and {Seifert}, M. and {Sherbert}, L.~E. and {Sherwood-Taylor}, H. and {Shih}, A.~Y. and {Sick}, J. and {Silbiger}, M.~T. and {Singanamalla}, S. and {Singer}, L.~P. and {Sladen}, P.~H. and {Sooley}, K.~A. and {Sornarajah}, S. and {Streicher}, O. and {Teuben}, P. and {Thomas}, S.~W. and {Tremblay}, G.~R. and {Turner}, J.~E.~H. and {Terr{\'o}n}, V. and {van Kerkwijk}, M.~H. and {de la Vega}, A. and {Watkins}, L.~L. and {Weaver}, B.~A. and {Whitmore}, J.~B. and {Woillez}, J. and {Zabalza}, V. and {Astropy Contributors}},
        title = "{The Astropy Project: Building an Open-science Project and Status of the v2.0 Core Package}",
      journal = {\aj},
         year = 2018,
        month = sep,
       volume = {156},
       number = {3},
          eid = {123},
        pages = {123},
          doi = {10.3847/1538-3881/aabc4f},
archivePrefix = {arXiv},
       eprint = {1801.02634},
 primaryClass = {astro-ph.IM},
       adsurl = {https://ui.adsabs.harvard.edu/abs/2018AJ....156..123A}
}

@ARTICLE{astropy:2022,
       author = {{Astropy Collaboration} and {Price-Whelan}, Adrian M. and {Lim}, Pey Lian and {Earl}, Nicholas and {Starkman}, Nathaniel and {Bradley}, Larry and {Shupe}, David L. and {Patil}, Aarya A. and {Corrales}, Lia and {Brasseur}, C.~E. and {N{\"o}the}, Maximilian and {Donath}, Axel and {Tollerud}, Erik and {Morris}, Brett M. and {Ginsburg}, Adam and {Vaher}, Eero and {Weaver}, Benjamin A. and {Tocknell}, James and {Jamieson}, William and {van Kerkwijk}, Marten H. and {Robitaille}, Thomas P. and {Merry}, Bruce and {Bachetti}, Matteo and {G{\"u}nther}, H. Moritz and {Aldcroft}, Thomas L. and {Alvarado-Montes}, Jaime A. and {Archibald}, Anne M. and {B{\'o}di}, Attila and {Bapat}, Shreyas and {Barentsen}, Geert and {Baz{\'a}n}, Juanjo and {Biswas}, Manish and {Boquien}, M{\'e}d{\'e}ric and {Burke}, D.~J. and {Cara}, Daria and {Cara}, Mihai and {Conroy}, Kyle E. and {Conseil}, Simon and {Craig}, Matthew W. and {Cross}, Robert M. and {Cruz}, Kelle L. and {D'Eugenio}, Francesco and {Dencheva}, Nadia and {Devillepoix}, Hadrien A.~R. and {Dietrich}, J{\"o}rg P. and {Eigenbrot}, Arthur Davis and {Erben}, Thomas and {Ferreira}, Leonardo and {Foreman-Mackey}, Daniel and {Fox}, Ryan and {Freij}, Nabil and {Garg}, Suyog and {Geda}, Robel and {Glattly}, Lauren and {Gondhalekar}, Yash and {Gordon}, Karl D. and {Grant}, David and {Greenfield}, Perry and {Groener}, Austen M. and {Guest}, Steve and {Gurovich}, Sebastian and {Handberg}, Rasmus and {Hart}, Akeem and {Hatfield-Dodds}, Zac and {Homeier}, Derek and {Hosseinzadeh}, Griffin and {Jenness}, Tim and {Jones}, Craig K. and {Joseph}, Prajwel and {Kalmbach}, J. Bryce and {Karamehmetoglu}, Emir and {Ka{\l}uszy{\'n}ski}, Miko{\l}aj and {Kelley}, Michael S.~P. and {Kern}, Nicholas and {Kerzendorf}, Wolfgang E. and {Koch}, Eric W. and {Kulumani}, Shankar and {Lee}, Antony and {Ly}, Chun and {Ma}, Zhiyuan and {MacBride}, Conor and {Maljaars}, Jakob M. and {Muna}, Demitri and {Murphy}, N.~A. and {Norman}, Henrik and {O'Steen}, Richard and {Oman}, Kyle A. and {Pacifici}, Camilla and {Pascual}, Sergio and {Pascual-Granado}, J. and {Patil}, Rohit R. and {Perren}, Gabriel I. and {Pickering}, Timothy E. and {Rastogi}, Tanuj and {Roulston}, Benjamin R. and {Ryan}, Daniel F. and {Rykoff}, Eli S. and {Sabater}, Jose and {Sakurikar}, Parikshit and {Salgado}, Jes{\'u}s and {Sanghi}, Aniket and {Saunders}, Nicholas and {Savchenko}, Volodymyr and {Schwardt}, Ludwig and {Seifert-Eckert}, Michael and {Shih}, Albert Y. and {Jain}, Anany Shrey and {Shukla}, Gyanendra and {Sick}, Jonathan and {Simpson}, Chris and {Singanamalla}, Sudheesh and {Singer}, Leo P. and {Singhal}, Jaladh and {Sinha}, Manodeep and {Sip{\H{o}}cz}, Brigitta M. and {Spitler}, Lee R. and {Stansby}, David and {Streicher}, Ole and {{\v{S}}umak}, Jani and {Swinbank}, John D. and {Taranu}, Dan S. and {Tewary}, Nikita and {Tremblay}, Grant R. and {de Val-Borro}, Miguel and {Van Kooten}, Samuel J. and {Vasovi{\'c}}, Zlatan and {Verma}, Shresth and {de Miranda Cardoso}, Jos{\'e} Vin{\'\i}cius and {Williams}, Peter K.~G. and {Wilson}, Tom J. and {Winkel}, Benjamin and {Wood-Vasey}, W.~M. and {Xue}, Rui and {Yoachim}, Peter and {Zhang}, Chen and {Zonca}, Andrea and {Astropy Project Contributors}},
        title = "{The Astropy Project: Sustaining and Growing a Community-oriented Open-source Project and the Latest Major Release (v5.0) of the Core Package}",
      journal = {\apj},
         year = 2022,
        month = aug,
       volume = {935},
       number = {2},
          eid = {167},
        pages = {167},
          doi = {10.3847/1538-4357/ac7c74},
archivePrefix = {arXiv},
       eprint = {2206.14220},
 primaryClass = {astro-ph.IM},
       adsurl = {https://ui.adsabs.harvard.edu/abs/2022ApJ...935..167A}
}

@article{Bruzual_2003,
   title={Stellar population synthesis at the resolution of 2003},
   volume={344},
   ISSN={1365-2966},
   url={http://dx.doi.org/10.1046/j.1365-8711.2003.06897.x},
   DOI={10.1046/j.1365-8711.2003.06897.x},
   number={4},
   journal={\mnras},
   publisher={Oxford University Press (OUP)},
   author={Bruzual, G. and Charlot, S.},
   year={2003},
   month=Oct, pages={1000–1028} }

@INPROCEEDINGS{Cepa_2002,
       author = {{Casta{\~n}eda}, H. and {Cepa}, J. and {Gonz{\'a}lez}, J.~J. and {Bland-Hawthorn}, J.},
        title = "{The Instrument OSIRIS at the Gran Telescopio Canarias}",
    booktitle = {Galaxies: the Third Dimension},
         year = 2002,
       editor = {{Rosada}, Margarita and {Binette}, Luc and {Arias}, Lorena},
       series = {ASP Conference Series},
       volume = {282},
        month = jan,
        pages = {395},
       adsurl = {https://ui.adsabs.harvard.edu/abs/2002ASPC..282..395C}
}

@ARTICLE{llerena_2026,
       author = {{Llerena}, M. and {Pentericci}, L. and {Amor{\'\i}n}, R. and {Ferrara}, A. and {Dickinson}, M. and {Arevalo-Gonzalez}, F. and {Calabr{\`o}}, A. and {Napolitano}, L. and {Mascia}, S. and {Arrabal Haro}, P. and {Begley}, R. and {Cleri}, N.~J. and {Davis}, K. and {Hu}, W. and {Kartaltepe}, J.~S. and {Koekemoer}, A.~M. and {Lucas}, R.~A. and {McGrath}, E. and {McLeod}, D.~J. and {Papovich}, C. and {Stanton}, T.~M. and {Taylor}, A.~J. and {Tripodi}, R. and {Wang}, X. and {Yung}, L.~Y.~A.},
        title = "{Extreme equivalent-width-selected low-mass starbursts at z = 4 {\ensuremath{-}} 9: Insights into their role in cosmic reionization}",
      journal = {\aap},
         year = 2026,
        month = apr,
       volume = {708},
          eid = {A152},
        pages = {A152},
          doi = {10.1051/0004-6361/202557897},
archivePrefix = {arXiv},
       eprint = {2510.25647},
 primaryClass = {astro-ph.GA},
       adsurl = {https://ui.adsabs.harvard.edu/abs/2026A&A...708A.152L}
}

@ARTICLE{MMT_Tang_2019,
       author = {{Tang}, Mengtao and {Stark}, Daniel P. and {Chevallard}, Jacopo and {Charlot}, St{\'e}phane},
        title = "{MMT/MMIRS spectroscopy of z = 1.3 - 2.4 extreme [O III] emitters: implications for galaxies in the reionization era}",
      journal = {\mnras},
         year = 2019,
        month = oct,
       volume = {489},
       number = {2},
        pages = {2572-2594},
          doi = {10.1093/mnras/stz2236},
archivePrefix = {arXiv},
       eprint = {1809.09637},
 primaryClass = {astro-ph.GA},
       adsurl = {https://ui.adsabs.harvard.edu/abs/2019MNRAS.489.2572T}
}

@ARTICLE{Nedkova_2021,
       author = {{Nedkova}, Kalina V. and {H{\"a}u{\ss}ler}, Boris and {Marchesini}, Danilo and {Dimauro}, Paola and {Brammer}, Gabriel and {Eigenthaler}, Paul and {Feinstein}, Adina D. and {Ferguson}, Henry C. and {Huertas-Company}, Marc and {Johnston}, Evelyn J. and {Kado-Fong}, Erin and {Kartaltepe}, Jeyhan S. and {Labb{\'e}}, Ivo and {Lange-Vagle}, Daniel and {Martis}, Nicholas S. and {McGrath}, Elizabeth J. and {Muzzin}, Adam and {Oesch}, Pascal and {Ordenes-Brice{\~n}o}, Yasna and {Puzia}, Thomas and {Shipley}, Heath V. and {Simmons}, Brooke D. and {Skelton}, Rosalind E. and {Stefanon}, Mauro and {van der Wel}, Arjen and {Whitaker}, Katherine E.},
        title = "{Extending the evolution of the stellar mass-size relation at z {\ensuremath{\leq}} 2 to low stellar mass galaxies from HFF and CANDELS}",
      journal = {\mnras},
         year = 2021,
        month = sep,
       volume = {506},
       number = {1},
        pages = {928-956},
          doi = {10.1093/mnras/stab1744},
archivePrefix = {arXiv},
       eprint = {2106.07663},
 primaryClass = {astro-ph.GA},
       adsurl = {https://ui.adsabs.harvard.edu/abs/2021MNRAS.506..928N}
}

@ARTICLE{atlas_OH,
       author = {{Rousselot}, P. and {Lidman}, C. and {Cuby}, J. -G. and {Moreels}, G. and {Monnet}, G.},
        title = "{Night-sky spectral atlas of OH emission lines in the near-infrared}",
      journal = {\aap},
         year = 2000,
        month = feb,
       volume = {354},
        pages = {1134-1150},
       adsurl = {https://ui.adsabs.harvard.edu/abs/2000A&A...354.1134R}
}

@ARTICLE{Keck_Tang_2022,
       author = {{Tang}, Mengtao and {Stark}, Daniel P. and {Ellis}, Richard S.},
        title = "{Stellar populations and star formation histories of the most extreme [O III] emitters at z = 1.3 - 3.7}",
      journal = {\mnras},
         year = 2022,
        month = jul,
       volume = {513},
       number = {4},
        pages = {5211-5223},
          doi = {10.1093/mnras/stac1280},
archivePrefix = {arXiv},
       eprint = {2202.04142},
 primaryClass = {astro-ph.GA},
       adsurl = {https://ui.adsabs.harvard.edu/abs/2022MNRAS.513.5211T}
}

@article{Cairos_2001,
doi = {10.1086/322532},
url = {https://doi.org/10.1086/322532},
year = {2001},
month = {oct},
publisher = {},
volume = {136},
number = {2},
pages = {393},
author = {Cairós, Luz M. and Caon, Nicola and Vílchez, José M. and González-Pérez, José N. and Muñoz-Tuñón, Casiana},
title = {Multiband Analysis of a Sample of Blue Compact Dwarf Galaxies.
II. Spatially Resolved and
Integrated Photometry},
journal = {\apjs}
}

@ARTICLE{Zalesky_2025,
       author = {{Euclid Collaboration} and {Zalesky}, L. and {McPartland}, C.~J.~R. and {Weaver}, J.~R. and {Toft}, S. and {Sanders}, D.~B. and {Mobasher}, B. and {Suzuki}, N. and {Szapudi}, I. and {Valdes}, I. and {Murphree}, G. and {Chartab}, N. and {Allen}, N. and {Taamoli}, S. and {Barrow}, S.~W.~J. and {Ch{\'a}vez Ortiz}, O. and {Finkelstein}, S.~L. and {Gwyn}, S. and {Sawicki}, M. and {McCracken}, H.~J. and {Stern}, D. and {Dannerbauer}, H. and {Altieri}, B. and {Andreon}, S. and {Auricchio}, N. and {Baccigalupi}, C. and {Baldi}, M. and {Bardelli}, S. and {Bender}, R. and {Bodendorf}, C. and {Bonino}, D. and {Branchini}, E. and {Brescia}, M. and {Brinchmann}, J. and {Camera}, S. and {Capobianco}, V. and {Carbone}, C. and {Carretero}, J. and {Casas}, S. and {Castander}, F.~J. and {Castellano}, M. and {Castignani}, G. and {Cavuoti}, S. and {Cimatti}, A. and {Colodro-Conde}, C. and {Congedo}, G. and {Conselice}, C.~J. and {Conversi}, L. and {Copin}, Y. and {Corcione}, L. and {Courbin}, F. and {Courtois}, H.~M. and {Da Silva}, A. and {Degaudenzi}, H. and {De Lucia}, G. and {Di Giorgio}, A.~M. and {Dinis}, J. and {Dubath}, F. and {Duncan}, C.~A.~J. and {Dupac}, X. and {Dusini}, S. and {Farina}, M. and {Farrens}, S. and {Ferriol}, S. and {Fotopoulou}, S. and {Frailis}, M. and {Franceschi}, E. and {Galeotta}, S. and {Garilli}, B. and {Gillard}, W. and {Gillis}, B. and {Giocoli}, C. and {G{\'o}mez-Alvarez}, P. and {Grazian}, A. and {Grupp}, F. and {Haugan}, S.~V.~H. and {Hoekstra}, H. and {Holmes}, W. and {Hook}, I. and {Hormuth}, F. and {Hornstrup}, A. and {Hudelot}, P. and {Jahnke}, K. and {Joachimi}, B. and {Keih{\"a}nen}, E. and {Kermiche}, S. and {Kiessling}, A. and {Kilbinger}, M. and {Kubik}, B. and {Kuijken}, K. and {K{\"u}mmel}, M. and {Kunz}, M. and {Kurki-Suonio}, H. and {Laureijs}, R. and {Ligori}, S. and {Lilje}, P.~B. and {Lindholm}, V. and {Lloro}, I. and {Mainetti}, G. and {Maino}, D. and {Maiorano}, E. and {Mansutti}, O. and {Marggraf}, O. and {Markovic}, K. and {Martinelli}, M. and {Martinet}, N. and {Marulli}, F. and {Massey}, R. and {Maurogordato}, S. and {Mei}, S. and {Mellier}, Y. and {Meneghetti}, M. and {Merlin}, E. and {Meylan}, G. and {Moresco}, M. and {Moscardini}, L. and {Munari}, E. and {Neissner}, C. and {Niemi}, S.-M. and {Nightingale}, J.~W. and {Padilla}, C. and {Paltani}, S. and {Pasian}, F. and {Pedersen}, K. and {Percival}, W.~J. and {Pettorino}, V. and {Pires}, S. and {Polenta}, G. and {Poncet}, M. and {Popa}, L.~A. and {Pozzetti}, L. and {Raison}, F. and {Rebolo}, R. and {Renzi}, A. and {Rhodes}, J. and {Riccio}, G. and {Romelli}, E. and {Roncarelli}, M. and {Rossetti}, E. and {Saglia}, R. and {Sakr}, Z. and {Sapone}, D. and {Scaramella}, R. and {Schirmer}, M. and {Schneider}, P. and {Schrabback}, T. and {Secroun}, A. and {Sefusatti}, E. and {Seidel}, G. and {Serrano}, S. and {Sirignano}, C. and {Sirri}, G. and {Stanco}, L. and {Steinwagner}, J. and {Tallada-Cresp{\'\i}}, P. and {Teplitz}, H.~I. and {Tereno}, I. and {Toledo-Moreo}, R. and {Torradeflot}, F. and {Tutusaus}, I. and {Valentijn}, E.~A. and {Valenziano}, L. and {Vassallo}, T. and {Verdoes Kleijn}, G. and {Veropalumbo}, A. and {Wang}, Y. and {Weller}, J. and {Zamorani}, G. and {Zucca}, E. and {Bolzonella}, M. and {Boucaud}, A. and {Bozzo}, E. and {Burigana}, C. and {Di Ferdinando}, D. and {Vigo}, J.~A. Escartin and {Farinelli}, R. and {Gracia-Carpio}, J. and {Mauri}, N. and {Nucita}, A.~A. and {Scottez}, V. and {Tenti}, M. and {Viel}, M. and {Wiesmann}, M. and {Akrami}, Y. and {Allevato}, V. and {Anselmi}, S. and {Ballardini}, M. and {Bethermin}, M. and {Blanchard}, A. and {Blot}, L. and {Borgani}, S. and {Bruton}, S. and {Cabanac}, R. and {Calabro}, A. and {Cappi}, A. and {Carvalho}, C.~S. and {Castro}, T. and {Chambers}, K.~C. and {Chary}, R. and {Contarini}, S.},
        title = "{Euclid preparation: LXI. Cosmic Dawn Survey: 'Pre-launch' multiwavelength catalogues for Euclid Deep Field North and Euclid Deep Field Fornax}",
      journal = {\aap},
         year = 2025,
        month = mar,
       volume = {695},
          eid = {A229},
        pages = {A229},
          doi = {10.1051/0004-6361/202451857},
archivePrefix = {arXiv},
       eprint = {2408.05296},
 primaryClass = {astro-ph.GA},
       adsurl = {https://ui.adsabs.harvard.edu/abs/2025A&A...695A.229E}
}

@ARTICLE{pgpg_2024,
       author = {{P{\'e}rez-Gonz{\'a}lez}, Pablo G. and {Barro}, Guillermo and {Rieke}, George H. and {Lyu}, Jianwei and {Rieke}, Marcia and {Alberts}, Stacey and {Williams}, Christina C. and {Hainline}, Kevin and {Sun}, Fengwu and {Pusk{\'a}s}, D{\'a}vid and {Annunziatella}, Marianna and {Baker}, William M. and {Bunker}, Andrew J. and {Egami}, Eiichi and {Ji}, Zhiyuan and {Johnson}, Benjamin D. and {Robertson}, Brant and {Rodr{\'\i}guez Del Pino}, Bruno and {Rujopakarn}, Wiphu and {Shivaei}, Irene and {Tacchella}, Sandro and {Willmer}, Christopher N.~A. and {Willott}, Chris},
        title = "{What Is the Nature of Little Red Dots and what Is Not, MIRI SMILES Edition}",
      journal = {\apj},
         year = 2024,
        month = jun,
       volume = {968},
       number = {1},
          eid = {4},
        pages = {4},
          doi = {10.3847/1538-4357/ad38bb},
archivePrefix = {arXiv},
       eprint = {2401.08782},
 primaryClass = {astro-ph.GA},
       adsurl = {https://ui.adsabs.harvard.edu/abs/2024ApJ...968....4P}
}

@INPROCEEDINGS{mosaic_2026,
       author = {{Pell{\'o}}, Roser and {Puech}, Mathieu and {Rodrigues}, Myriam and {Sanchez-Janssen}, Rub{\'e}n. and {Dalton}, Gavin B. and {Augusteijn}, Thomas and {El Hadi}, Kacem and {Garc{\'\i}a-Vargas}, Mar{\'\i}sa L. and {Larrieu}, Marie and {Meoli}, Alessandro and {Schnetler}, Hermine and {Sola}, Alondra and {Stephan}, Jay and {Surace}, Christian and {de Frondat}, F{\'a}tima and {Bharmal}, Nazim A. and {Chapuis}, Diane and {Dohlen}, Kjetil and {Hottier}, Cl{\'e}ment and {Jaquet}, Marc and {Kelz}, Andreas and {Martin}, Laurent and {Mohamed}, Meghna and {Morris}, Tim and {P{\'e}rez-Calpena}, Ana and {Seifert}, Walter and {Xu}, Wenli and {Kaper}, Lex and {Gallego}, Jes{\'u}s and {Afonso}, Jose and {Barbuy}, Beatriz and {Contini}, Thierry and {Evans}, Chistopher and {Finoguenov}, Alexis and {Kassin}, Susan and {Matthee}, Jorryt and {Miller}, Christopher and {Ostlin}, G{\"o}ran and {Pentericci}, Laura and {Schaerer}, Daniel and {Steinmetz}, Matthias and {Ziegler}, Bodo and {Caillier}, Patrick and {Cortes}, Angela and {Mainieri}, Vincenzo and {Pfuhl}, Oliver and {Ramsay}, Suzanne and {Araujo}, Ricardo and {Brynnel}, Joar and {Butterley}, Tim and {Castilho}, Bruno and {Conselice}, Christopher J. and {Cvetojevic}, Nick and {Davison}, Christopher and {Dejonghe}, Julien and {Dessauges-Zavadsky}, Mirka and {Dupieux}, Michel and {Ferreira}, D{\'e}cio and {Flores}, H{\'e}ctor and {Franzetti}, Paolo and {Frey}, Steffen and {Gil de Paz}, Armando and {Gon{\c{c}}alves}, Thiago S. and {Guinouard}, Isabelle and {Hayes}, Matthew J. and {Haigron}, Regis and {Hirschmann}, Michaela and {Ivanisenko}, Yevgeniy and {Johnson}, Sean D. and {Kehrig}, Carol and {King}, Jennifer and {Kneib}, Jean-Paul and {Krajnovi{\'c}}, Davor and {Lanotte}, Audrey A. and {Laporte}, Nicolas and {Laporte}, Philippe and {Larsen}, Soren and {Lewis}, Ian and {Nardetto}, Nicolas and {Pancino}, Elena and {Pannetier}, Cyril and {Pieri}, Matthew M. and {Porquet}, Delphine and {Quirrenbach}, Andreas and {Rombach}, Maxime and {Thurneysen}, Markus and {Vergani}, Susanna and {Wildi}, Fran{\c{c}}ois and {Alvarez Moreno}, Fernando and {Artan}, Raziye and {Barr}, David and {Beaulieu}, Mathilde and {Besada}, Eva and {Bhatt}, Anjali and {Bik}, Arjan and {Bond}, Charlotte and {Bouri}, Mohamed and {Boy}, J{\'e}r{\'e}mie and {Bozier}, Alexandre and {Bramall}, David and {Brands}, Sarah and {Braulio}, Antonio and {Cabello}, Cristina and {Calero de Ory}, Marina and {Calvo}, Rocio and {Castillo Morales}, Africa and {Challita}, Zalpha and {Chittik}, Stephen and {Curto Maldonado}, Andr{\'e}s. and {Ducret}, Franck and {Dijkstra}, Elfi and {Elswijk}, Eddy and {Fasola}, Gilles and {Feiz}, Carmen and {Fialho}, Fabio and {Floriot}, Johan and {Fumana}, Marco and {Gabarra}, Louis and {Garcia}, Lia and {Garc{\'\i}a}, Oscar and {Gargiulo}, Adriana and {Gaudemard}, Julien and {Giannone}, Domenico and {Gill}, Polly and {Gimenez}, Jean-Luc and {G{\'o}mez-Gutierrez}, Alicia and {Gouvret}, Carole and {Guenther}, Alan and {Hahfauline}, Alexandre T. and {Harvey}, Douglas and {Ib{\'a}{\~n}ez}, Jos{\'e} Miguel and {Iglesias}, Jorge and {Ives}, Dereck and {Jouve}, Estelle and {King}, David and {Kuncarayakti}, Hanindyo and {Kunst}, Peter and {Kwast}, Sander and {Lagarde}, St{\'e}phane and {Leschinski}, Kieran and {Li}, Jiang-Tao and {Licausi}, Gianluca and {Ligori}, Sebastiano and {Lowe}, Adam and {Lopez Orozco}, Jos{\'e} Antonio and {Macintosh}, Mike and {Magan}, H{\'e}ctor and {Maldonado}, Manuel and {Marquart}, Thomas and {Martins}, Lucimara and {Melara}, Marisole and {Melinder}, Jens and {Molema}, Jeannet and {Montgomery}, David and {M{\'u}gica}, Emma and {Najarro}, Francisco and {Navarro}, Ramon and {Ottomani}, Antoine and {Pamplona}, Tony and {Parr-Burman}, Phil and {Pascual}, Sergio and {Pe{\~n}ataro}, Mar{\'\i}a. and {P{\'e}rez Grande}, Isabel and {Peterzon}, Jan Rinze and {Pichon}, Thomas and {Piqueras}, Javier and {Piskunov}, Nicolai and {Rodriguez Cardoso}, Ram{\'o}n. and {Rodriguez Venzal}, Sergio and {Romp}, Rick and {Rostami}, Hossein and {Royer}, Fr{\'e}d{\'e}ric and {Sablowski}, Daniel and {S{\'a}nchez}, Ainhoa and {S{\'a}nchez Blanco}, Ernesto and {Schallig}, Ellen and {Schmoll}, Jurgen and {Schwartz}, Noah and {Taburet}, Sylvestre and {Terrett}, David and {Torralbo}, Ignacio and {Tromp}, Niels and {Veredas}, Gerardo and {Yang}, Yanbin and {York}, Alec and {Zeilinger}, Werner},
        title = "{MOSAIC at ELT: overview and current status}",
    booktitle = {Ground-based and Airborne Instrumentation for Astronomy XI},
         year = 2026,
       editor = {{Vernet}, Jo{\"e}l. D.~R. and {Bryant}, Julia J. and {Motohara}, Kentaro},
       series = {SPIE Conference Series},
       volume = {14149},
        month = aug,
          eid = {1414917},
        pages = {1414917},
          doi = {10.1117/12.3101780},
       adsurl = {https://ui.adsabs.harvard.edu/abs/2026SPIE14149E..17P}
}

\begin{appendix}
\onecolumn

\section{Spectroscopic subsample: emission-line measurements and properties}

\begin{table}[h!]
    \caption{Emission-line fluxes and derived properties for the spectroscopic subsample.}
\centering
\footnotesize

\setlength{\tabcolsep}{4pt}
\begin{tabular}{cccccccccc}
\hline\hline
ID & \z{spec} & F(\Hbeta) & F(\OIIIline{4959}) & F(\OIIIline{5007}) & F(\Halpha) & log M$_*$/M$_\odot$& SFR(100Myr) & SFR(10Myr) & Facility/Instrument \\
           &            &    10$^{-18}$cgs     &   10$^{-18}$cgs        &  10$^{-18}$cgs   &   10$^{-18}$cgs           & [\massunit]      &   [\sfrunit]         & [\sfrunit]  &       \\
        (1)   &    (2)        &     (3)      & (4)  &  (5)  &   (6)  &  (7) &  (8) & (9)  &  (10) \\           
     \hline\hline
        11982 & 1.72 & 16.70(5.30) & - & 120.00(5.90) & - & 8.22(0.43) & 4.30 & 4.44(1.41) & HST/WFC3  \\
        12893 & 0.93 & - & - & - & 46.50(5.40) & 8.66(0.22) & 0.35 & 0.96(0.11) & HST/WFC3  \\
        21026 & 1.79 & 11.20(5.10) & - & 42.00(6.80) & - & 8.15(0.19) & 1.16 & 3.27(1.49) & HST/WFC3  \\
        25968 & 1.72 & 32.60(7.30) & - & 249.00(9.00) & - & 9.09(0.19) & 12.61 & 8.66(1.94) & HST/WFC3  \\
        7247 & 1.70 & 19.90(5.80) & - & 43.20(7.10) & - & 8.23(0.25) & 1.33 & 5.13(1.49) & HST/WFC3  \\
        9642 & 1.72 & 27.00(5.00) & - & 125.00(4.90) & - & 8.45(0.53) & 5.49 & 7.16(1.33) & HST/WFC3 \\
        26937 & 1.78 & 27.00(6.10) & - & 186.00(7.60) & - & 8.99(0.20) & 15.16 & 7.81(1.77) & HST/WFC3  \\
        31683 & 1.71 & 42.20(4.90) & - & 351.00(6.90) & - & 8.11(0.33) & 0.42 & 11.02(1.28) & HST/WFC3 \\
        3483 & 1.40 & <17.93(0.90) & 32.39(9.87) & 73.48(7.44) & 99.50(5.50) & 8.00(0.29) & 1.35 & 15.00(2.35) & MMT/MMIRS \\
        8850 & 1.67 & 52.44(3.90) & 99.55(1.54) & 298.64(4.62) & 159.38(10.63) & 6.35(0.53) & - & 15.16(3.04) & MMT/MMIRS \\
        9112 & 1.61 & <8.17(0.41) & <20.42(3.38) & 44.90(3.45) & 21.20(3.87) & 8.40(0.25) & 2.81 & 1.67(0.30) & MMT/MMIRS \\
        11705 & 1.61 & <20.39(1.54) & <32.36(1.58) & 58.13(6.19) & - & 7.99(0.25) & 1.32 & 4.64(0.35) & MMT/MMIRS \\
        24065 & 1.60 & 23.95(2.24) & 37.00(2.83) & 102.09(2.19) & 58.42(17.51) & 8.94(0.16) & 10.15 & 4.53(1.36) & MMT/MMIRS \\
        1948 & 1.69 & <15.81(0.72) & - & 44.39(3.77) & - & 8.13(0.46) & 1.84 & 4.01(0.18) & Keck/MOSFIRE  \\
        13912 & 1.63 & 28.31(2.07) & 32.06(0.92) & 96.19(2.75) & - & 8.81(0.17) & 7.25 & 6.59(0.48) & Keck/MOSFIRE  \\
        15001 & 1.68 & 8.41(0.36) & 4.09(0.44) & 6.89(1.86) & - & 7.60(0.45) & 0.99 & 2.13(0.09) & Keck/MOSFIRE \\
        26607 & 1.63 & 25.61(3.62) & <10.58(1.15) & 18.99(3.23) & 10.09(2.81) & 8.01(0.21) & 0.90 & 0.82(0.23) & Keck/MOSFIRE  \\
        19403 & 1.59 & - & - & 35.90(1.10) & - & 7.86(0.64) & 0.54 & - & JWST/NIRSpec  \\
        20900 & 1.59 & 2.42(1.37) & - & 27.00(1.44) & - & 8.00(0.37) & 0.85 & 0.54(0.31) & JWST/NIRSpec  \\
        31049 & 1.57 & 4.17(0.45) & - & 55.10(0.98) & - & 8.01(0.20) & 1.28 & 0.89(0.10) & JWST/NIRSpec  \\
        5883 &  1.10 & - & - & - & - & 7.82(0.28) & 1.32 & - & CANDELS  \\
        8926 &  1.10 & - & - & - & - & 8.27(0.39) & 2.22 & - & CANDELS  \\
        11143 & 0.80 & - & - & - & - & 8.57(0.08) & 3.13 & - & CANDELS  \\
        13282 & 0.77 & - & - & - & - & 8.55(0.37) & 3.38 & - & CANDELS  \\
        27409 & 1.10 & - & - & - & - & 8.12(0.35) & 1.41 & - & CANDELS  \\
        32856 & 0.96 & - & 1.08(0.33)  & 3.24(0.10) & - & 7.25(0.27) & 0.36 & - & GTC/OSIRIS+ \\
        34094 & 0.83 & - & - & 5.90(0.12) & -  & 7.25(0.29) & 0.28 & - & GTC/OSIRIS+ \\
         \\
     \hline\hline
    \end{tabular}

\tablefoot{Columns: (1) object ID; (2) spectroscopic redshift; (3) \Hbeta\ flux; (4) \OIIIline{4959} flux; (5) \OIIIline{5007} flux; (6) \Halpha\ flux; (7) stellar mass from SED fitting; (8) SFR averaged over 100~Myr from the SED fitting; (9) instantaneous SFR from the \citet{Kennicutt_1998} \Halpha\ calibration; (10) telescope and instrument. Fluxes are in cgs units of \fluxlinesunit. ``$<$'' indicates a $2\sigma$ upper limit.}
\label{tab:fluxes}
\end{table}

\begin{figure}[htbp]
\centering
\includegraphics[width=\textwidth]{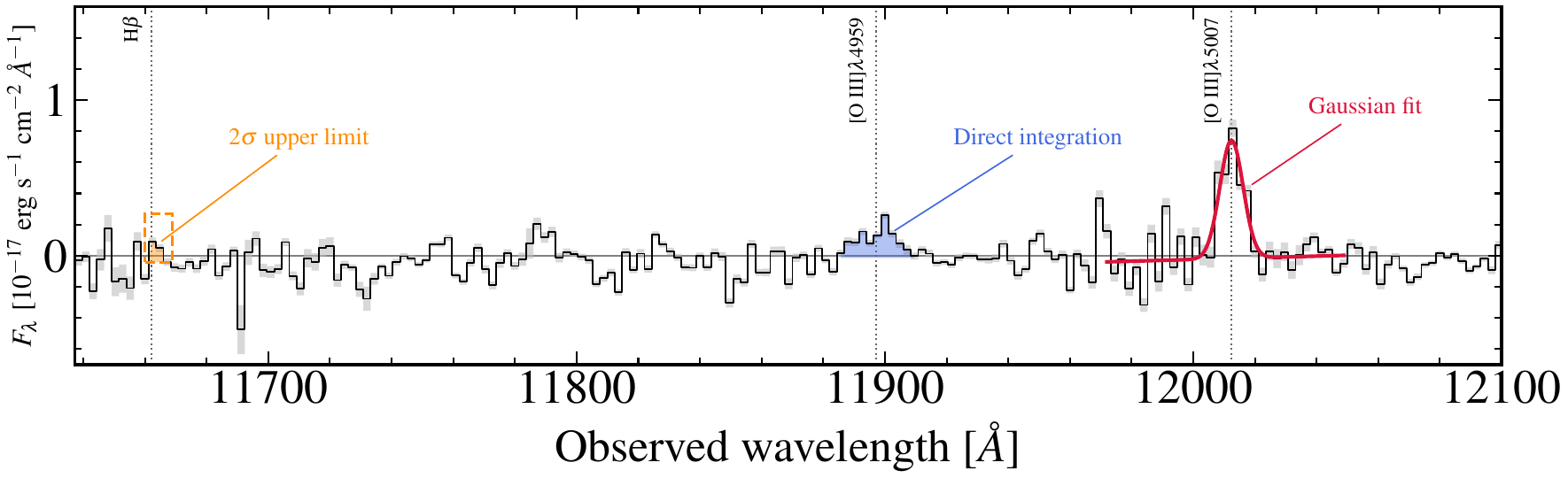}
\caption{Example of the three emission-line measurement regimes described in Sect.~\ref{sec:line_fitting_and_results}, applied to the spectrum of EGS-3483. The black line shows the observed spectrum, and the gray shaded region represents its $1\sigma$ uncertainty. Vertical dotted lines mark the expected positions of \Hbeta, \OIIIline{4959}, and \OIIIline{5007}. Lines with a peak S/N $> 5$ are measured via a Gaussian fit plus a flat continuum (red curve). Lines with a peak S/N between 2 and 5 ($2 < \text{S/N} < 5$) are measured through direct integration (blue shaded area). Finally, lines with a peak S/N $< 2$ are reported as $2\sigma$ upper limits (orange dashed rectangle, whose area is the estimated upper limit to the flux).}
\label{fitting_example}
\end{figure}

\begin{figure}[htbp]
\centering
\includegraphics[width=\textwidth]{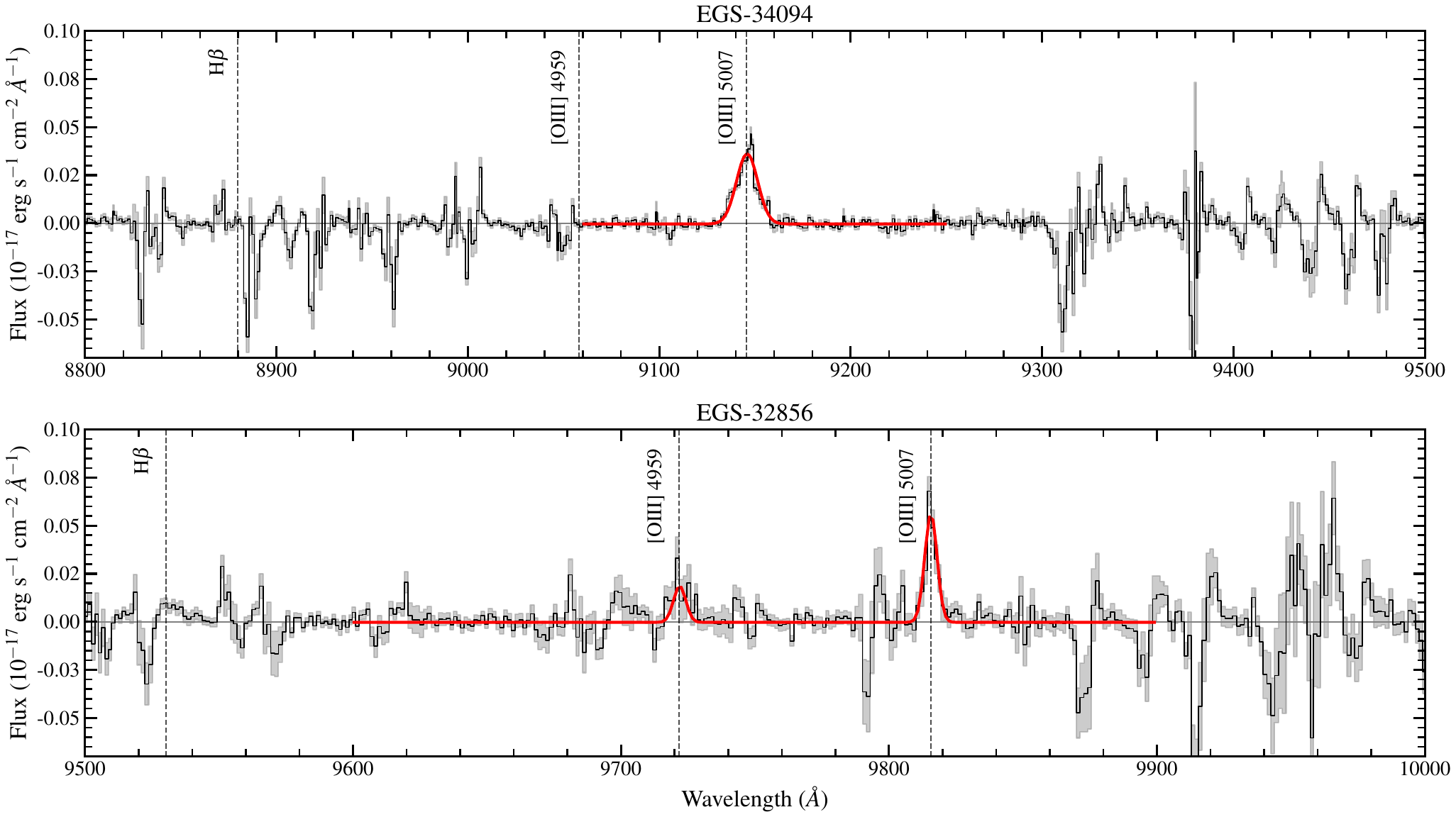}
\caption{Spectra of the two sources detected in our GTC/OSIRIS+ observing campaign. The dashed lines indicate the positions of the \Hbeta, \OIIIline{4959}, and \OIIIline{5007} lines. The gray region represents the $1\sigma$ uncertainty.}
\label{osiris_espectro}
\end{figure}

\section{Size-mass relation}

\begin{figure}[H]
\centering
\includegraphics[width=\textwidth]{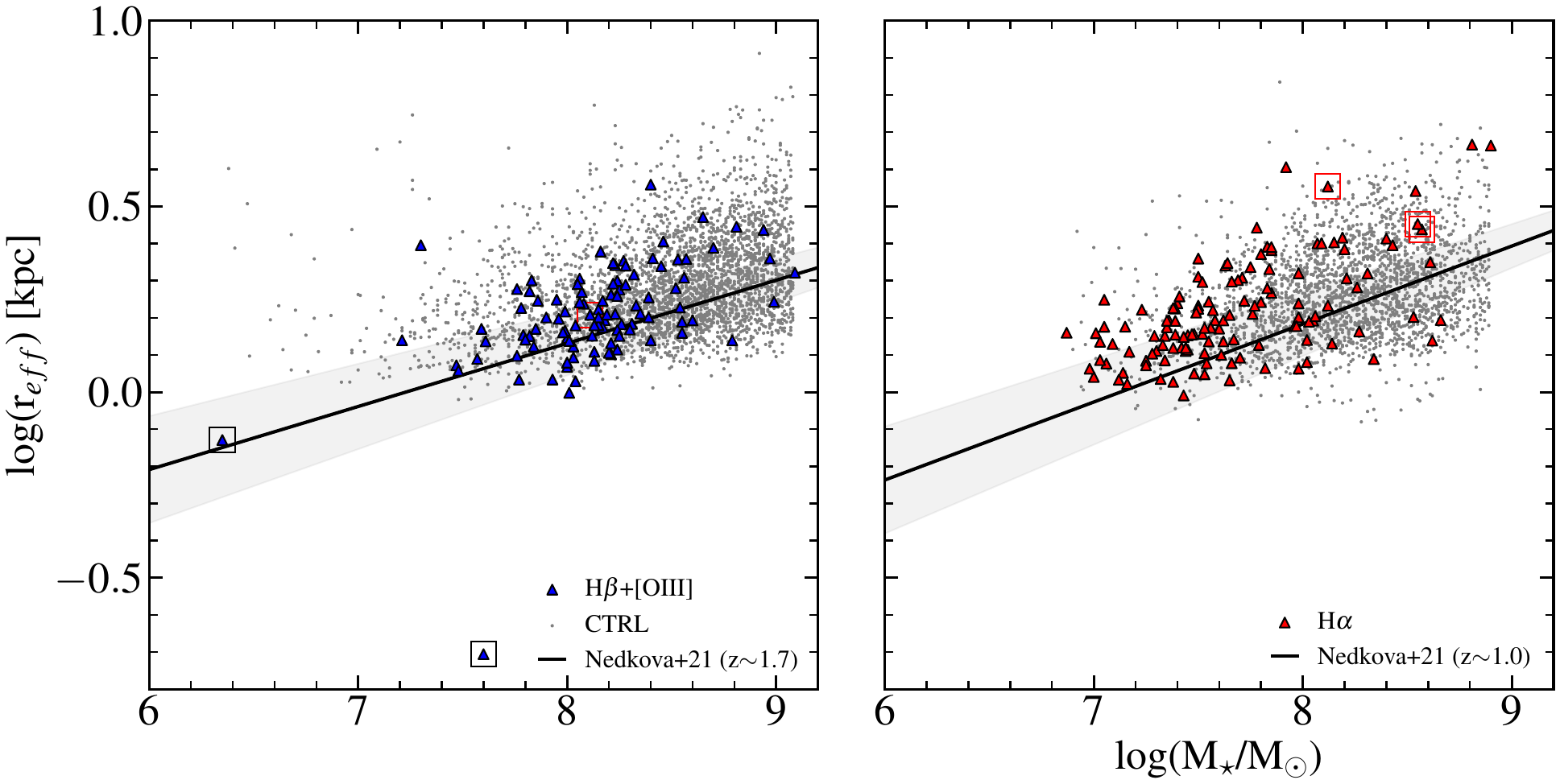}
\caption{Effective radius in the F160W band as a function of stellar mass for the \OIIIHbeta\ (blue) and \Halpha\ (red) samples, alongside a control sample (gray). The solid black lines represent the size-mass relations at $z = 1.0$ and $z = 1.7$ from \citet{Nedkova_2021}, with the shaded regions indicating the 3$\sigma$ scatter.}
\label{radios}
\end{figure}

\section{Morphology classification}

\begin{figure*}[!h]
\centering
\includegraphics[width=\textwidth]{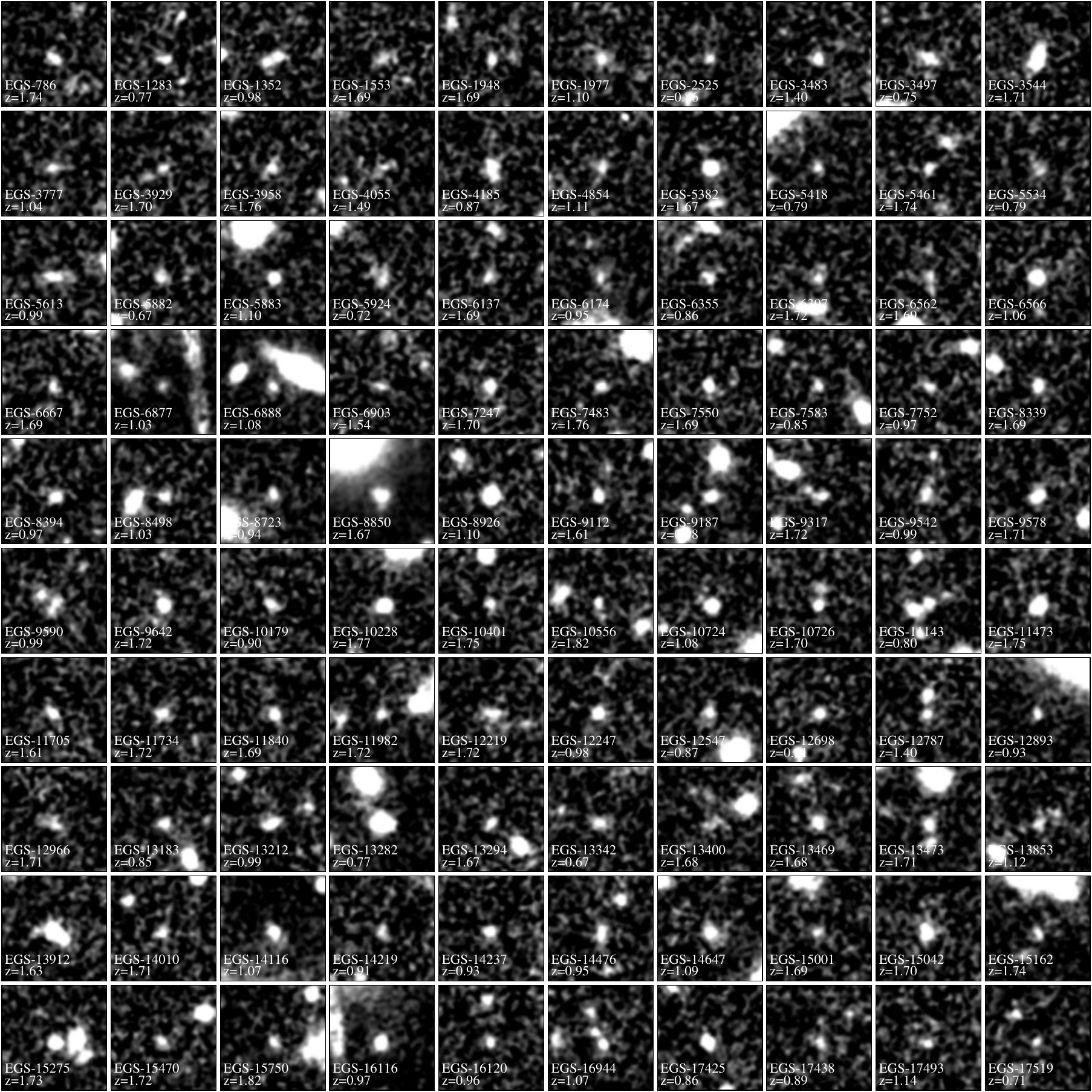}
\caption{Black-and-white image cutouts of the 100 brightest F160W EELG candidates in our sample from CANDELS\protect\footnotemark. Each image has a size of $4^{\prime\prime} \times 4^{\prime\prime}$.}
\label{mosaic}
\end{figure*}
\footnotetext{\url{https://archive.stsci.edu/hlsp/candels}}

\clearpage

\begin{longtable}{lccccl}

\caption{Morphological classification of the EELGs sample. Columns: (1) Object ID; (2) Right Ascension (RA); (3) Declination (Dec); (4) Redshift; (5) Classification based on the criteria of \citet{Huertas_Company_2015}; (6) Visual classification following \citet{Cairos_2001}.} \\

\hline
\hline
ID & R.A. & Dec. & z & Deep-learning   & Interaction/Structure \\
   & [deg] & [deg] &  &    & \\
(1) & (2) & (3) & (4) &  (5) & (6) \\
\hline
\hline

EGS-786 & 215.06694 & 52.90063 & 1.737 & Spheroid & Merger/Interacting \\
EGS-1283 & 215.21164 & 53.00604 & 0.767 & Spheroid & Round/Nucleated\\
EGS-1352 & 215.20414 & 53.00104 & 0.985 & Spheroid & Merger/Interacting\\
EGS-1553 & 215.23667 & 53.02542 & 1.692 & Spheroid & Clumpy/Chain \\
EGS-1948 & 215.19972 & 53.00182 & 1.687 & Spheroid & Round/Nucleated\\
EGS-1977 & 215.15554 & 52.97096 & 1.103 & Spheroid & Round/Nucleated\\
EGS-2525 & 214.78617 & 52.71203 & 0.861 & Spheroid & Round/Nucleated\\
EGS-3483 & 214.89949 & 52.79856 & 1.399 & Spheroid & Round/Nucleated\\
EGS-3497 & 214.78317 & 52.71564 & 0.750 & Spheroid & Round/Nucleated\\
EGS-3544 & 214.85027 & 52.76334 & 1.705 & Spheroid & Cometary/Tadpole\\
EGS-3777 & 214.91452 & 52.81101 & 1.042 & Spheroid & Cometary/Tadpole\\
EGS-3929 & 215.10634 & 52.94766 & 1.699 & Spheroid & Round/Nucleated\\
EGS-3958 & 214.88410 & 52.79037 & 1.763 & Spheroid & Merger/Interacting\\
EGS-4055 & 214.75382 & 52.69786 & 1.489 & Spheroid & Round/Nucleated\\
EGS-4185 & 215.14136 & 52.97363 & 0.871 & Spheroid & Clumpy/Chain\\
EGS-4854 & 215.27736 & 53.07294 & 1.108 & Spheroid & Round/Nucleated\\
EGS-5382 & 215.18611 & 53.01173 & 1.669 & Spheroid & Round/Nucleated\\
EGS-5418 & 214.92594 & 52.82833 & 0.791 & Spheroid & Round/Nucleated\\
EGS-5461 & 215.18465 & 53.01131 & 1.736 & Spheroid & Round/Nucleated\\
EGS-5534 & 215.17228 & 53.00304 & 0.786 & Spheroid & Clumpy/Chain\\
EGS-5613 & 215.05244 & 52.91894 & 0.986 & Spheroid & Cometary/Tadpole\\
EGS-5882 & 214.98093 & 52.86992 & 0.673 & Spheroid & Round/Nucleated\\
EGS-5883 & 215.19708 & 53.02227 & 1.103 & Spheroid & Round/Nucleated\\
EGS-5924 & 214.85992 & 52.78417 & 0.718 & Spheroid & Clumpy/Chain\\
EGS-6137 & 214.94451 & 52.84562 & 1.693 & Spheroid & Cometary/Tadpole\\
EGS-6174 & 215.25665 & 53.06586 & 0.954 & Spheroid & Round/Nucleated\\
EGS-6355 & 214.91574 & 52.82648 & 0.860 & Spheroid & Round/Nucleated\\
EGS-6397 & 214.87744 & 52.79943 & 1.717 & Spheroid & Clumpy/Chain\\
EGS-6562 & 215.14971 & 52.99304 & 1.691 & Spheroid & Merger/Interacting\\
EGS-6566 & 215.01249 & 52.89611 & 1.062 & Spheroid & Round/Nucleated\\
EGS-6667 & 214.94175 & 52.84669 & 1.692 & Spheroid & Round/Nucleated\\
EGS-6877 & 215.03828 & 52.91636 & 1.032 & Spheroid & Cometary/Tadpole\\
EGS-6888 & 215.08329 & 52.94811 & 1.080 & Spheroid & Round/Nucleated\\
EGS-6903 & 214.93980 & 52.84674 & 1.537 & Spheroid & Round/Nucleated\\
EGS-7247 & 214.89786 & 52.81903 & 1.697 & Spheroid & Round/Nucleated\\
EGS-7483 & 215.11099 & 52.97124 & 1.765 & Spheroid & Cometary/Tadpole\\
EGS-7550 & 215.01840 & 52.90634 & 1.691 & Spheroid & Cometary/Tadpole\\
EGS-7583 & 214.93359 & 52.84656 & 0.852 & Spheroid & Round/Nucleated\\
EGS-7752 & 215.13083 & 52.98685 & 0.969 & Spheroid & Round/Nucleated\\
EGS-8339 & 215.24997 & 53.07380 & 1.688 & Spheroid & Merger/Interacting\\
EGS-8394 & 215.18495 & 53.02854 & 0.972 & Spheroid & Round/Nucleated\\
EGS-8498 & 215.17979 & 53.02546 & 1.035 & Spheroid & Merger/Interacting\\
EGS-8723 & 214.78826 & 52.74943 & 0.941 & Spheroid & Cometary/Tadpole\\
EGS-8850 & 214.86161 & 52.80122 & 1.672 & Irregular & Clumpy/Chain\\
EGS-8926 & 215.07091 & 52.95085 & 1.097 & Spheroid & Round/Nucleated \\
EGS-9112 & 214.89088 & 52.82458 & 1.607 & Spheroid & Round/Nucleated\\
EGS-9187 & 215.16012 & 53.01506 & 0.977 & Spheroid & Merger/Interacting\\
EGS-9317 & 215.10624 & 52.97807 & 1.719 & Spheroid & Merger/Interacting\\
EGS-9542 & 214.99315 & 52.89940 & 0.989 & Spheroid & Round/Nucleated\\
EGS-9578 & 215.04747 & 52.93793 & 1.707 & Spheroid & Clumpy/Chain\\
EGS-9590 & 214.80226 & 52.76388 & 0.993 & Spheroid & Merger/Interacting\\
EGS-9642 & 214.75074 & 52.72731 & 1.717 & Spheroid & Round/Nucleated\\
EGS-10179 & 214.85611 & 52.80527 & 0.899 & Spheroid & Round/Nucleated\\
EGS-10228 & 215.16608 & 53.02457 & 1.771 & Spheroid & Round/Nucleated \\
EGS-10401 & 214.78156 & 52.75326 & 1.753 & Spheroid & Round/Nucleated\\
EGS-10556 & 214.93222 & 52.86130 & 1.818 & Spheroid & Merger/Interacting\\
EGS-10724 & 215.13070 & 53.00236 & 1.080 & Spheroid & Round/Nucleated\\
EGS-10726 & 215.02394 & 52.92722 & 1.698 & Spheroid & Round/Nucleated\\
EGS-11143 & 214.85395 & 52.80870 & 0.797 & Spheroid & Merger/Interacting\\
EGS-11473 & 215.17702 & 53.03878 & 1.750 & Spheroid & Clumpy/Chain\\
EGS-11705 & 214.86224 & 52.81743 & 1.613 & Spheroid & Round/Nucleated\\
EGS-11734 & 214.80649 & 52.77783 & 1.717 & Spheroid & Round/Nucleated\\
EGS-11840 & 215.18995 & 53.04960 & 1.686 & Spheroid & Round/Nucleated \\
EGS-11982 & 214.82243 & 52.79041 & 1.720 & Spheroid & Round/Nucleated\\
EGS-12219 & 215.18163 & 53.04574 & 1.718 & Spheroid & Merger/Interacting\\
EGS-12247 & 214.92320 & 52.86343 & 0.978 & Spheroid & Round/Nucleated\\
EGS-12547 & 214.99217 & 52.91418 & 0.874 & Spheroid & Round/Nucleated\\
EGS-12698 & 215.06736 & 52.96816 & 0.905 & Spheroid & Round/Nucleated\\
EGS-12787 & 214.83973 & 52.80714 & 1.395 & Spheroid & Merger/Interacting\\
EGS-12893 & 214.74000 & 52.73652 & 0.935 & Spheroid & Round/Nucleated\\
EGS-12966 & 214.82262 & 52.79594 & 1.712 & Spheroid & Cometary/Tadpole\\
EGS-13183 & 215.04508 & 52.95492 & 0.848 & Spheroid & Round/Nucleated\\
EGS-13212 & 214.85578 & 52.82063 & 0.988 & Spheroid & Merger/Interacting\\
EGS-13282 & 214.76335 & 52.75471 & 0.773 & Irregular & Clumpy/Chain\\
EGS-13294 & 214.67801 & 52.69416 & 1.674 & Spheroid & Merger/Interacting\\
EGS-13342 & 214.80427 & 52.78485 & 0.669 & Spheroid & Cometary/Tadpole\\
EGS-13400 & 215.00962 & 52.93108 & 1.679 & Spheroid & Merger/Interacting\\
EGS-13469 & 214.98478 & 52.91384 & 1.675 & Spheroid & Round/Nucleated\\
EGS-13473 & 215.15147 & 53.03149 & 1.705 & Spheroid & Merger/Interacting\\
EGS-13853 & 215.09894 & 52.99657 & 1.116 & Spheroid & Round/Nucleated\\
EGS-13912 & 215.10723 & 53.00235 & 1.628 & Spheroid & Cometary/Tadpole\\
EGS-14010 & 214.96661 & 52.90388 & 1.710 & Spheroid & Round/Nucleated\\
EGS-14116 & 215.18836 & 53.06060 & 1.067 & Spheroid & Round/Nucleated\\
EGS-14219 & 215.03573 & 52.95389 & 0.908 & Spheroid & Round/Nucleated\\
EGS-14237 & 214.95304 & 52.89548 & 0.928 & Spheroid & Round/Nucleated\\
EGS-14476 & 215.05739 & 52.97053 & 0.953 & Spheroid & Merger/Interacting\\
EGS-14647 & 215.08346 & 52.98977 & 1.089 & Spheroid & Clumpy/Chain\\
EGS-15001 & 214.98754 & 52.92417 & 1.685 & Spheroid & Clumpy/Chain\\
EGS-15042 & 214.92908 & 52.88272 & 1.704 & Spheroid & Clumpy/Chain\\
EGS-15162 & 215.03757 & 52.96042 & 1.743 & Spheroid & Round/Nucleated\\
EGS-15275 & 215.20163 & 53.07640 & 1.730 & Spheroid & Merger/Interacting\\
EGS-15470 & 214.82452 & 52.81085 & 1.717 & Spheroid & Round/Nucleated\\
EGS-15750 & 215.09927 & 53.00691 & 1.816 & Spheroid & Round/Nucleated\\
EGS-16116 & 214.83154 & 52.81839 & 0.973 & Disk     & Round/Nucleated\\
EGS-16120 & 214.87891 & 52.85272 & 0.965 & Spheroid & Round/Nucleated\\
EGS-16944 & 215.13000 & 53.03451 & 1.075 & Spheroid & Merger/Interacting\\
EGS-17425 & 214.93339 & 52.89829 & 0.864 & Spheroid & Clumpy/Chain\\
EGS-17438 & 215.06519 & 52.99153 & 0.885 & Spheroid & Round/Nucleated\\
EGS-17493 & 214.85344 & 52.84186 & 1.143 & Spheroid & Round/Nucleated\\
EGS-17519 & 214.84829 & 52.83840 & 0.708 & Spheroid & Round/Nucleated\\
EGS-17596 & 215.05567 & 52.98578 & 1.572 & Spheroid & Round/Nucleated\\
EGS-17794 & 215.03062 & 52.96904 & 1.755 & Spheroid & Merger/Interacting\\
EGS-17929 & 215.12737 & 53.03784 & 1.717 & Spheroid & Merger/Interacting\\
EGS-18000 & 214.97894 & 52.93354 & 0.990 & Spheroid & Round/Nucleated\\
EGS-18325 & 215.06920 & 52.99900 & 0.808 & Spheroid & Merger/Interacting\\
EGS-18405 & 214.95156 & 52.91627 & 0.851 & Spheroid & Merger/Interacting\\
EGS-18496 & 214.85145 & 52.84559 & 1.080 & Spheroid & Round/Nucleated\\
EGS-18510 & 215.15926 & 53.06320 & 1.721 & Spheroid & Round/Nucleated\\
EGS-18938 & 214.89874 & 52.88130 & 0.736 & Spheroid & Merger/Interacting\\
EGS-19049 & 215.05898 & 52.99499 & 0.863 & Spheroid & Round/Nucleated\\
EGS-19289 & 214.90368 & 52.88665 & 0.931 & Spheroid & Round/Nucleated\\
EGS-19403 & 215.02253 & 52.97134 & 1.593 & Spheroid & Round/Nucleated\\
EGS-20107 & 214.99384 & 52.95456 & 1.062 & Spheroid & Round/Nucleated\\
EGS-20318 & 215.11687 & 53.04251 & 0.999 & Spheroid & Round/Nucleated\\
EGS-20470 & 214.74351 & 52.77830 & 1.031 & Spheroid & Round/Nucleated\\
EGS-20483 & 215.07997 & 53.01719 & 0.940 & Spheroid & Round/Nucleated\\
EGS-20574 & 214.72416 & 52.76517 & 1.722 & Spheroid & Merger/Interacting\\
EGS-20675 & 214.70519 & 52.75205 & 1.008 & Spheroid & Cometary/Tadpole\\
EGS-20879 & 214.79635 & 52.81822 & 1.058 & Spheroid & Merger/Interacting\\
EGS-20889 & 214.97894 & 52.94798 & 1.704 & Spheroid & Round/Nucleated\\
EGS-20900 & 215.07337 & 53.01465 & 1.578 & Spheroid & Round/Nucleated\\
EGS-21026 & 214.96081 & 52.93577 & 1.790 & Spheroid & Round/Nucleated\\
EGS-21853 & 215.06533 & 53.01382 & 1.898 & Spheroid & Round/Nucleated\\
EGS-21919 & 214.84480 & 52.85793 & 1.708 & Spheroid & Clumpy/Chain\\
EGS-21972 & 215.13222 & 53.06136 & 1.013 & Spheroid & Round/Nucleated\\
EGS-22600 & 215.08770 & 53.03317 & 0.977 & Spheroid & Round/Nucleated\\
EGS-22870 & 215.02367 & 52.98964 & 0.734 & Spheroid & Round/Nucleated\\
EGS-22874 & 214.81003 & 52.83819 & 0.976 & Spheroid & Round/Nucleated\\
EGS-22961 & 214.93847 & 52.92990 & 1.003 & Spheroid & Merger/Interacting\\
EGS-22988 & 214.92297 & 52.91875 & 0.981 & Spheroid & Cometary/Tadpole\\
EGS-23042 & 214.73932 & 52.78863 & 1.731 & Spheroid & Clumpy/Chain\\
EGS-23492 & 215.10521 & 53.05033 & 1.012 & Spheroid & Round/Nucleated\\
EGS-23688 & 215.10647 & 53.05208 & 1.701 & Spheroid & Clumpy/Chain\\
EGS-23876 & 215.00968 & 52.98505 & 0.952 & Spheroid & Cometary/Tadpole\\
EGS-23918 & 214.88883 & 52.89969 & 1.726 & Spheroid & Cometary/Tadpole\\
EGS-23969 & 215.13970 & 53.07716 & 0.997 & Spheroid & Round/Nucleated\\
EGS-24065 & 214.82275 & 52.85274 & 1.597 & Disk     & Clumpy/Chain\\
EGS-24178 & 215.13529 & 53.07520 & 1.701 & Spheroid & Round/Nucleated\\
EGS-24330 & 214.74473 & 52.79935 & 1.718 & Spheroid & Merger/Interacting\\
EGS-24376 & 214.97370 & 52.96219 & 1.735 & Spheroid & Merger/Interacting\\
EGS-24656 & 215.12202 & 53.06807 & 1.071 & Spheroid & Merger/Interacting\\
EGS-24744 & 214.86496 & 52.88693 & 1.098 & Spheroid & Round/Nucleated\\
EGS-24745 & 214.69628 & 52.76660 & 1.084 & Spheroid & Round/Nucleated\\
EGS-24857 & 215.14431 & 53.08482 & 0.672 & Spheroid & Round/Nucleated\\
EGS-24910 & 214.73762 & 52.79702 & 0.858 & Spheroid & Round/Nucleated\\
EGS-25082 & 215.06394 & 53.02946 & 0.993 & Spheroid & Round/Nucleated\\
EGS-25156 & 214.74353 & 52.80256 & 1.546 & Spheroid & Merger/Interacting\\
EGS-25332 & 215.12239 & 53.07190 & 1.707 & Spheroid & Round/Nucleated\\
EGS-25403 & 214.94293 & 52.94567 & 1.788 & Spheroid & Cometary/Tadpole\\
EGS-25731 & 214.80946 & 52.85263 & 0.680 & Spheroid & Round/Nucleated\\
EGS-25877 & 215.00684 & 52.99352 & 1.716 & Spheroid & Round/Nucleated\\
EGS-25968 & 214.99040 & 52.98193 & 1.718 & Disk     & Round/Nucleated\\
EGS-26066 & 214.91799 & 52.93156 & 1.144 & Spheroid & Merger/Interacting\\
EGS-26088 & 215.01362 & 52.99929 & 1.753 & Spheroid & Merger/Interacting\\
EGS-26163 & 214.95035 & 52.95495 & 1.017 & Spheroid & Clumpy/Chain\\
EGS-26607 & 214.99671 & 52.98996 & 1.629 & Spheroid & Round/Nucleated\\
EGS-26690 & 215.11102 & 53.07083 & 1.703 & Spheroid & Merger/Interacting\\
EGS-26863 & 214.74625 & 52.81353 & 0.786 & Spheroid & Round/Nucleated\\
EGS-26937 & 214.97429 & 52.97534 & 1.779 & Disk     & Merger/Interacting\\
EGS-26972 & 214.72241 & 52.79699 & 0.714 & Spheroid & Round/Nucleated\\
EGS-27121 & 214.84116 & 52.88241 & 0.791 & Spheroid & Merger/Interacting\\
EGS-27130 & 215.03138 & 53.01716 & 1.748 & Spheroid & Merger/Interacting\\
EGS-27409 & 214.74728 & 52.81672 & 1.103 & Spheroid & Cometary/Tadpole\\
EGS-27531 & 214.97163 & 52.97711 & 1.728 & Spheroid & Round/Nucleated\\
EGS-29006 & 214.63382 & 52.75316 & 1.080 & Spheroid & Round/Nucleated\\
EGS-29469 & 214.70855 & 52.79390 & 1.677 & Spheroid & Round/Nucleated\\
EGS-29523 & 214.94654 & 52.97027 & 0.923 & Spheroid & Round/Nucleated\\
EGS-29657 & 214.77453 & 52.84822 & 0.998 & Spheroid & Merger/Interacting\\
EGS-29694 & 214.97821 & 52.99470 & 1.730 & Spheroid & Cometary/Tadpole\\
EGS-29796 & 214.88733 & 52.92822 & 0.987 & Spheroid & Cometary/Tadpole\\
EGS-29809 & 214.71704 & 52.80901 & 0.939 & Spheroid & Round/Nucleated\\
EGS-30120 & 214.99164 & 53.00120 & 1.764 & Spheroid & Cometary/Tadpole\\
EGS-30213 & 214.95057 & 52.97010 & 0.911 & Spheroid & Cometary/Tadpole\\
EGS-30349 & 214.84182 & 52.89323 & 0.693 & Spheroid & Merger/Interacting\\
EGS-30351 & 214.74700 & 52.82871 & 0.899 & Spheroid & Merger/Interacting\\
EGS-30444 & 214.91361 & 52.94676 & 1.736 & Spheroid & Merger/Interacting\\
EGS-30509 & 215.05090 & 53.04325 & 1.116 & Spheroid & Round/Nucleated\\
EGS-30797 & 215.01221 & 53.01412 & 1.706 & Spheroid & Merger/Interacting\\
EGS-30951 & 214.72515 & 52.80622 & 1.712 & Spheroid & Round/Nucleated\\
EGS-31049 & 214.89323 & 52.92736 & 1.573 & Spheroid & Round/Nucleated\\
EGS-31098 & 215.13213 & 53.09565 & 0.857 & Spheroid & Round/Nucleated\\
EGS-31100 & 214.78450 & 52.84855 & 1.025 & Spheroid & Round/Nucleated\\
EGS-31366 & 215.00665 & 53.00546 & 0.843 & Spheroid & Round/Nucleated\\
EGS-31368 & 214.86321 & 52.90492 & 0.949 & Spheroid & Round/Nucleated\\
EGS-31683 & 214.99043 & 52.99759 & 1.706 & Spheroid & Merger/Interacting\\
EGS-32273 & 215.05281 & 52.89567 & 0.964 & Spheroid & Merger/Interacting\\
EGS-32276 & 214.92135 & 52.80252 & 1.743 & Spheroid & Merger/Interacting\\
EGS-32387 & 214.95309 & 52.82755 & 0.679 & Spheroid & Round/Nucleated\\
EGS-32503 & 215.06361 & 52.90833 & 1.723 & Spheroid & Round/Nucleated\\
EGS-32584 & 215.15772 & 52.97596 & 0.948 & Spheroid & Round/Nucleated\\
EGS-32631 & 214.85345 & 52.76127 & 0.774 & Spheroid & Merger/Interacting\\
EGS-32680 & 214.97412 & 52.84806 & 1.700 & Spheroid & Merger/Interacting\\
EGS-32856 & 214.92269 & 52.81529 & 0.961 & Spheroid & Round/Nucleated\\
EGS-33058 & 214.99928 & 52.87307 & 1.745 & Spheroid & Round/Nucleated\\
EGS-33143 & 214.76499 & 52.70773 & 0.755 & Spheroid & Round/Nucleated\\
EGS-33312 & 215.14655 & 52.98084 & 0.955 & Spheroid & Round/Nucleated\\
EGS-33332 & 214.90886 & 52.81323 & 1.718 & Spheroid & Round/Nucleated\\
EGS-33355 & 215.20460 & 53.02249 & 1.707 & Spheroid & Clumpy/Chain\\
EGS-33562 & 214.84545 & 52.77190 & 0.817 & Spheroid & Merger/Interacting\\
EGS-33613 & 215.15219 & 52.99022 & 1.073 & Spheroid & Merger/Interacting\\
EGS-33866 & 214.90005 & 52.81633 & 0.925 & Spheroid & Round/Nucleated\\
EGS-34094 & 214.87629 & 52.80369 & 0.827 & Spheroid & Round/Nucleated\\
EGS-34998 & 215.16762 & 53.02664 & 0.856 & Spheroid & Round/Nucleated\\
EGS-35029 & 215.09818 & 52.97828 & 0.938 & Spheroid & Merger/Interacting\\
EGS-35263 & 215.04234 & 52.94301 & 1.743 & Spheroid & Round/Nucleated\\
EGS-35539 & 215.15075 & 53.02425 & 1.743 & Spheroid & Round/Nucleated\\
EGS-35697 & 214.88399 & 52.83834 & 1.703 & Spheroid & Round/Nucleated\\
EGS-35941 & 215.10953 & 53.00218 & 0.763 & Spheroid & Round/Nucleated\\
EGS-36392 & 215.11869 & 53.01596 & 0.807 & Spheroid & Cometary/Tadpole\\
EGS-36453 & 215.20436 & 53.07720 & 1.691 & Spheroid & Round/Nucleated\\
EGS-36485 & 214.98050 & 52.92007 & 0.852 & Spheroid & Merger/Interacting\\
EGS-36763 & 214.91109 & 52.87651 & 0.758 & Spheroid & Round/Nucleated\\
EGS-36806 & 215.16088 & 53.05354 & 0.906 & Spheroid & Cometary/Tadpole\\
EGS-36854 & 215.08215 & 52.99873 & 1.706 & Spheroid & Merger/Interacting\\
EGS-36913 & 214.78062 & 52.78595 & 0.885 & Spheroid & Round/Nucleated\\
EGS-37253 & 214.78543 & 52.79533 & 1.767 & Spheroid & Merger/Interacting\\
EGS-37416 & 215.14961 & 53.05580 & 1.705 & Spheroid & Clumpy/Chain\\
EGS-37622 & 215.14482 & 53.05584 & 1.713 & Spheroid & Round/Nucleated\\
EGS-37750 & 214.91420 & 52.89507 & 1.017 & Spheroid & Round/Nucleated\\
EGS-37956 & 214.96486 & 52.93477 & 0.777 & Spheroid & Round/Nucleated\\
EGS-37964 & 214.94900 & 52.92368 & 0.810 & Spheroid & Round/Nucleated\\
EGS-38207 & 215.02109 & 52.97858 & 1.598 & Spheroid & Merger/Interacting\\
EGS-38830 & 214.86511 & 52.87920 & 0.706 & Spheroid & Round/Nucleated\\
EGS-39021 & 214.98125 & 52.96504 & 1.748 & Spheroid & Round/Nucleated\\
EGS-39184 & 215.03866 & 53.00814 & 1.021 & Spheroid & Cometary/Tadpole\\
EGS-39211 & 214.73329 & 52.79179 & 0.913 & Spheroid & Cometary/Tadpole\\
EGS-39692 & 215.03580 & 53.01449 & 1.676 & Spheroid & Round/Nucleated\\
EGS-39781 & 214.89020 & 52.91300 & 0.840 & Spheroid & Round/Nucleated\\
EGS-41324 & 214.77685 & 52.84846 & 0.763 & Spheroid & Cometary/Tadpole\\
\hline
\hline

\label{tab:morfologia}
\end{longtable}

\end{appendix}

\end{document}